\documentclass[12pt,a4paper]{article} 
\usepackage[colorlinks]{hyperref}
\usepackage{amsmath}
\usepackage{amsthm}
\usepackage{amsfonts}
\usepackage{amssymb}
\usepackage{mathabx}
\usepackage{bussproofs}
\usepackage{lineno}
\theoremstyle{plain} 
\newtheorem{thm}{Theorem}[section]

\usepackage{graphics}
\usepackage{mathrsfs}
\usepackage{booktabs,array}\newcolumntype{P}[1]{>{\raggedright\arraybackslash}p{\dimexpr#1\linewidth-2\tabcolsep}}

\newtheorem{lem}[thm]{Lemma}

\newtheorem{cor}[thm]{Corollary}
\theoremstyle{definition}
\newtheorem{dfn}[thm]{Definition}
\newtheorem{exam}[thm]{Example}
\newtheorem{rem}[thm]{Remark}

\newtheorem{convention}[thm]{Convention}

\def\FL_e{\mathrm{FL_e}}

\def\CPC{\mathrm{CPC}}

\def\4D{\mathsf{4D}}

\def\S4{\mathsf{S4}}
\def\IPC{\mathsf{IPC}}
\def\CPC{\mathsf{CPC}}

\newcommand{\tGamma}{\tilde{\Gamma}}
\newcommand{\tSigma}{\tilde{\Sigma}}
\newcommand{\tPi}{\tilde{\Pi}}
\newcommand{\tDelta}{\tilde{\Delta}}
\newcommand{\tLambda}{\tilde{\Lambda}}
\def\phi{\varphi}
\def\phi{\varphi}
\newcommand{\Am}{\forall\!_{{\cal M}} \hspace{-.08mm}p\hspace{.2mm}} 
\newcommand{\Aemp}{\forall\!_{{\emptyset}} \hspace{-.08mm}p\hspace{.2mm}} 
\newcommand{\Aa}{\forall\!_{{\cal A}} \hspace{-.08mm}p\hspace{.2mm}} 
\newcommand{\Ab}{\forall_{{\Box}} p\hspace{.2mm}}
\newcommand{\ALR}{\forall\!_{{\cal L}} p\hspace{.2mm}}
\newcommand{\ARR}{\forall\!_{{\cal R}} p\hspace{.2mm}}
\newcommand{\Em}{\exists\!_{{\cal M}} \hspace{-.08mm}p\hspace{.2mm}} 
\newcommand{\EA}{\exists\!_{{\cal A}} \hspace{-.08mm}p\hspace{.2mm}} 
\newcommand{\ER}{\exists\hspace{-.20mm}_{{\cal L}} \hspace{-.00mm}p\hspace{.2mm}}

\newcommand{\calL}{\mathcal{L}}

\begin{document}

\title{Universal Proof Theory: Semi-analytic Rules and Uniform Interpolation} 

\author{Amirhossein Akbar Tabatabai \footnote{amir.akbar@gmail.com}, Raheleh Jalali \footnote{rahele.jalali@gmail.com}\\
\small{$^*$Bernoulli Institute, University of Groningen, Groningen, The Netherlands}\\
\small{$^\dagger$Department of Computer Science, University of Bath, Bath, The UK}}

\date{}

\maketitle

\begin{abstract}
In \cite{Craig}, we introduced a syntactically defined and highly general class of calculi known as \emph{semi-analytic}. We then demonstrated that any sufficiently strong (modal) substructural logic with a semi-analytic calculus must satisfy the Craig interpolation property.
In this paper, we show that if the calculus is also terminating in a certain formal sense, then its logic has the Uniform Interpolation Property (UIP). 
This result has significant applications. On the positive side, it provides a uniform and modular method for proving UIP for various logics, including $\mathsf{FL_e}$, $\mathsf{FL_{ew}}$, $\mathsf{CFL_e}$, $\mathsf{CFL_{ew}}$, and their $K$, $D$, and $T$-type modal extensions, as well as $\mathsf{CPC}$, $\mathsf{K}$, and $\mathsf{KD}$. However, its more striking consequence lies in the negative direction. It extends the negative results of \cite{Craig} to logics with CIP but without UIP. In particular, it shows that the modal logics $\mathsf{K4}$ and $\mathsf{S4}$ do not have a terminating semi-analytic calculus. \\

\noindent \textbf{keywords:} Uniform interpolation, Sequent calculi, Substructural logics, Modal logics, Subexponential modalities

\end{abstract}

\newpage  \tableofcontents \newpage 

\section{Introduction}
Proof systems are central \emph{tools} in proof theory and, like any other tool, they are often regarded as technical apparatuses designed for specific purposes, in this case, to study particular logics.
This perspective is reminiscent of the pre-modern approach to abstract mathematical structures such as groups and rings. 
Historically, mathematicians studied these structures, like the ring of integers or the symmetry groups of geometric shapes primarily as \emph{tools} for understanding the underlying concrete objects of interest, namely, numbers and geometric figures. Outside of these practical applications, there was little interest in treating algebraic structures as independent mathematical entities, or in exploring their \emph{generic} properties in a systematic way.

Fortunately, in recent years, a new approach to proof systems has emerged, focusing on their \emph{generic} behavior rather than on specific instances or technical applications \cite{Cia, Iemhoff1, Iemhoff, Craig, DP}. In \cite{Craig}, this emerging field was named \emph{Universal Proof Theory} (UPT, for short),\footnote{We are grateful to Masoud Memarzadeh for this elegant terminological suggestion.} evoking the term \emph{universal algebra}, the field that studies the \emph{generic} behavior of algebraic structures.

UPT aims to address the following fundamental problems:

\vspace{4pt}
\noindent $(i)$
\emph{Existence problem}: Investigating the existence of a proof system of a given type for a specific logic or mathematical theory. Examples include analytic, terminating, and normalizable proof systems.

\vspace{4pt}
\noindent $(ii)$
\emph{Equivalence problem}: Exploring natural notions of equivalence between proof systems for a given logic or mathematical theory. This problem extends Hilbert's twenty-fourth problem, expanding the study from the equivalence of individual mathematical proofs to the equivalence of entire proof systems.

\vspace{4pt}
\noindent $(iii)$
\emph{Characterization problem}: Investigating possible characterizations of proof systems for a given logic or mathematical theory, through a specified equivalence relation, as discussed in $(ii)$.
\vspace{4pt}

So far, universal proof theory has primarily focused on the first of these problems, yielding many successful and surprising results, particularly in proving the \emph{non-existence} of broad types of proof systems. Following the approach in \cite{Iemhoff1, Iemhoff}, the key idea behind establishing such non-existence results is to identify a \emph{proof-independent} and purely logical \emph{invariant property} such that, if a theory possesses a proof system of a given type, it must also have this property. Then, by showing that the theory lacks this property, we can establish the non-existence of a proof system of the specified form. To illustrate with a simple example of an invariant property, consider decidability, and for a given form of a proof system, consider termination: since the existence of a terminating calculus for a logic guarantees its decidability, an undecidable logic cannot possess a terminating calculus.

\subsection*{Uniform Interpolation and the Origins of UPT}

Pitts' seminal proof-theoretic argument \cite{Pitts} establishes that $\mathsf{IPC}$ satisfies the Uniform Interpolation Property (UIP). In his proof, Pitts employs Dyckhoff’s terminating calculus for $\mathsf{IPC}$ \cite{Dyck} to construct uniform interpolants.
The first sparks of UPT emerged when Iemhoff made the crucial observation \cite{Iemhoff1,Iemhoff} that Pitts’ argument was, in fact, sufficiently flexible to apply to any axiom and rule satisfying a specific, yet sufficiently general form. In short, pure syntactical form implies logical properties, in this case, UIP. She termed these axioms and rules \emph{focused axioms} and \emph{focused rules}, respectively.

The focused axioms are essentially the axioms of the intuitionistic sequent calculus $\mathbf{LJ}$. Focused rules, on the other hand, are roughly characterized by having a single main formula in their conclusion and being  ``side-preserving" and ``occurrence-preserving". The former means that if the main formula appears on the left (or right) side of the conclusion, then all active formulas in any of the premises must also appear on the same side. The occurrence-preserving condition ensures that any atom appearing in any of the active formulas in each premise must also appear in the main formula in the conclusion. Prototype examples of focused rules include the usual conjunction and disjunction rules. Clear counterexamples are implication rules and the cut rule: the former do not respect the position of the main formula, while in the latter, the atoms in the cut formula may not appear in the conclusion.  

Having identified focused axioms and rules, Iemhoff proceeded to show that the logic of any terminating extension of Dyckhoff’s calculus in the single-conclusion case, or $\mathbf{G3c}$ in the multi-conclusion case, by focused axioms, focused rules, and certain modal rules, enjoys UIP \cite{Iemhoff1,Iemhoff}. This result led to the proof of UIP for some (intuitionistic) modal logics, notably $\mathsf{IK}$ and $\mathsf{IKD}$. On the other hand, it also established the non-existence of such calculi for any (modal) superintuitionistic logic lacking UIP. The latter includes all superintuitionistic logics (except for at most seven), as well as the logics $\mathsf{K4}$ and $\mathsf{S4}$, and all extensions of $\mathsf{S4}$ (except for at most 37 of them).

\subsection*{Semi-analytic Rules and Craig Interpolation}

In \cite{Craig}, we extended Iemhoff’s results in several directions. First, we lowered the base logic from $\mathsf{IPC}$ to the basic substructural logic $\mathsf{FL_e}$, thereby broadening the applicability of our results to substructural logics. Second, we generalized the concept of focused axioms into a more flexible family, while still referring to them as \emph{focused axioms}. Third, we introduced a broad class of rules, termed (single-conclusion) multi-conclusion \emph{semi-analytic rules}, which encompasses a wide range of rules, including focused rules, implication rules, and any combination of multiplicative and additive settings in substructural logics. We called a calculus \emph{multi-conclusion semi-analytic} if it consists of focused axioms, multi-conclusion semi-analytic rules, and certain specific modal rules. A similar definition applies to \emph{single-conclusion semi-analytic calculi}.  
Fourth, we weakened the invariant property from UIP to the Craig Interpolation Property (CIP), broadening the scope of our negative results, as described in the following.

With all these generalizations in place, we then showed that if a (modal) substructural logic (including superintuitionistic logics) extending $\mathsf{CFL_e}$ (resp. $\mathsf{FL_e}$) has a multi-conclusion (resp. single-conclusion) semi-analytic sequent calculus, it must satisfy CIP. Since CIP is a weaker property than UIP and semi-analytic rules are more general than focused rules, our negative results significantly broaden the scope of those presented in \cite{Iemhoff1, Iemhoff}.
Using this framework, we presented a uniform and modular method for proving CIP across a broad range of modal substructural logics. Moreover, by leveraging the rarity of CIP in substructural, superintuitionistic, and modal logics, we established several non-existence results. For instance, we showed that monoidal t-norm logic $\mathsf{MTL}$, all superintuitionistic logics except at most seven, and all consistent extensions of Hájek’s basic fuzzy logic $\mathsf{BL}$ except at most three, do not have a single-conclusion semi-analytic calculus. Similarly, the relevant logic $\mathsf{R}$ and all extensions of $\mathsf{S4}$, except at most thirty-seven, do not have a multi-conclusion semi-analytic calculus. The full list of these non-existence results is beyond the scope of this introduction (see \cite{Craig}).

\subsection*{Our Contribution}

In this paper, we return to the origins of UPT, focusing once again on uniform interpolation as the invariant property. The main result of this work is showing that any (modal) substructural logic extending $\mathsf{CFL_e}$ (resp. $\mathsf{FL_e}$) that possesses a \emph{terminating} multi-conclusion (resp. single-conclusion) semi-analytic calculus\footnote{Our use of semi-analytic calculus in this paper differs from the one in \cite{Craig}, as we allow fewer modal rules, see Definition \ref{def: semi-analytic calculus}.} satisfies the UIP. On the positive side, our result provides a uniform and modular method for proving UIP by simply checking whether the sequent calculus of the logic exhibits a specific syntactical form and whether it is terminating. Logics with this property include various substructural logics, such as $\mathsf{FL_e}$, $\mathsf{FL_{ew}}$, $\mathsf{CFL_e}$, $\mathsf{CFL_{ew}}$, and their $K$, $D$, and $T$-type modal extensions, as well as $\mathsf{CPC}$, $\mathsf{K}$, and $\mathsf{KD}$. Although UIP for purely substructural or purely modal logics is known \cite{Aliz,Bil,visser1996bisimulations,visser1996uniform,shavrukov1993subalgebras}, our paper provides the first proof of UIP for their combinations in modal substructural logics, as far as we know.

While the positive applications of our main theorem are certainly noteworthy, as already mentioned, the primary motivation for our main theorem is its negative use in proving the non-existence of terminating semi-analytic calculi, leveraging the rarity of UIP. For logics that do not even satisfy CIP, such as all superintuitionistic logics (except at most seven) and all extensions of $\mathsf{S4}$ (except at most 37), it is already known that they cannot have a semi-analytic sequent calculus. Here, we establish a weaker result, showing that these logics cannot have a terminating semi-analytic calculus. However, our proof follows an entirely different approach from previous ones and may therefore be of independent interest. 

For logics that have CIP but not UIP, or for those where only the lack of UIP is known, we provide a new family of non-existence results.  
For example, the modal logics $\mathsf{K4}$ and $\mathsf{S4}$ satisfy CIP but do not have UIP \cite{Bil,Ghil1}.
As a result, we conclude that these logics do not have any terminating single-conclusion or multi-conclusion semi-analytic calculus.

\section{Preliminaries}\label{sec: Prelim}


In this section, we introduce some basic notions that will be used throughout the paper. For further details, the reader is referred to \cite{troelstra}, \cite{Chagrov}, and \cite{Ono}, for proof theory, modal logics, and substructural logics, respectively.

Consider the languages $\calL_p=\{\wedge, \vee, *, \to, \top, \bot, 1, 0 \}$ and $\calL_\Box=\calL_p \cup \{\Box\}$. If we want to refer to either of these languages, we use the notation $\calL \in \{\calL_p, \calL_\Box\}$. The $\calL_\Box$-formulas are defined by the grammar
\[
F ::= p \mid  0 \mid 1 \mid \bot \mid \top \mid F_1 \wedge F_2 \mid F_1 \vee F_2 \mid F_1 \to F_2 \mid F_1 * F_2 \mid \Box F
\]
$\calL_p$-formulas are defined similarly, except that the clause $\Box F$ is omitted, as $\Box \notin \calL_p$. When $\mathcal{L} \in \{\mathcal{L}_p, \mathcal{L}_\Box\}$ is clear from the context and there is no risk of ambiguity, we refer to $\mathcal{L}$-formulas simply as ``formulas". \emph{Propositional variables} (also called \emph{atomic formulas} or \emph{atoms}) are denoted by small Roman letters $p,q, \dots$, formulas by capital Roman letters $A, B, \dots$ and sometimes by small Greek letters $\phi, \psi, \dots$.

For formulas $A$ and $B$, define $\neg A:=A \to 0$, $A + B := \neg (\neg A * \neg B)$ and $A^n$ by $A^0:=1$ and $A^{n+1}:= A * A^n$, for any $n \geq 0$. The set of \emph{variables} of a formula $A$, denoted $\mathcal{V}(A)$, is defined recursively: $\mathcal{V}(c)=\emptyset$, for $c \in \{0, 1, \bot,  \top\}$, $\mathcal{V}(p)=\{p\}$, if $p$ is an atomic formula, $\mathcal{V}(A \circ B)= \mathcal{V}(A) \cup \mathcal{V}(B)$, for $\circ \in \{\wedge, \vee, \to , *\}$, and $\mathcal{V}(\Box A)=\mathcal{V}(A)$. For any atom $p$, a formula $A$ is called \emph{$p$-free} if $p \notin \mathcal{V}(A)$, i.e., $p$ does not occur in the formula $A$.

\emph{Multisets of formulas} are defined as usual and we assume that they are always finite. We use capital Greek letters $\Gamma, \Delta, \dots$, as well as the bar notation as in $\overline{\phi}, \overline{\psi}, \overline{A}, \overline{B}, \dots$ to refer to multisets of formulas. By $|\Gamma|$, we mean the number of formulas (counting the multiplicity) in the multiset $\Gamma$. For any two multisets $\Gamma$ and $\Pi$, by $(\Gamma, \Pi)$ or $\Gamma \cup \Pi$, we mean the union of $\Gamma$ and $\Pi$ as multisets. If $\Gamma= \{A_1,  \dots, A_n\}$, define $\bigast \Gamma:= A_1 * \dots * A_n$, $\bigplus \Gamma:= A_1 + \dots + A_n$, $\bigast \emptyset :=1$, $\bigplus \emptyset :=0$, $\Box \Gamma:= \{\Box A_1 , \dots , \Box A_n\}$, and $\Box \emptyset:= \emptyset$. For a multiset $\Gamma$, define $\mathcal{V}(\Gamma):=\bigcup_{A \in \Gamma} \mathcal{V}(A)$. For any atom $p$, the multiset $\Gamma$ is called \emph{$p$-free} if $p \notin \mathcal{V}(\Gamma)$, i.e., $p$ does not occur in any of the formulas in $\Gamma$.

We will also use \emph{multiset variables} (also called \emph{contexts}) as variables that refer to multisets, as is usual in proof theory. Multiset variables are denoted by $\tilde{\Gamma}, \tilde{\Delta}, \dots$.
Fixing  $\mathcal{L} \in \{\mathcal{L}_p, \mathcal{L}_{\Box}\}$, a \emph{substitution} $\sigma$ is a map that assigns $\mathcal{L}$-formulas to atomic formulas and multisets of $\mathcal{L}$-formulas to multiset variables. Then,  $\sigma(A(p_1, \dots, p_n))$ means $A(\sigma(p_1), \dots, \sigma(p_n))$, where $p_1, \dots, p_n$ are all the atoms in $A$. 
By a \emph{local substitution} we mean a substitution whose domain is restricted to a finite set. For a  substitution $\sigma$ and a multiset $\Gamma$, define $\sigma(\Gamma):=\{\sigma(A) \mid A \in \Gamma\}$.

\subsection{Logics and Sequent Calculi}
In this subsection, we will introduce what we mean by a logic and a sequent calculus, as well as their connection. 

\begin{dfn} \label{DfnLogic} 
Let $\calL \in \{\calL_p, \calL_\Box\}$. A \emph{logic} $L$ over the language $\mathcal{L}$ is a set of $\mathcal{L}$-formulas closed under substitution and the following rules:

\item[] $\bullet$
the \emph{modus ponens} rule, i.e., if $\phi, \phi \to \psi \in L$, then $\psi \in L$, and

\item[] $\bullet$
the \emph{adjunction} rule, i.e., if $\phi, \psi \in L$, then $\phi \wedge \psi \in L$,

\item[] $\bullet$
and if $\calL=\calL_\Box$, also the \emph{necessitation} rule, i.e., if $\phi \in L$, then $\Box \phi \in L$.

Denote the language of a logic $L$ by $\mathcal{L}_L$. We say $L$ \emph{extends} the logic $L'$ when $\mathcal{L}_{L'} \subseteq \mathcal{L}_L$ and $L' \subseteq L$. A logic $L$ over $\mathcal{L}_p$ is called \emph{superintuitionistic} when $\IPC \subseteq L \subseteq \CPC$. For a set of $\mathcal{L}$-formulas $\{A_i\}_{i \in I}$, by $L + \{A_i\}_{i \in I}$ we mean the smallest logic over $\mathcal{L}$, extending $L$ and containing all the formulas in $\{A_i\}_{i \in I}$.
\end{dfn}

Now we move to sequent calculus. Let $\calL \in \{\calL_p, \calL_\Box\}$. A \emph{multi-conclusion sequent} (\emph{sequent}, for short) over $\mathcal{L}$ is an expression $S$ = $(\Gamma \Rightarrow \Delta)$, where $\Gamma$, the \emph{antecedent}, and $\Delta$, the \emph{succedent}, are finite multisets of $\mathcal{L}$-formulas. We sometimes denote $\Gamma$ by $S^a$ and $\Delta$ by $S^s$. We call $S$ a \emph{single-conclusion sequent} when $|\Delta| \leq 1$. The \emph{interpretation} of the sequent $S$ is the formula $I(S):=(\bigast \Gamma \to \bigplus \Delta)$. For two sequents $S$ and $T$ by $S \cdot T$ we mean $S^a \cup T^a \Rightarrow S^s \cup T^s$ and by $\mathcal{V}(S)$, we mean $\mathcal{V}(S^a) \cup \mathcal{V}(S^s)$. For a substitution $\sigma$ and a sequent $S=(\Gamma \Rightarrow \Delta)$, by $\sigma(S)$ we mean the sequent $\sigma(\Gamma) \Rightarrow \sigma(\Delta)$. The sequent $(\Gamma' \Rightarrow \Delta')$ is called a \emph{subsequent} of $(\Gamma \Rightarrow \Delta)$, if $\Gamma' \subseteq \Gamma$ and $\Delta' \subseteq \Delta$. It is called \emph{proper} if either $\Gamma' \subsetneqq \Gamma$ or $\Delta' \subsetneqq \Delta$. 

A \emph{multi-conclusion meta-sequent} (\emph{meta-sequent}, for short) $\mathcal{S}$ over $\mathcal{L}$ is an expression of the form
\begin{center}
    $\Box\tGamma_1, \ldots, \Box\tGamma_m, \tPi_{1}, \ldots, \tPi_n, \overline{\mu} \Rightarrow \bar{\nu}, \Box\tDelta_1, \ldots, \Box\tDelta_r, \tLambda_1, \ldots, \tLambda_s \qquad (1)$
\end{center}
where $\overline{\mu}$ and $\overline{\nu}$ are multisets of $\mathcal{L}$-formulas, called the \emph{formulas} of $\mathcal{S}$, and 
$\tGamma_i$'s, $\tPi_j$'s, $\tDelta_k$'s and $\tLambda_l$'s 
are multiset variables. We denote the left-hand side (resp. right-hand side) of $(\Rightarrow)$ in $(1)$ as $\mathcal{S}^a$ (resp. $\mathcal{S}^s$), called the \emph{antecedent} (resp. \emph{succedent}) of $\mathcal{S}$. For a substitution $\sigma$ and a meta-sequent $\mathcal{S}$, by $\sigma(\mathcal{S})$, we mean the sequent resulting from applying $\sigma$ to any formula and multiset variable in $\mathcal{S}$.
The meta-sequent $\mathcal{S}$ is \emph{single-conclusion} if either $\mathcal{S}^s=\overline{\nu}$ and $|\overline{\nu}| \leq 1$ or $\mathcal{S}^s=\Box\tDelta_1, \ldots, \Box\tDelta_r, \tLambda_1, \ldots, \tLambda_s$. 
We will use a notion called the \emph{single-conclusion version} of a meta-sequent $\mathcal{S}$ as in $(1)$, defined as follows:

\vspace{2pt}
$\bullet$
If $|\overline{\nu}|>1$, then $\mathcal{S}$ does not have a single-conclusion version.

$\bullet$
If $|\overline{\nu}|=1$, then the single-conclusion version of $\mathcal{S}$ is 
\begin{center}
$\Box\tGamma_1, \ldots, \Box\tGamma_m, \tPi_{1}, \ldots, \tPi_n, \overline{\mu} \Rightarrow \bar{\nu}.$
\end{center}

$\bullet$
If $|\overline{\nu}|=\emptyset$, then the single-conclusion version of $\mathcal{S}$ is $\mathcal{S}$.

\vspace{2pt}
\noindent A \emph{multi-conclusion rule} (\emph{rule}, for short) over $\mathcal{L}$ is an expression 
\begin{center}
\AxiomC{$\mathcal{S}_1 \ldots \mathcal{S}_n$}
\UnaryInfC{$\mathcal{S}_0$}
\DisplayProof \qquad (2)
\end{center}
where $\mathcal{S}_i$'s are meta-sequents over $\mathcal{L}$. We call $\mathcal{S}_1, \ldots, \mathcal{S}_n$ the \emph{premises} and $\mathcal{S}_0$ the \emph{conclusion} of the rule. An \emph{axiom} is a rule with no premises. The formulas in the premises are \emph{active} formulas, and the formulas in the conclusion are the \emph{main} formulas. We usually work with rules with \emph{one} main formula. A \emph{single-conclusion rule} is a rule where all its meta-sequents are single-conclusion. 
Define the \emph{single-conclusion version} of a rule $R$ of the form $(2)$ as follows:

$\bullet$
if at least one $\mathcal{S}_i$ for $0 \leq i \leq n$ does not have a single-conclusion version, then $R$ does not have a single-conclusion version, 

$\bullet$
otherwise, replace each $\mathcal{S}_i$, for $0 \leq i \leq n$ with its single-conclusion version. Then, if a multiset variable appears in the conclusion and does not appear in any of the premises, delete it. 

\begin{exam}
The following rule has no single-conclusion version:
\begin{center}
\AxiomC{$\tGamma\Rightarrow p,q, \tDelta$}
 \RightLabel{\footnotesize$(R+)$} 
 \UnaryInfC{$\tGamma \Rightarrow p + q, \tDelta$}
 \DisplayProof
\end{center}
However, the single-conclusion version of the rule $(R \wedge)$ is $(R \wedge)'$:
\begin{center}
\begin{tabular}{c c}
\AxiomC{$\tGamma\Rightarrow p, \tDelta$}
\AxiomC{$\tGamma\Rightarrow q, \tDelta$}
 \RightLabel{\footnotesize$(R \wedge)$} 
 \BinaryInfC{$\tGamma \Rightarrow p \wedge q, \tDelta$}
 \DisplayProof
 
\hspace{5pt} & \hspace{5pt}
\AxiomC{$\tGamma\Rightarrow p$}
\AxiomC{$\tGamma\Rightarrow q$}
 \RightLabel{\footnotesize$(R \wedge)'$} 
 \BinaryInfC{$\tGamma \Rightarrow p \wedge q$}
 \DisplayProof
\end{tabular}
\end{center}
\end{exam}
Let $R$ be a rule. By a \emph{local substitution for $R$}, we mean a local substitution whose domain is exactly the set of multiset variables and atomic formulas appearing in $R$.
For a rule $R$ of the form $(2)$ and a local substitution $\sigma$ for $R$, by the $\sigma$-instance of $R$, we mean $\sigma(\mathcal{S}_1) \dots \sigma(\mathcal{S}_n) / \sigma(\mathcal{S}_0)$. 
When clear from context and emphasis on $\sigma$ is unnecessary, we simply call it an \emph{instance} of $R$. For a single-conclusion rule, the instance is \emph{single-conclusion} if, additionally, each of $\sigma(\mathcal{S}_0), \sigma(\mathcal{S}_1), \dots, \sigma(\mathcal{S}_n)$ is single-conclusion. 

A \emph{multi-conclusion sequent calculus} (or simply a \emph{calculus}) over $\mathcal{L}$ is a \emph{finite} set of rules over $\mathcal{L}$. A calculus is \emph{single-conclusion} if all of its rules are single-conclusion. Denote the language of a calculus $G$ by $\mathcal{L}_G$. The \emph{single-conclusion version} of a calculus $G$ consists of the single-conclusion versions of the rules in $G$, provided they exist, i.e., it excludes rules that lack a single-conclusion version. For a (single-conclusion) sequent calculus $G$ and a finite set $X$ of (single-conclusion rules), by $G+X$, we mean the (single-conclusion) calculus consisting of all the rules in $G \cup X$.

A (single-conclusion) sequent $S$ is \emph{provable} from a set of (single-conclusion) sequents $\mathcal{A}$ in the (single-conclusion) calculus $G$ if there is a finite tree where the nodes are labeled by (single-conclusion) sequents and the root is labeled by $S$; the leaves are labeled either by (single-conclusion) instances of the axioms of $G$ or by the elements of $\mathcal{A}$; and, the label of each internal node together with the labels of its children form a (single-conclusion) instance of a rule in $G$.
The tree is called a \emph{proof} of $S$ from $\mathcal{A}$ in $G$. The \emph{depth} of a proof is the maximum length of the branches in the tree, where the \emph{length of a branch} is the number of nodes in the branch minus $1$. If $\mathcal{A}=\emptyset$, we write $G \vdash^{\pi} S$, when $\pi$ is a proof of $S$ from $\mathcal{A}$ in $G$. We write $G \vdash S$ when there is a tree $\pi$ such that $G \vdash^{\pi} S$.
Formulas $\phi$ and $\psi$ are \emph{provably equivalent} in a calculus $G$ if $\phi \Rightarrow \psi$ and $\psi \Rightarrow \phi$ are provable in $G$. 

We finally define the connection between logics and sequent calculi:
\begin{dfn} \label{dfnLogicOfG}
Let $L$ be a logic and $G$ a (single-conclusion) calculus over $\mathcal{L}$. Then, $G$ is called a \emph{calculus for} $L$, or $L$ is the \emph{logic of} $G$, when $G \vdash \Gamma \Rightarrow \Delta$ if{f} $\bigast \Gamma \to \bigplus \Delta \in L$, for any (single-conclusion) sequent $\Gamma \Rightarrow \Delta$ over $\mathcal{L}$.
\end{dfn}
Note that if $G$ is a calculus for $L$, then $G \vdash \phi \Rightarrow \psi$ if{f} $\phi \to \psi \in L$, for any formulas $\phi$ and $\psi$.

\begin{dfn} \label{CalculusExtension}
Let $G$ and $H$ be two (single-conclusion) calculi such that $\mathcal{L}_G \subseteq \mathcal{L}_{H}$. We say that $H$ is an \emph{extension} of $G$ if the following is satisfied:
\begin{itemize}
\item[$\bullet$] every (single-conclusion) instance of the axioms of $G$ is provable in $H$,

\item[$\bullet$] every rule of $G$ is \emph{admissible} in $H$, i.e., for any (single-conclusion) instance of a rule of $G$, if the premises are provable in $H$, then so is its conclusion.
\end{itemize}
\end{dfn}

\subsection{Substructural Logics and Calculi}
We recall several substructural calculi and their corresponding logics that will be used throughout the paper. First, we define the calculus $\mathbf{CFL_e}$ over $\mathcal{L}_p$ as the set of rules and axioms presented in Figure \ref{tableFLe}. It is straightforward to verify that the following rules are provable in $\mathbf{CFL_e}$ (see \cite{Ono}):
\begin{center}
 \AxiomC{$\tGamma, p \Rightarrow \tDelta$}
 \AxiomC{$\tSigma, q \Rightarrow \tLambda$}
 \RightLabel{\scriptsize{$(L+)$}} 
 \BinaryInfC{$\tGamma, \tSigma , p + q \Rightarrow \tDelta, \tLambda$}
 \DisplayProof
 \quad \quad \quad \quad 
 \AxiomC{$\tGamma\Rightarrow p, q, \tDelta$}
 \RightLabel{\scriptsize{$(R+)$}} 
 \UnaryInfC{$\tGamma \Rightarrow p + q, \tDelta$}
 \DisplayProof
\end{center}
This means that for any instance of the rule, the conclusion is provable from the set of premises of the rule.
\begin{figure}[t]
{\footnotesize
\begin{center}
\begin{tabular}{c c c c c}
$p \Rightarrow p \; \footnotesize{(id)} $ \hspace{10pt}
&
 $\Rightarrow 1$ \hspace{10pt}
 &
 $0 \Rightarrow$ \hspace{10pt}
 &
 $\tGamma \Rightarrow \top, \tDelta$ \hspace{10pt}
 &
 $\tGamma, \bot \Rightarrow \tDelta$ \\[2ex]
\end{tabular}
\begin{tabular}{cc}
\AxiomC{$\tGamma \Rightarrow \tDelta$}
 \RightLabel{\scriptsize{$(1 w)$}} 
 \UnaryInfC{$\tGamma, 1 \Rightarrow \tDelta$}
 \DisplayProof
&
\small \AxiomC{$\tGamma \Rightarrow \tDelta$}
 \RightLabel{\scriptsize{$(0 w)$}} 
 \UnaryInfC{$\tGamma \Rightarrow 0 , \tDelta$}
 \DisplayProof\\[3ex]
 \end{tabular}
\begin{tabular}{ccc}
 \AxiomC{$\tGamma, p \Rightarrow \tDelta$}
 \RightLabel{\scriptsize{$(L \wedge_1)$} }
 \UnaryInfC{$\tGamma, p \wedge q \Rightarrow \tDelta$}
 \DisplayProof
&

\AxiomC{$\tGamma, q \Rightarrow \tDelta$}
 \RightLabel{\scriptsize{$(L \wedge_2)$} }
 \UnaryInfC{$\tGamma, p \wedge q \Rightarrow \tDelta$}
\DisplayProof
 & 

\AxiomC{$\tGamma \Rightarrow p, \tDelta$}
 \AxiomC{$\tGamma \Rightarrow q, \tDelta$}
 \RightLabel{\scriptsize{$(R \wedge)$} }
 \BinaryInfC{$\tGamma \Rightarrow p \wedge q, \tDelta$}
  \DisplayProof \\[3ex]
\end{tabular}
\begin{tabular}{ccc}
  \AxiomC{$\tGamma, p \Rightarrow \tDelta$}
 \AxiomC{$\tGamma, q \Rightarrow \tDelta$}
 \RightLabel{\scriptsize{$(L \vee)$} }
 \BinaryInfC{$\tGamma, p \vee q \Rightarrow \tDelta$}
 \DisplayProof
& 

 \AxiomC{$\tGamma \Rightarrow p, \tDelta$}
 \RightLabel{\scriptsize{$(R \vee_1)$} }
 \UnaryInfC{$\tGamma \Rightarrow p \vee q, \tDelta$}
\DisplayProof
 & 
\AxiomC{$\tGamma \Rightarrow q, \tDelta$}
 \RightLabel{\scriptsize{$(R \vee_2)$} }
 \UnaryInfC{$\tGamma \Rightarrow p \vee q, \tDelta$}
 \DisplayProof \\[3ex]
\end{tabular}
\begin{tabular}{cc}
\AxiomC{$\tGamma, p, q \Rightarrow \tDelta$}
 \RightLabel{\scriptsize{$(L*)$}}
 \UnaryInfC{$\tGamma, p * q \Rightarrow \tDelta$}
\DisplayProof
&
 \AxiomC{$\tGamma \Rightarrow p, \tDelta$}
 \AxiomC{$\tSigma \Rightarrow q, \tLambda$}
 \RightLabel{\scriptsize{$(R*)$} }
 \BinaryInfC{$\tGamma, \tSigma \Rightarrow p * q, \tDelta, \tLambda$}
\DisplayProof \\[3ex]
\end{tabular}
\begin{tabular}{cc}
\AxiomC{$\tGamma \Rightarrow p, \tDelta$}
 \AxiomC{$\tSigma, q \Rightarrow \tLambda$}
 \RightLabel{\scriptsize{$(L\!\to)$} }
 \BinaryInfC{$\tGamma, \tSigma, p \to q \Rightarrow \tDelta, \tLambda$}
\DisplayProof
&
\AxiomC{$\tGamma, p \Rightarrow q, \tDelta$}
 \RightLabel{\scriptsize{$(R\!\to)$} }
 \UnaryInfC{$\tGamma \Rightarrow p \to q, \tDelta$}
\DisplayProof
\end{tabular}
\end{center}
}
\caption{The system $\mathbf{CFL_e}$. When the specific index is not of primary concern, we use the notation $(L\wedge)$ to refer to both $(L\wedge_1)$ and $(L\wedge_2)$, and similarly, we write $(R\vee)$ to encompass $(R\vee_1)$ and $(R\vee_2)$.  \label{tableFLe}}
\end{figure} 
The calculus $\mathbf{FL_e}$ is the single-conclusion version of $\mathbf{CFL_e}$. Adding some or all of the following \emph{structural rules}
\begin{center}
Weakening rules: \ \  
 \AxiomC{$\tGamma \Rightarrow \tDelta$}
 \RightLabel{\scriptsize{$(L w)$} }
\UnaryInfC{$\tGamma, p \Rightarrow \tDelta$}
 \DisplayProof
 \quad \quad 
 \AxiomC{$\tGamma \Rightarrow \tDelta$}
 \RightLabel{\scriptsize{$(R w)$} }
 \UnaryInfC{$\tGamma \Rightarrow p, \tDelta$}
 \DisplayProof
 \vspace{7pt}\\
Contraction rules:  \ \ 
 \AxiomC{$\tGamma, p, p \Rightarrow \tDelta$}
 \RightLabel{\scriptsize{$(L c)$} }
\UnaryInfC{$\tGamma, p \Rightarrow \tDelta$}
 \DisplayProof
\quad \quad 
 \AxiomC{$\tGamma\Rightarrow  p, p, \tDelta$}
 \RightLabel{\scriptsize{$(R c)$} }
 \UnaryInfC{$\tGamma \Rightarrow p, \tDelta$}
 \DisplayProof
\end{center}
to $\mathbf{CFL_e}$ or their single-conclusion versions (if they exist) to $\mathbf{FL_e}$, we get new calculi:
 $\mathbf{CFL_{ew}}=\mathbf{CFL_{e}}+\{Lw, Rw\}$, $\mathbf{CFL_{ec}}=\mathbf{CFL_{e}}+\{Lc, Rc\}$,  $\mathbf{CFL_{ewc}}=\mathbf{CFL_{ew}}+\{Lc, Rc\}$, $\mathbf{FL_{ew}}=\mathbf{FL_{e}}+\{Lw, Rw\}$, $\mathbf{FL_{ec}}=\mathbf{FL_{e}}+\{Lc\}$, and  $\mathbf{FL_{ewc}}=\mathbf{FL_{ew}}+\{Lc\}$.
Note that since $(R c)$ does not have a single-conclusion version, it is allowed only in multi-conclusion calculi. It is well known that the cut rule:
\begin{center}
\begin{tabular}{c}
\AxiomC{$\tGamma \Rightarrow p, \tDelta$}
 \AxiomC{$\tSigma, p \Rightarrow \tLambda$}
 \RightLabel{\scriptsize{$(cut)$} }
 \BinaryInfC{$\tGamma, \tSigma \Rightarrow \tDelta, \tLambda$}
 \DisplayProof
\end{tabular}
\end{center}
is admissible in all the calculi introduced above (see \cite{Ono}). 

For any of the aforementioned calculi $G$, define the set $L_G = \{ A \mid G \vdash \, \Rightarrow A \}$. By the admissibility of the cut rule in $G$, it follows easily that $L_G$ is a logic in the sense of Definition \ref{DfnLogic} and, in fact, is the logic of $G$ in the sense of Definition \ref{dfnLogicOfG}. To maintain a clear connection between a calculus $G$ and its associated logic $L_G$, we use boldface letters for calculi and the corresponding sans-serif notation for their logics. For example, the logic of $G = \mathbf{FL_e}$ is denoted by $L_G = \mathsf{FL_e}$.

\begin{rem}
In the standard definition of substructural logics, the propositional constants of the language typically include only $1$ and $0$, while $\bot$ and $\top$ are not considered. However, the presence of $\bot$ and $\top$ is crucial for establishing the uniform interpolation property. Hence, in alignment with \cite{Aliz}, we incorporate these constants into the definition.
\end{rem}

\begin{rem}
If the weakening rules are present, $\top$ (resp. $\bot$) is provably equivalent to $1$ (resp. $0$). If both the weakening and contraction rules are included, the formulas $\phi \wedge \psi$ and $\phi * \psi$ are provably equivalent.  
Thus, $\mathbf{CFL_{ewc}}$ (resp. $\mathbf{FL_{ewc}}$) is essentially equivalent to the standard calculus $\mathbf{LK}$ (resp. $\mathbf{LJ}$) for classical (resp. intuitionistic) logic $\CPC$ (resp. $\IPC$), with the difference that the language includes additional yet redundant constants $0$, $1$, and connective $*$. To maintain a unified presentation, we treat $\CPC$ and $\IPC$ as if they are defined over the extended language $\mathcal{L}_p$, and we use the calculi $\mathbf{CFL_{ecw}}$ and $\mathbf{FL_{ecw}}$ to define these logics, respectively.
\end{rem}

For classical logic $\mathsf{CPC}$ over $\mathcal{L}_p$, there exists another calculus that is particularly relevant to our study, as it is terminating in a formal sense that we will elaborate on later. This calculus, denoted by $\mathbf{Gcs}$, is defined in Figure \ref{Fig. G3c}.  
Notably, the axioms for ($1$ and $\top$), ($0$ and $\bot$), as well as the rules for ($\wedge$ and $*$), are identical to each other and also coincide with their counterparts in the standard calculus $\mathbf{G3c}$ \cite{troelstra}. The duplication of these rules is merely a technical device to extend the usual $\mathbf{G3c}$ to accommodate the language $\mathcal{L}_p$.  
One can easily establish that for any sequent $S$ over $\mathcal{L}_p$, we have $\mathbf{CFL_{ecw}} \vdash S$ if and only if $\mathbf{Gcs} \vdash S$. Consequently, $\mathbf{Gcs}$ also serves as a calculus for classical logic $\mathsf{CPC}$ over $\mathcal{L}_p$. 

\begin{figure}[t]
{\footnotesize
\begin{center}
\begin{tabular}{c c c c c}
$p \Rightarrow p $ \hspace{10pt}
&
 $\Rightarrow 1$ \hspace{10pt}
 &
 $0 \Rightarrow$ \hspace{10pt}
 &
 $\tGamma \Rightarrow \top, \tDelta$ \hspace{10pt}
 &
 $\tGamma, \bot \Rightarrow \tDelta$ \\[2ex]
\end{tabular}
\begin{tabular}{cc}
 \AxiomC{$\tGamma \Rightarrow \tDelta$}
 \RightLabel{\scriptsize{$(L w)$} }
\UnaryInfC{$\tGamma, p \Rightarrow \tDelta$}
 \DisplayProof
 \quad \quad 
 \AxiomC{$\tGamma \Rightarrow \tDelta$}
 \RightLabel{\scriptsize{$(R w)$} }
 \UnaryInfC{$\tGamma \Rightarrow p, \tDelta$}
 \DisplayProof\\[4ex]
 \end{tabular}
\begin{tabular}{cccc}
\AxiomC{$\tGamma, p, q \Rightarrow \tDelta$}
 \UnaryInfC{$\tGamma, p * q \Rightarrow \tDelta$}
\DisplayProof
&
 \AxiomC{$\tGamma \Rightarrow p, \tDelta$}
 \AxiomC{$\tGamma \Rightarrow q, \tDelta$}
 \BinaryInfC{$\tGamma \Rightarrow p * q, \tDelta$}
 \DisplayProof 
 &
\AxiomC{$\tGamma, p, q \Rightarrow \tDelta$}
 \UnaryInfC{$\tGamma, p \wedge q \Rightarrow \tDelta$}
\DisplayProof
&
 \AxiomC{$\tGamma \Rightarrow p, \tDelta$}
 \AxiomC{$\tGamma \Rightarrow q, \tDelta$}
 \BinaryInfC{$\tGamma \Rightarrow p \wedge q, \tDelta$}
 \DisplayProof \\[4ex]
\end{tabular}
\begin{tabular}{cccc}
 \AxiomC{$\tGamma, p \Rightarrow \tDelta$}
 \AxiomC{$\tGamma, q \Rightarrow \tDelta$}
 \BinaryInfC{$\tGamma, p \vee q \Rightarrow \tDelta$}
 \DisplayProof
&
 \AxiomC{$\tGamma \Rightarrow p, q, \tDelta$}
 \UnaryInfC{$\tGamma \Rightarrow p \vee q, \tDelta$}
 \DisplayProof
 &
 \AxiomC{$\tGamma \Rightarrow p, \tDelta$}
 \AxiomC{$\tGamma, q \Rightarrow \tDelta$}
 \BinaryInfC{$\tGamma, p \to q \Rightarrow \tDelta$}
 \DisplayProof
&
\AxiomC{$\tGamma, p \Rightarrow q, \tDelta$}
 \UnaryInfC{$\tGamma \Rightarrow p \to q, \tDelta$}
 \DisplayProof
\end{tabular}
\end{center}
}
\caption{The system $\mathbf{Gcs}$.\label{Fig. G3c}}
\end{figure} 

Now, let us move to modal substructural logics and their calculi. Consider the rules in Figure \ref{tableModalAxiom}. Here are some remarks. First, typically, the standard presentation of the rule $(D)$ does not involve $p$ and $\Box p$ and it has the form
\begin{center}
\AxiomC{$\tGamma \Rightarrow $}
\UnaryInfC{$\Box \tGamma \Rightarrow $}
 \DisplayProof
\end{center}
Our version appears to be weaker, as it explicitly assumes that the antecedent of the premises cannot be empty. The only substantive difference between these two versions is that in our formulation of rule $(D)$, deriving the empty sequent from itself is not allowed. However, since the logics we consider are all consistent and thus in their calculi, we never derive the empty sequent, this modification imposes no real restriction on capturing $D$-type modal logics within sequent calculi. Second, note that the rule $(Lw_\Box)$ and some natural rules for contracting two boxed formulas are used in the linear logic literature. For instance, the rule:
\begin{center}
\AxiomC{$\tGamma, \Box p, \Box p \Rightarrow \tDelta$}
\UnaryInfC{$\tGamma, \Box p \Rightarrow \tDelta$}
 \DisplayProof
\end{center}
and $(Lw_\Box)$ are used in Linear Logic when $\Box$ is denoted by `$!$' \cite{girard1987linear}. However, we do not consider contraction-like rules, such as the one mentioned above, in this work, as our focus is on terminating sequent calculi, and the inclusion of contraction rules typically poses challenges to termination.

Working over $\mathcal{L}_{\Box}$, one can add any subset of the rules in Figure \ref{tableModalAxiom} to any of the substructural calculi introduced above to obtain a new calculus. We always pick $(K)$ when we pick $(D)$ to have a reasonable system. To emphasize this fact, we write $(KD)$ as the combination of the two rules $(K)$ and $(D)$. Therefore, for any $X \subseteq \{K, KD, T, Lw_{\Box}\}$ and any of the substructural calculi $G$ defined above, we can define the calculus $GX$ consisting of the rules in $G$ and $X$. For instance, $\mathbf{CFL_{e}}X$ is the result of adding the rules in $X$ to $\mathbf{CFL_e}$.

For any of these systems except for $\mathbf{Gcs}X$, cut admissibility is proved in \cite{Ono,Ono90,onoketab,kiriyama} for substructural and \cite{LLL,affine,kanovich,tro92,dosen} for modal substructural calculi. For 
a uniform treatment of cut admissibility see \cite{elaine,Craig}. For $\mathbf{Gcs}X$, it is easy to prove the equivalence between $\mathbf{CFL_{ecw}}X$ and $\mathbf{Gcs}X$, which proves cut admissibility for the latter.
Using the cut admissibility in $GX$, we can define $L_{GX}=\{A \mid GX \vdash \, \Rightarrow A\}$ and ensure that $L_{GX}$ is actually the logic of $GX$. Similar to what we had before, we use the sans-serif font to denote the logic $L_{GX}$. For instance, $\mathsf{FL_{ew}}X$ is the logic of the calculus $\mathbf{FL_{ew}}X$. However, we denote the logics of $\mathbf{Gcs}$, $\mathbf{Gcs}\{K\}$ and $\mathbf{Gcs}\{K, D\}$ by their usual names, i.e., $\mathsf{CPC}$, $\mathsf{K}$, and $\mathsf{KD}$, respectively.
Moreover, we define the modal logic $\mathsf{K4}$ as $\CPC + \{\Box p \to \Box \Box p, \Box (p \to q) \to (\Box p \to \Box q)\}$ and $\mathsf{S4}$ as $\mathsf{K4}+\{\Box p \to p\}$.

\begin{figure}[!ht]
\footnotesize{\begin{center}
\begin{tabular}{ c c c c }
name & rule & name & rule\\
\midrule
$K$ &  \AxiomC{$\tGamma \Rightarrow p$}
 \UnaryInfC{$\Box \tGamma \Rightarrow \Box p$}
 \DisplayProof & $D$ &  \AxiomC{$\tGamma, p \Rightarrow $} 
 \UnaryInfC{$\Box \tGamma, \Box p \Rightarrow $}
 \DisplayProof \\ 
\midrule
 $T$ & \AxiomC{$\tGamma, p \Rightarrow \tDelta$}
\UnaryInfC{$\tGamma, \Box p \Rightarrow \tDelta$}
 \DisplayProof & $Lw_\Box$ & \AxiomC{$\tGamma \Rightarrow \tDelta$}
\UnaryInfC{$\tGamma, \Box p \Rightarrow \tDelta$}
 \DisplayProof 
\end{tabular}
\caption{\small{Modal rules. The rules $T$, $Lw_{\Box}$ can be read as single- or multi-conclusion.}}\label{tableModalAxiom}
\end{center}}
\end{figure}

In \cite{Craig}, we proved the following lemma connecting the extensions of $\mathsf{FL_e}$ (resp. $\mathsf{CFL_e}$) on the logical side to the extensions of $\mathbf{FL_e}$ (resp. $\mathbf{CFL_e}$) on the calculus side. The proof is straightforward. However, for the sake of completeness, we repeat it here.

\begin{lem}\label{FLAdmissibility}
Let $L \in \{\mathsf{FL_e}, \mathsf{CFL_e}\}$ and $G_L \in \{\mathbf{FL_e}, \mathbf{CFL_e}\}$ be its corresponding calculus. Then, any calculus $G$ for any logic $M \supseteq L$ over any $\mathcal{L} \in \{\mathcal{L}_p, \mathcal{L}_{\Box}\}$ extends $G_L$.
\end{lem}
\begin{proof}
First, observe that $M$ is closed under fusion, i.e., if $\phi, \psi \in M$ then $\phi * \psi \in M$. To see why, note that $\phi \to (\psi \to \phi * \psi) \in \mathsf{FL_e} \subseteq L \subseteq M$ and as $M$ is closed under modus ponens, we get $\phi * \psi \in M$. 

Now, we need to show that each instance of each axiom of $G_L$ is provable and each instance of each rule of $G_L$ is admissible in $G$. If $S$ is an instance of an axiom in $G_L$, as $L$ is the logic of $G_L$, we have $I(S) \in L \subseteq M$. Therefore, $G \vdash S$, as $G$ is a calculus for $M$. For the rules, let $R$ be a rule in $G_L$ and
\AxiomC{$S_1, \ldots, S_n$}
 \UnaryInfC{$S$}
 \DisplayProof
be an instance of $R$. Assume that $G \vdash S_i$ for any $1 \leq i \leq n$. Then, we should prove $G \vdash S$. There are two cases to consider:
    \item[] $\bullet$ 
If $R$ is a conjunction or disjunction rule, it is clear that
$[\bigwedge_{i=1}^n I(S_i) \to I(S)] \in L \subseteq M$. As $G \vdash S_i$ for all $1 \leq i \leq n$, and $G$ is a calculus for $M$, we get $I(S_i) \in M$. Hence, as any logic is closed under conjunction, we have $ \bigwedge_{i=1}^n I(S_i) \in M$ and as $M$ is closed under modus ponens, we reach $I(S) \in M$ which implies $G \vdash S$, as $M$ is the logic of $G$.

\item[] $\bullet$ 
If $R$ is the fusion or implication rules or $(0w)$ or $(1w)$, it is easy to see that $[\bigast_{i=1}^n I(S_i) \to I(S)] \in L \subseteq M$. As $G \vdash S_i$ for all $1 \leq i \leq n$, we have $I(S_i) \in M$ which implies $ \bigast_{i=1}^n I(S_i) \in M$, by the closure of $M$ under fusion. Then, since $M$ is closed under modus ponens, $I(S) \in M$ which implies $G \vdash S$, as $M$ is the logic of $G$.
\end{proof}

\section{Semi-analytic Calculi and Termination} \label{sec: semi-analytic}
In \cite{Craig}, we introduced a general family of \emph{focused axioms} and (single-conclusion) multi-conclusion \emph{semi-analytic} rules as a sufficiently general family of axioms and rules, respectively. In this section, we first recall them and then we introduce the notion of \emph{termination} for a sequent calculus.

\begin{dfn} \label{Dfn Focused Axioms}
Let $\tGamma$ and $\tDelta$ be distinct multiset variables,  $\bar{\mu}$, $\bar{\nu}$, and $\bar{\rho}$ be multisets of formulas and $\mu$ a formula. A meta-sequent is a \emph{focused axiom}, if it has one of the following forms:
\begin{center}
 $\mu \Rightarrow \mu \quad \quad  \quad \quad \Rightarrow \bar{\nu} \quad \quad  \quad \quad  \bar{\rho} \Rightarrow \quad \quad  \quad \quad \tGamma, \bar{\mu} \Rightarrow \tDelta \quad \quad  \quad \quad  \tGamma \Rightarrow \bar{\mu}, \tDelta$   
\end{center} 
where $\mathcal{V}(\nu_i)= \mathcal{V}(\nu_j)$, for any $\nu_i, \nu_j \in \bar{\nu}$, $\mathcal{V}(\rho_i)= \mathcal{V}(\rho_j)$, for any $\rho_i, \rho_j \in \bar{\rho}$, and $\mathcal{V}(\mu_i)= \mathcal{V}(\mu_j)$, for any $\mu_i, \mu_j \in \bar{\mu}$. By a \emph{single-conclusion focused axiom} we mean a focused axiom satisfying the following conditions: $|\bar{\nu}| \leq 1$ in the axiom $\Rightarrow \bar{\nu}$, and instead of the axiom $\tGamma \Rightarrow \bar{\mu}, \tDelta$, we use $\tGamma \Rightarrow \bar{\mu}$, where $|\bar{\mu}| \leq 1$.
\end{dfn}

\begin{exam}\label{ExampleSemi-analyticRulesI}
Axioms in Figure \ref{tableFLe} are focused. More generally, any meta-sequent in the form $(\Rightarrow \nu)$, $(\rho \Rightarrow \,)$, $(\tGamma, \mu \Rightarrow \tDelta)$ and $(\tGamma \Rightarrow \mu, \tDelta)$ is focused, where $\nu$, $\rho$, and $\mu$ are each a \emph{single} formula. Additionally, some less common but useful axioms, such as $(p, \neg p \Rightarrow 0)$ and $(\tGamma \Rightarrow p, \neg p, \tDelta)$, are also focused.
As for non-examples, the axiom $(p, \neg p, q \Rightarrow \,)$, where $p$ and $q$ are distinct atoms, is not focused because $\mathcal{V}(p) \neq \mathcal{V}(q)$.
\end{exam}

\begin{rem}
For any two formulas $\mu_1$ and $\mu_2$, if $\mathcal{V}(\mu_1) = \mathcal{V}(\mu_2)$, then for any substitution $\sigma$, we have $\mathcal{V}(\sigma(\mu_1)) = \mathcal{V}(\sigma(\mu_2))$. Therefore, by this observation, the variable condition on focused axioms extends to all of their instances.
\end{rem}

A rule is called \emph{occurrence-preserving} if it has one main formula, and any variable appearing in any of the active formulas in the premises also appears in the main formula. Note that occurrence-preservation is weaker than the analyticity property found in analytic rules: a rule is \emph{analytic} if the active formulas in the premises are subformulas of the main formula.

\begin{exam}
The following rule is occurrence-preserving
\small\begin{center}
 \AxiomC{$\{\tGamma_i, \bar{\mu}_{ir} \Rightarrow \bar{\nu}_{ir}, \tDelta_i \mid 1 \leq i \leq n, 1 \leq r \leq k_i \}$}
 \UnaryInfC{$\tGamma_1, \dots, \tGamma_n, \mu \Rightarrow \tDelta_1, \dots, \tDelta_n $}
 \DisplayProof
\end{center}
\normalsize if 
$\bigcup_{i, r} [\mathcal{V}(\bar{\mu}_{ir}) \cup \mathcal{V}(\bar{\nu}_{ir})] \subseteq \mathcal{V}(\mu)$. Additionally, all the rules in Figures \ref{tableFLe}, \ref{Fig. G3c}, and \ref{tableModalAxiom} are occurrence-preserving. For an instance of an occurrence-preserving rule that is not analytic, consider:
\begin{center}
 \AxiomC{$\tGamma \Rightarrow p, \tDelta$} 
 \AxiomC{$\tGamma \Rightarrow p \to q, \tDelta$}
 \BinaryInfC{$\tGamma \Rightarrow p \wedge q, \tDelta$}
 \DisplayProof
\end{center}
\end{exam}

\begin{dfn} \label{semi-analyticRules}
Let $\tGamma_i$'s, $\tDelta_i$'s, and $\tPi_j$'s be pairwise disjoint families of pairwise distinct multiset variables, for $1 \leq i \leq n$ and $1 \leq j \leq m$. Assume  $\bar{\mu}_{ir}$'s, $\bar{\nu}_{js}$'s, and $\bar{\rho}_{js}$'s are multisets of formulas and $\mu$ a formula, where $1 \leq r \leq k_i$ and $1 \leq s \leq l_j$. If we choose the value $0$ for $n$, $m$, $k_i$, or $l_j$, we mean $i$, $j$, $r$, or $s$ range over the empty set, respectively, and if there is no fear of confusion, we omit the domain of these indices. A \emph{single-conclusion semi-analytic rule} is a rule that is occurrence-preserving and has one of the following forms:\\

\noindent $\bullet$
\emph{left single-conclusion semi-analytic}:
\small\begin{center}
 \AxiomC{$\{\tPi_j , \bar{\nu}_{js} \Rightarrow \bar{\rho}_{js} \mid 1 \leq j \leq m, 1 \leq s \leq l_j \}$} 
 \AxiomC{$\{\tGamma_i , \bar{\mu}_{ir} \Rightarrow \tDelta_i \mid 1 \leq i \leq n, 1 \leq r \leq k_i \}$}
 \BinaryInfC{$\tPi_1, \dots, \tPi_m, \tGamma_1, \dots, \tGamma_n, \mu \Rightarrow \tDelta_1, \dots, \tDelta_n $}
 \DisplayProof
\end{center}
\normalsize where 
$|\bar{\rho}_{js}| \leq 1$, for each $j$ and $s$. As the rule is single-conclusion, at most one of $\tDelta_i$'s in each instance of the rule can be substituted by a formula and the rest are empty. If $n=0$ (resp. $m=0$), there is no premise of the right (resp. left) branch and the conclusion is of the form $(\tPi_1, \dots, \tPi_m, \mu \Rightarrow \,)$ (resp. $\tGamma_1, \dots, \tGamma_n, \mu \Rightarrow \tDelta_1, \dots, \tDelta_n$). Note that if $m=n=0$, the rule has no premise and its conclusion is $(\mu \Rightarrow \,)$. Notice that such a rule is actually a focused axiom, by Example \ref{ExampleSemi-analyticRulesI}.\\

\noindent $\bullet$
\emph{right single-conclusion semi-analytic}:
\small\begin{center}
 \AxiomC{$\{\tGamma_i , \bar{\mu}_{ir} \Rightarrow \bar{\nu}_{ir} \mid 1 \leq i \leq n, 1 \leq r \leq k_i\}$}
 \UnaryInfC{$\tGamma_1, \dots, \tGamma_n \Rightarrow \mu $}
 \DisplayProof
\end{center}
\normalsize where $|\bar{\nu}_{ir}| \leq 1$, for each $i$ and $r$. Note that if $n=0$, the rule has no premise and its conclusion is $(\, \Rightarrow \mu)$. Such a rule is a focused axiom, by Example \ref{ExampleSemi-analyticRulesI}.\\

A \emph{multi-conclusion semi-analytic rule} is a rule that is occurrence-preserving and has one of the following forms:\\

\noindent $\bullet$
\emph{left multi-conclusion semi-analytic}:
\small \begin{center}
 \AxiomC{$\{ \tGamma_i , \bar{\mu}_{ir} \Rightarrow \bar{\nu}_{ir}, \tDelta_i \mid 1 \leq i \leq n, 1 \leq r \leq k_i \}$}
 \UnaryInfC{$\tGamma_1, \dots, \tGamma_n, \mu \Rightarrow \tDelta_1, \dots, \tDelta_n $}
 \DisplayProof
\end{center}
\normalsize 
$\bullet$
\emph{right multi-conclusion semi-analytic}:
\small \begin{center}
 \AxiomC{$\{ \tGamma_i , \bar{\mu}_{ir} \Rightarrow \bar{\nu}_{ir}, \tDelta_i \mid 1 \leq i \leq n, 1 \leq r \leq k_i\}$}
 \UnaryInfC{$\tGamma_1, \dots, \tGamma_n \Rightarrow \mu, \tDelta_1, \dots, \tDelta_n $}
 \DisplayProof
\end{center} 
\normalsize If $n=0$, the conclusion of these rules has the form $(\mu \Rightarrow \,)$ and $(\, \Rightarrow \mu)$, respectively. Notice that both are focused axioms. Whenever we do not want to emphasize the fact that a rule is single-conclusion or multi-conclusion semi-analytic, we simply call the rule \emph{semi-analytic}.
\end{dfn}

\begin{exam}\label{ExampleSemi-analyticRules}
The rules in Figure \ref{tableFLe} and \ref{Fig. G3c} are multi-conclusion semi-analytic, and their single-conclusion versions, if they exist, are single-conclusion semi-analytic. The same applies to the weakening and contraction rules, $(L+)$, $(R+)$, $(T)$, and $(Lw_\Box)$.
As for non-examples, the cut rule is not multi-conclusion semi-analytic because the cut formula may include atoms that do not appear in the conclusion, violating the occurrence-preserving condition. Similarly, the single-conclusion version of cut is not single-conclusion semi-analytic. Likewise, the rules $(K)$ and $(D)$ are neither single-conclusion nor multi-conclusion semi-analytic because in these rules, the context $\tGamma$ transforms into $\Box \tGamma$ which alters the structure in a way that breaks the semi-analytic criteria.
For further explanations and examples, see \cite{Craig}.
\end{exam}

\begin{rem}\label{BackwardFiniteness}
Let $S$ be a multi-conclusion (single-conclusion) sequent and $R$ be a multi-conclusion (single-conclusion) semi-analytic rule. Then, there are finitely many local substitutions $\sigma$ for $R$ such that the $\sigma$-instance of $R$ has the conclusion $S$. 
o prove this, note that for any multiset variable $\tGamma$ appearing in $R$, $\sigma(\tGamma)$ must be a submultiset of $S^a \cup S^s$, and hence it has finitely many choices. Moreover, for any atomic formula $p$ occurring in $R$, since $R$ is occurrence-preserving, $p$ must also appear in the main formula, and hence $\sigma(p)$ is a subformula of a formula in $S$. Therefore, $\sigma(p)$ also has finitely many possibilities.
\end{rem}

\begin{dfn}\label{def: semi-analytic calculus}
A calculus $G$ over $\mathcal{L}_p$ is called \emph{multi-conclusion} (\emph{single-conclusion}) \emph{semi-analytic} if it consists of (single-conclusion versions of) focused axioms and (single-conclusion) multi-conclusion semi-analytic rules. For calculi greater than $\mathcal{L}_\Box$, the definition remains the same, except that the rules $(K)$ and $(KD)$ are also allowed in $G$.
\end{dfn}

\begin{rem}
In the definition of single-conclusion or multi-conclusion semi-analytic calculi, we allow only the rules $(K)$ or $(KD)$. This is more restrictive than the definition used in \cite{Craig}, where additional modal rules such as  
\begin{center}
\begin{tabular}{cc}
\AxiomC{$\Box \Gamma \Rightarrow p $}
\UnaryInfC{$\Box \Gamma \Rightarrow \Box p$}
\DisplayProof \qquad \qquad
&
\AxiomC{$\Box \Gamma, \Gamma \Rightarrow p $}
\UnaryInfC{$\Box \Gamma \Rightarrow \Box p$}
\DisplayProof
\end{tabular}
\end{center}
were allowed. The reason for this restriction is simply that, among the modal rules studied in \cite{Craig}, only $(K)$ and $(KD)$ can be handled to prove UIP. Recall that the logics $\mathsf{K4}$ and $\mathsf{S4}$ do not have UIP.
\end{rem}


One crucial notion employed in this paper is that of terminating calculi. Here is the formal definition:

\begin{dfn}\label{Dfn:TerminatingCalculi}
Let $\mathcal{L} \in \{\mathcal{L}_p, \mathcal{L}_{\Box}\}$, $G$ be a multi-conclusion sequent calculus over $\mathcal{L}$, and $\preceq$ be a well-founded order on the sequents over $\mathcal{L}$. We say that $G$ is \emph{terminating with respect to $\preceq$} if the following are strictly below $S$ according to the order $\preceq$:

\vspace{2pt}
$\bullet$
the premises of any instance of any rule in $G$ where the conclusion is $S$,

$\bullet$
proper subsequents of $S$, and

$\bullet$
any $S'=(\Gamma \Rightarrow \Delta)$, where $\Gamma \cup \Delta \neq \emptyset$, if $S=(\Box \Gamma\Rightarrow \Box \Delta)$.

\noindent For single-conclusion sequent calculi, the definition remains the same, except that $\preceq$ is defined over single-conclusion sequents, and $S$ ranges over single-conclusion sequents.
A multi-conclusion (resp. single-conclusion) calculus $G$ is called \emph{terminating} if there exists a well-founded order $\preceq$ such that $G$ is terminating with respect to $\preceq$.
\end{dfn}

\begin{thm}\label{Thm: terminating calculi}
All contraction-free sequent calculi introduced in the previous subsection are terminating. 
\end{thm}
\begin{proof}
We only explain the multi-conclusion case. The single-conclusion case is identical. Define the \emph{length} of a formula as the number of symbols in it. Then, for any sequent $S$, define a \emph{replacement} applied on $S$, as the action of deleting \emph{one} formula from $S$ and adding a \emph{finite set} (possibly empty) of \emph{shorter} formulas to either of its side.
Now, define the order $\preceq$ on sequents over $\mathcal{L}_{\Box}$ (and consequently $\mathcal{L}_p$) as follows: $S \preceq T$ if there exists a sequence $(S_0, \ldots, S_n)$ of sequents from $T$ to $S$, i.e., $S_0 = T$ and $S_n = S$, such that each $S_{i+1}$ is obtained by a replacement applied on $S_i$. Notably, by taking the singleton sequence $(S)$ with $n=0$, we obtain $S \preceq S$, ensuring that $\preceq$ is reflexive. It is also clear that $\preceq$ is transitive.

We aim to prove that $\preceq$ is also a partial order and well-founded. The most straightforward approach is to map sequents to ordinals in a way that aligns $\preceq$ with the natural order $\leq$ on ordinals below $\omega^\omega$. For that purpose, to any sequent $S$, we assign an ordinal $o(S) < \omega^\omega$ as follows: if $S$ is empty, set $o(S) = 0$. Otherwise, define  
\[
o(S) = \omega^{l_1} e_1 + \omega^{l_2} e_2 + \dots + \omega^{l_m} e_m,
\]
where $l_1 > l_2 > \dots > l_m$ are the lengths of formulas in $S$, and $e_i \in \mathbb{N}$ is the number of occurrences of formulas with length $l_i$ in $S$. The key observation is that if a sequence $(S_0, \ldots, S_n)$ of replacements exists that goes from $T$ to $S$ with $n \geq 1$, then $o(S) < o(T)$. As a consequence, if $S \prec T$, i.e., $S \preceq T$ and $S \neq T$, then $o(S) < o(T)$. The reason for the latter is that if $S \prec T$, then there is a sequence of replacements from $T$ to $S$ with length $n$. If $n=0$, then $S=T$, which is impossible. Hence, $n \geq 1$, which implies $o(S) < o(T)$.

The above observation and its consequence guarantee that $\preceq$ is a well-founded partial order, as illustrated below: To establish anti-symmetry, assume $S \preceq T$ and $T \preceq S$, meaning there exist sequences of replacements from $T$ to $S$ and vice versa. Let $m$ and $n$ be the lengths of these sequences. Then, we obtain a  sequence from $S$ to itself with length $m+n$. If either $m$ or $n$ is nonzero, then so is the length. By the above observation, this implies $o(S) < o(S)$ which is a contradiction. Hence, $m = n = 0$, implying $S = T$.
For well-foundedness, suppose there exists an infinite strictly decreasing sequence $\{S_{i+1} \prec S_i\}_{i=0}^{\infty}$ of sequents. Therefore, $\{o(S_{i+1}) < o(S_i)\}_{i=0}^{\infty}$ must be a strictly decreasing sequence of ordinals, which is impossible.

Now, we aim to prove that any contraction-free calculus introduced in the previous subsection is terminating with respect to $\preceq$. To establish this, first observe that any proper subsequent $S_1$ of a sequent $S$ is strictly smaller than $S$, and removing any occurrence of $\Box$ from a formula in $S$ results in a strictly smaller sequent $S_2$. The reason is that both $S_1$ and $S_2$ are reachable by at least one replacement of a formula by a multiset (possibly empty) of shorter formulas. Second, note that for any rule $R$ in $\mathbf{CFL_{ew}}$, $\mathbf{Gcs}$, or the set $\{K, KD, T, Lw_{\Box}\}$, and for any instance of $R$, each premise is smaller than the conclusion. This follows from the fact that, in every instance of these rules, the premises are derived from the conclusion by replacing a formula with a multiset (possibly empty) of shorter formulas. Consequently, all the calculi $\mathbf{FL_e}X$, $\mathbf{CFL_e}X$, $\mathbf{FL_{ew}}X$, $\mathbf{CFL_{ew}}X$, as well as the calculi $\mathbf{Gcs}$, $\mathbf{Gcs}\{K\}$, and $\mathbf{Gcs}\{K, D\}$ are terminating for any $X \subseteq \{K, KD, T, Lw_{\Box}\}$.
\end{proof}

\section{Main Theorem and Its Applications}\label{sec: UIP}

In this section, we present our main theorem and discuss its applications. Then, in Section \ref{Sec: TheProof}, we proceed with its proof. First, we recall the definition of uniform interpolation.

\begin{dfn} \label{DfnUniformInterpolation}
Let $\mathcal{L}\in \{\mathcal{L}_p, \mathcal{L}_{\Box}\}$. A logic $L$ over $\mathcal{L}$ has the \emph{uniform interpolation property} (UIP), if for any $\mathcal{L}$-formula $\phi$ and any atom $p$, there are $p$-free $\mathcal{L}$-formulas $\exists p \phi$ and $\forall p \phi$ such that: 
\begin{enumerate}
\item[$(var)$] $\mathcal{V}(\exists p \phi) \subseteq \mathcal{V}(\phi)$ and $\mathcal{V}(\forall p \phi) \subseteq \mathcal{V}(\phi)$,
\item[$(i)$]
$L \vdash \phi \to \exists p \phi$,
\item[$(ii)$]
for any $p$-free formula $\psi$ if $L \vdash \phi \to \psi$ then $L \vdash \exists p \phi \to \psi$,
\item[$(iii)$]
$L \vdash \forall p \phi \to \phi$, and
\item[$(iv)$]
for any $p$-free formula $\psi$ if $L \vdash \psi \to \phi$ then $L \vdash \psi \to \forall p \phi$.
\end{enumerate}
\end{dfn}

\begin{rem}
Since $\forall p \phi$ and $\exists p \phi$ are provably unique, we are allowed to use the functional notations as in Definition \ref{DfnUniformInterpolation}.
\end{rem}

In the following, we present the main theorem, which constitutes the central contribution of our paper. 

\begin{thm}[Main Theorem]\label{MainTheorem} 
Let $\mathcal{L}\in \{\mathcal{L}_p, \mathcal{L}_{\Box}\}$ and $L$ be a logic over $\mathcal{L}$:
\begin{itemize}
\item 
If $L \supseteq \mathsf{FL_e}$ and it has a terminating single-conclusion semi-analytic calculus, or
\item 
if $L \supseteq \mathsf{CFL_e}$ and it has a terminating multi-conclusion semi-analytic calculus,
\end{itemize}
then $L$ has the uniform interpolation property.
\end{thm}

Theorem \ref{MainTheorem} has two types of applications. On the positive side, it provides a uniform and modular method for proving the uniform interpolation property for a wide range of (modal) substructural logics, simply by checking the syntactic form of the axioms and rules in their sequent calculi and ensuring that these calculi are terminating. Using this approach, we extend the results of \cite{Aliz,Bil,visser1996bisimulations,visser1996uniform} from purely substructural or purely modal settings to the broader context of modal substructural logics.

\begin{cor}\label{FLeUniformInterpolation}
The logics $\mathsf{FL_e} X$, $\mathsf{FL_{ew}} X$, $\mathsf{CFL_e} X$, $\mathsf{CFL_{ew}} X$, where $X \subseteq \{K, T, KD, Lw_\Box\}$, as well as $\mathsf{CPC}$, $\mathsf{K}$, and $\mathsf{KD}$ have the UIP.
\end{cor}
\begin{proof}
For $\mathsf{FL_e} X$ and $\mathsf{FL_{ew}} X$ (resp. $\mathsf{CFL_e} X$ and $\mathsf{CFL_{ew}} X$), the calculi introduced in Section \ref{sec: Prelim} are single-conclusion (resp. multi-conclusion) semi-analytic. These calculi are terminating by Theorem \ref{Thm: terminating calculi}. Thus, the result follows from Theorem \ref{MainTheorem}.  
For $\mathsf{CPC}$, $\mathsf{K}$, and $\mathsf{KD}$, we rely on the calculi $\mathbf{Gcs}$, $\mathbf{Gcs}\{K\}$, and $\mathbf{Gcs}\{K, D\}$, respectively. Clearly, these calculi are multi-conclusion semi-analytic and by Theorem \ref{Thm: terminating calculi}, they are also terminating. Consequently, UIP also holds for these logics.
\end{proof} 

As for negative type of applications of Theorem \ref{MainTheorem}, in \cite{Craig} we showed that a sufficiently strong logic without CIP cannot have a semi-analytic calculus. Using this result, we demonstrated that many families of logics, such as all superintuitionistic logics (except for at most seven), and all extensions of $\mathsf{S4}$ (except at most 37) lack a semi-analytic calculus. 
Here, we extend this negative result by showing that even if we weaken the assumption from the absence of CIP to the lack of UIP, the logic still faces a structural limitation: it cannot have a \emph{terminating} semi-analytic calculus. This broadens the scope of our previous findings, demonstrating that even a weaker failure of interpolation imposes strict constraints on the possible sequent calculi for the logic. 

In general, for modal and substructural logics, CIP is strictly weaker than UIP. Thus, our new negative result has its own specific family of applications. For example, although $\mathsf{K4}$ and $\mathsf{S4}$ satisfy CIP, they fail to satisfy UIP. Hence, by Theorem \ref{MainTheorem}, we have:



\begin{cor}\label{Cor: modal}
The logics $\mathsf{K4}$ and $\mathsf{S4}$ lack a terminating single-conclusion or multi-conclusion semi-analytic sequent calculus.
\end{cor}

\section{Proof of the Main Theorem} \label{Sec: TheProof}

In this section, we will prove the main theorem, Theorem \ref{MainTheorem}. Our strategy begins by defining a suitable version of UIP for a sequent calculus, which requires formulas of the form $\exists p T$ and $\forall p S$ with certain properties for any sequents $S$ and $T$ and any atom $p$. Then, as the calculus is terminating, it must be terminating with respect to a well-founded order $\preceq$. We will use recursion on this order to define $\exists p T$ and $\forall p S$, and apply induction on $\preceq$ to prove that they satisfy the expected properties. 

First, let us begin with the sequent-style versions of UIP.

\begin{dfn}\label{DfnRelativeInterpolation}
Let $\mathcal{L}\in \{\mathcal{L}_p, \mathcal{L}_{\Box}\}$. A multi-conclusion calculus $G$ over $\mathcal{L}$ has \emph{multi-conclusion uniform interpolation}, if for any sequents $T$ and $S$ over $\mathcal{L}$ and any atom $p$, there exist $p$-free $\mathcal{L}$-formulas $\exists p T$ and $\forall p S$ such that:
\begin{enumerate}
\item[$(var)$]\label{var}  $\mathcal{V}(\forall p S) \subseteq \mathcal{V}(S)$ and $\mathcal{V}(\exists p T) \subseteq \mathcal{V}(T)$,
\item[$(i)$]\label{i}  
$T \cdot (\Rightarrow \exists p T)$ is derivable in $G$,
\item[$(ii)$]\label{ii} 
for any (possibly empty) $p$-free multisets of $\mathcal{L}$-formulas $\bar{C}$ and $\bar{D}$, if $T \cdot (\bar{C} \Rightarrow \bar{D})$ is derivable in $G$, then ($\exists p T, \bar{C} \Rightarrow \bar{D}$) is derivable in $G$,
\item[$(iii)$]\label{iii}
$S \cdot (\forall p S \Rightarrow)$ is derivable in $G$, and
\item[$(iv)$]\label{iv} 
for any (possibly empty) 
$p$-free multisets of $\mathcal{L}$-formulas $\bar{C}$ and $\bar{D}$, if $S \cdot (\bar{C} \Rightarrow \bar{D})$ is derivable in $G$ then ($\bar{C} \Rightarrow \forall p S, \bar{D}$) is derivable in $G$.
\end{enumerate}
For a single-conclusion calculus, having the \emph{single-conclusion uniform interpolation} is defined as above, except that $T$ ranges over the sequents with empty succedents, $S$ ranges over single-conclusion sequents, in $(iv)$ we assume $\bar{D}=\emptyset$, and in $(ii)$ we quantify over $|\bar{D}| \leq 1$.
\end{dfn}

\begin{rem}
Similarly to the definition of UIP for logics, since $\exists p T$ and $\forall p S$ are provably unique, we are allowed to use the functional notation as employed in Definition \ref{DfnRelativeInterpolation}.
\end{rem}

As expected, the two different notions of UIP must be equivalent:

\begin{thm} \label{SequentImpliesLogic} 
Let $\mathcal{L} \in \{\mathcal{L}_p, \mathcal{L}_{\Box}\}$, $L$ be a logic over $\mathcal{L}$, and $G$ be a single-conclusion (multi-conclusion) calculus over $\mathcal{L}$ for $L$. If $G$ has single-conclusion (multi-conclusion) uniform interpolation, then $L$ has UIP.
\end{thm}

\begin{proof}
Suppose $G$ is a single-conclusion calculus for $L$. Let $\phi$ be an $\mathcal{L}$-formula, $p$ an atom, and let $T = (\phi \Rightarrow \,)$. By Definition \ref{DfnRelativeInterpolation}, there exists a $p$-free formula $\exists p T$ such that $\mathcal{V}(\exists p T) \subseteq \mathcal{V}(\phi)$, $G \vdash T \cdot (\, \Rightarrow \exists p T) = (\phi \Rightarrow \exists p T)$, and for any $p$-free multiset of $\mathcal{L}$-formulas $\bar{D}$, where $|\bar{D}| \leq 1$, if $G \vdash T \cdot (\, \Rightarrow \bar{D}) = (\phi \Rightarrow \bar{D})$, then $G \vdash \exists p T \Rightarrow \bar{D}$. Set $\exists p \phi$ as $\exists p T$. Therefore, $\exists p \phi$ is $p$-free, and $\mathcal{V}(\exists p \phi) \subseteq \mathcal{V}(\phi)$. For the other properties, since $G$ is a calculus for $L$, by Definition \ref{dfnLogicOfG}, we have $L \vdash \phi \to \psi$ if and only if $G \vdash \phi \Rightarrow \psi$. Therefore, $L \vdash \phi \to \exists p T$, and for any $p$-free formula $\psi$, if $L \vdash \phi \to \psi$, then $L \vdash \phi \to \exists p T$. Hence, properties $(i)$ and $(ii)$ of Definition \ref{DfnUniformInterpolation} are satisfied. A similar argument also applies for $\forall p \phi$ defined as $\forall p S$, where $S = (\, \Rightarrow \phi)$. The case where $G$ is multi-conclusion is analogous.
\end{proof}

\begin{convention} \label{convention}
To simplify the presentation of proofs and for the sake of brevity, we introduce the following notation for the remainder of the paper. For any substitution $\sigma$, we denote the results of the substitution as follows:
\[
\sigma(\bar{\mu}_{ir}) = \bar{\phi}_{ir}, \quad \sigma(\bar{\nu}_{ir}) = \bar{\psi}_{ir}, \quad \sigma(\bar{\nu}_{js}) = \bar{\psi}_{js}, \quad \sigma(\bar{\rho}_{js}) = \bar{\theta}_{js}, \quad \sigma(\mu) = \phi.
\]
Moreover, throughout this section, we fix a language $\mathcal{L} \in \{\mathcal{L}_p, \mathcal{L}_{\Box}\}$ and refer to it only when necessary.
\end{convention}

\subsection{Focused Axioms}

To simplify the presentation of the proof of the main theorem proof, we first prove a similar claim that focuses solely on the axioms.

\begin{thm} \label{UniformInterpolationForAxiom}
Let $\mathcal{A}$ be a finite set of focused axioms, and $G$ be a multi-conclusion calculus extending $\mathbf{CFL_e}+ \mathcal{A}$. Then, for any sequents $S$ and $T$ and any atom $p$, there exist $p$-free formulas $\exists_{\!\mathcal{A}} p T$ and $\forall_{\!\mathcal{A}} p S$ such that:
\begin{enumerate}
\item[$(var)$]\label{var}  $\mathcal{V}(\exists_{\!\mathcal{A}} p T) \subseteq \mathcal{V}(T)$ and $\mathcal{V}(\forall_{\!\mathcal{A}} p S) \subseteq \mathcal{V}(S)$,
\item[$(i)$]\label{i}  
$T \cdot (\Rightarrow \exists_{\!\mathcal{A}} p T)$ is derivable in $G$,
\item[$(ii)$]\label{ii} 
for any (possibly empty) $p$-free multisets of formulas $\bar{C}$ and $\bar{D}$, if $T \cdot (\bar{C} \Rightarrow \bar{D})$ is an instance of an axiom in $\mathcal{A}$,  then $(\exists_{\!\mathcal{A}} p T, \bar{C} \Rightarrow \bar{D})$ is derivable in $G$,
\item[$(iii)$]\label{iii}
$S \cdot (\forall_{\!\mathcal{A}} p S \Rightarrow)$ is derivable in $G$,    
\item[$(iv)$]\label{iv} 
for any (possibly empty) 
$p$-free multisets of formulas $\bar{C}$ and $\bar{D}$, if $S \cdot (\bar{C} \Rightarrow \bar{D})$ is an instance of an axiom in $\mathcal{A}$, then $(\bar{C} \Rightarrow \forall_{\!\mathcal{A}} p S, \bar{D})$ is derivable in $G$.
\end{enumerate}
A similar claim holds for a finite set of single-conclusion focused axioms and a single-conclusion calculus extending $\mathbf{FL_e}+\mathcal{A}$, provided that $T$ ranges over sequents with empty succedents, $S$ ranges over single-conclusion sequents, $\bar{D} = \emptyset$ in $(iv)$, and $|\bar{D}| \leq 1$ in $(ii)$.
\end{thm}

\begin{proof}
We first prove the single-conclusion case, which is a bit more complicated than the multi-conclusion case. Let $\mathcal{A}$ be a finite set of single-conclusion focused axioms, and let $G$ be a single-conclusion calculus extending $\mathbf{FL_e}$ that proves all instances of the axioms in $\mathcal{A}$. Therefore, w.l.o.g., we can assume that $\mathbf{FL_e} + \mathcal{A} \subseteq G$.
First, for any sequent $T$ such that $T^s = \emptyset$, we present a formula $\exists_{\!\mathcal{A}} p T$ and prove its claimed properties. Let $T= (\Pi \Rightarrow \,)$. Define:
\begin{center}
$\exists_{\!\mathcal{A}} p T := (\bigast \Pi_p * A_1) \wedge A_2 \wedge A_3,$   
\end{center}
where $\Pi_p$ is the multiset of all $p$-free formulas in $\Pi$ and for any $i \in \{1, 2, 3\}$, the formula $A_i$ is defined as $\top$, except in the following cases:
\begin{equation*}
  \begin{cases}
   \text{$A_1=1$} & \text{if $\Pi=\Pi_p$}\\
     \text{$A_2=0$} & \text{if $T$ is an instance of an axiom in the form $(\bar{\rho} \Rightarrow \,) \in \mathcal{A}$} \\
     \text{$A_3=\bot$} & \text{if $T$ is an instance of an axiom in the form $(\tGamma , \bar{\mu} \Rightarrow \tDelta) \in \mathcal{A}$} \\
  \end{cases}
\end{equation*}
\normalsize 
Clearly, $\exists_{\!\mathcal{A}} p T$ is a $p$-free formula and $\mathcal{V}(\exists_{\!\mathcal{A}} p T) \subseteq \mathcal{V}(T)$. To prove $(i)$, i.e., $G \vdash \Pi \Rightarrow \exists_{\!\mathcal{A}} p T$, it is enough to prove $G \vdash \Pi \Rightarrow c$, for each conjunct $c$ of $\exists_{\!\mathcal{A}} p T$. For the first conjunct, note that $\Pi=\Pi_p \cup (\Pi - \Pi_p)$. For any $\psi \in \Pi_p$, we have $G \vdash \psi \Rightarrow \psi$, by the axiom $(id)$ of $\mathbf{FL_e}$. Applying the rule $(R*)$ for $|\Pi_p|$-many times, we get $G \vdash \Pi_p \Rightarrow \bigast \Pi_p$. If $\Pi= \Pi_p$, by the axiom $(\Rightarrow 1)$ and $(R*)$, we get $G \vdash \Pi \Rightarrow \bigast \Pi_p *1$. If $\Pi \neq \Pi_p$, by the axiom $(\tGamma \Rightarrow \top)$ in $\mathbf{FL_e}$, we get $G \vdash \Pi - \Pi_p \Rightarrow \top$ and by $(R*)$, we reach $G \vdash \Pi_p, \Pi - \Pi_p \Rightarrow (\bigast \Pi_p) * \top$ or equivalently $G \vdash \Pi \Rightarrow (\bigast \Pi_p) * \top$. Therefore, in any case, we have $G \vdash \Pi \Rightarrow (\bigast \Pi_p) * A_1$.

For the conjunct $A_i$, where $i \in \{2, 3\}$, if $A_i=\top$, it is clear that $G \vdash \Pi \Rightarrow A_i$. Therefore, it is enough to check the exceptional cases.
For $A_2$, if $T$ is an instance of an axiom in the form $(\bar{\rho} \Rightarrow) \in \mathcal{A}$, then $A_2=0$. As $\mathcal{A} \subseteq G$, we have $G \vdash (\Pi \Rightarrow \,)$. Then, by $(0 w)$, we get $G \vdash \Pi \Rightarrow 0$. Hence, $G \vdash \Pi \Rightarrow A_2$. 

For $A_3$, assume that $T= (\Pi \Rightarrow)$ is a $\sigma$-instance of an axiom $(\tGamma, \bar{\mu} \Rightarrow \tDelta) \in \mathcal{A}$. Then, $A_3=\bot$, by definition. It is easy to see that $\Pi \Rightarrow \bot$ is an instance of the same axiom, using the substitution $\sigma$ for $\tGamma$ and $\bar{\mu}$, while substituting $\tDelta$ by $\{\bot\}$. As $\mathcal{A} \subseteq G$, any instance of $(\tGamma , \bar{\mu} \Rightarrow \tDelta)$, such as $\Pi \Rightarrow \bot$, is provable in $G$. Hence, $G \vdash \Pi \Rightarrow A_3$. This completes the proof of $(i)$. 

For $(ii)$, let $\bar{C}$ and $\bar{D}$ be $p$-free multisets of formulas such that $|\bar{D}| \leq 1$ and $\Pi, \bar{C} \Rightarrow \bar{D}$ is an instance of a meta-sequent in $\mathcal{A}$. We want to prove that $G \vdash \exists_{\!\mathcal{A}} p T, \bar{C} \Rightarrow \bar{D}$. There are five possibilities to consider, according to the form of the used meta-sequent in $\mathcal{A}$:

\item  $\bullet$ If
$(\Pi, \bar{C} \Rightarrow \bar{D}) = (\phi \Rightarrow \phi)$ is an instance of the axiom $(\mu \Rightarrow \mu)$, then $\phi$ is $p$-free since $\bar{D}=\{\phi\}$. Now, there are two cases to consider: either $\Pi=\{\phi\}$ or $\Pi=\emptyset$. If $\Pi=\{\phi\}$, then $\bar{C} = \emptyset$ and $\Pi = \Pi_p=\{\phi\}$. As $\Pi=\Pi_p$, we have $A_1=1$. Therefore, as $G \vdash \phi * 1 \Rightarrow \phi$, we reach $G \vdash \bigast \Pi_p * A_1 \Rightarrow \phi$. Finally, by $(L \wedge)$, we get $G \vdash \exists_{\!\mathcal{A}} p T, \bar{C} \Rightarrow \bar{D}$.

If $\Pi = \emptyset$, then $\bar{C} =\{\phi\}$ and $\Pi = \Pi_p= \emptyset$. Hence, by definition, $\bigast \Pi_p =1$ and $A_1=1$. As $(\bar{C} \Rightarrow \bar{D})$ is an instance of the axiom $(id)$ in $G$, we have $G \vdash \bar{C} \Rightarrow \bar{D}$. Now, by $(1 w)$ and $(L *)$, we get $G \vdash \bigast \Pi_p *A_1, \bar{C} \Rightarrow \bar{D}$ and by $(L \wedge)$, we reach $G \vdash \exists_{\!\mathcal{A}} p T , \bar{C} \Rightarrow \bar{D}$.

\item[] $\bullet$ 
If $(\Pi , \bar{C} \Rightarrow \bar{D})=(\Rightarrow \bar{\psi})$ is an instance of the axiom $(\, \Rightarrow \bar{\nu}) \in \mathcal{A}$, then $\Pi = \bar{C} = \emptyset$ and $\bar{\psi}$ is $p$-free, as $\bar{D}=\bar{\psi}$. As $\mathcal{A} \subseteq G$, we have $G \vdash \, \Rightarrow \bar{\psi}$. As $\Pi=\Pi_p=\emptyset$, by definition $\bigast \Pi_p=A_1=1$. Therefore, by $(1w)$ and $(L*)$, we have $G \vdash \bigast \Pi_p *A_1 \Rightarrow \bar{\psi}$. Finally, by $(L\wedge)$, we reach $G \vdash \exists_{\!\mathcal{A}} p T \Rightarrow \bar{\psi}$ or equivalently $G \vdash \exists_{\!\mathcal{A}} p T, \bar{C} \Rightarrow \bar{D}$.

\item[] $\bullet$ 
If $(\Pi , \bar{C} \Rightarrow \bar{D})=(\bar{\theta} \Rightarrow)$ is an instance of the axiom $(\bar{\rho} \Rightarrow) \in \mathcal{A}$, then there are two cases to consider: either $\Pi=\bar{\theta}$ or $\Pi \subsetneq \bar{\theta}$. If $\Pi=\bar{\theta}$, then $\bar{C}= \bar{D}= \emptyset$. Moreover, as $T$ is an instance of the axiom $(\bar{\rho} \Rightarrow) \in \mathcal{A}$, we have $A_2=0$. By the axiom $(0 \Rightarrow \,)$, we get $G \vdash 0, \bar{C} \Rightarrow \bar{D}$ or equivalently $G \vdash A_2, \bar{C} \Rightarrow \bar{D}$. Therefore, by $(L \wedge)$, we get $G \vdash \exists_{\!\mathcal{A}} p T, \bar{C} \Rightarrow \bar{D}$.

If $\Pi \subsetneq \bar{\theta}$, as $\bar{\theta}= \Pi \cup \bar{C}$, we have $\bar{C} \neq \emptyset$. As $\bar{C}$ is $p$-free and by Definition \ref{Dfn Focused Axioms}, any two formulas in $\bar{\theta}$ have the same variables, every formula in $\Pi$ is $p$-free and hence $\Pi=\Pi_p$ which implies $A_1=1$. As $\mathcal{A} \subseteq G$, we have $G \vdash \Pi, \bar{C} \Rightarrow \bar{D}$ or equivalently $G \vdash \Pi_p, \bar{C} \Rightarrow \bar{D}$. Therefore, by $(1w)$ and $(L *)$, we get $G \vdash \bigast \Pi_p * A_1, \bar{C} \Rightarrow \bar{D}$ and by $(L \wedge)$, we reach $G \vdash \exists_{\!\mathcal{A}} p T, \bar{C} \Rightarrow \bar{D}$.

\item[] $\bullet$ 
Assume that $(\Pi , \bar{C} \Rightarrow \bar{D})$ is the $\sigma$-instance of the axiom $(\tGamma, \bar{\mu} \Rightarrow \tDelta) \in \mathcal{A}$ for a substitution $\sigma$, and set $\sigma(\bar{\mu})=\bar{\phi}$. Then, there are two cases to consider: either $\bar{\phi} \subseteq \Pi$ or $\bar{\phi} \nsubseteq \Pi$. 
If $\bar{\phi} \subseteq \Pi$, then $T=(\Pi \Rightarrow \,)$ is also an instance of the axiom $(\tGamma, \bar{\mu} \Rightarrow \tDelta) \in \mathcal{A}$. Hence, $A_3=\bot$. As $G \vdash \bar{C}, \bot \Rightarrow \bar{D}$ or equivalently $G \vdash \bar{C}, A_3 \Rightarrow \bar{D}$, by $(L \wedge)$, we get $G \vdash \exists_{\!\mathcal{A}} p T , \bar{C} \Rightarrow \bar{D}$.

If $\bar{\phi} \nsubseteq \Pi$, then at least one of the elements of $\bar{\phi}$ is in $\bar{C}$ and hence it is $p$-free. Since by Definition \ref{Dfn Focused Axioms} any two formulas in $\bar{\phi}$ have the same variables, $\bar{\phi}$ is $p$-free. Hence, $\bar{\phi} \subseteq \Pi_p \cup \bar{C}$. Therefore, the sequent $\Pi_p, \bar{C}, A_1 \Rightarrow \bar{D}$ is an instance of the axiom $(\tGamma, \bar{\mu} \Rightarrow \tDelta) \in \mathcal{A}$.
Now, as $\mathcal{A} \subseteq G$, we obtain $G \vdash \Pi_p, \bar{C}, A_1 \Rightarrow \bar{D}$. By $(L *)$, we get $G \vdash \bigast \Pi_p *A_1, \bar{C} \Rightarrow \bar{D}$ and by $(L \wedge)$, we reach $G \vdash \exists_{\!\mathcal{A}} p T , \bar{C} \Rightarrow \bar{D}$.

\item[] $\bullet$ 
Assume that $(\Pi , \bar{C} \Rightarrow \bar{D})$ is the $\sigma$-instance of the axiom $(\tGamma \Rightarrow \bar{\mu})  \in \mathcal{A}$ for a substitution $\sigma$, and set $\sigma(\bar{\mu})=\bar{\phi}$. Then, as $\bar{\phi} = \bar{D}$, the sequent $\exists_{\!\mathcal{A}} p T , \bar{C} \Rightarrow \bar{D}$ is an instance of the same axiom. As $\mathcal{A} \subseteq G$, we reach $G \vdash \exists_{\!\mathcal{A}} p T , \bar{C} \Rightarrow \bar{D}$.\\

Now, for any single-conclusion sequent $S=(\Sigma \Rightarrow \Lambda)$, we present a formula $\forall_{\!\mathcal{A}} p S$ satisfying the claimed properties. 
Define: 
\begin{center}
$\forall_{\!\mathcal{A}} p S := (\bigast \Sigma_p \to B_1) \vee B_2 \vee B_3 \vee B_4,$   
\end{center}
where $\Sigma_p$ is the multiset of all $p$-free formulas in $\Sigma$ and for any $i \in \{1, 2, 3, 4\}$, the formula $B_i$ is defined as $\bot$, except in the following cases:
\begin{equation*}
  \begin{cases}
   \text{$B_1=0$} & \text{if $\Sigma=\Sigma_p$ and $\Lambda=\emptyset$}\\
\text{$B_2=\phi$} & \text{if $\Sigma=\emptyset$ and $\Lambda=\{\phi\}$ is $p$-free} \\
      \text{$B_2=0$} & \text{if $\Sigma=\Lambda=\emptyset$} \\
     \text{$B_3=1$} & \text{if $S$ is an instance of $(\mu \Rightarrow \mu)$ or some axiom in the form} \\
     &
     \text{$(\Rightarrow \bar{\nu})$ or $(\bar{\rho} \Rightarrow)$ in $\mathcal{A}$} \\
     \text{$B_4=\top$} & \text{if $S$ is an instance of an axiom in the form} \\
     & \text{$(\tGamma, \bar{\mu}\Rightarrow \tDelta) \in \mathcal{A}$ or $(\tGamma \Rightarrow \bar{\mu}) \in \mathcal{A}$}
  \end{cases}
\end{equation*}
\normalsize 
Clearly, $\forall_{\!\mathcal{A}} p S$ is a $p$-free formula and $\mathcal{V}(\forall_{\!\mathcal{A}} p S) \subseteq \mathcal{V}(\Sigma \cup \Lambda)$. For $(iii)$, i.e., $G \vdash S \, . \, (\forall_{\!\mathcal{A}} p S \Rightarrow) = (\Sigma, \forall_{\!\mathcal{A}} p S \Rightarrow \Lambda)$, it is enough to prove $G \vdash \Sigma, d \Rightarrow \Lambda$, for any disjunct $d$ in the definition of $\forall_{\!\mathcal{A}} p S$.  
For the first disjunct, if $\Sigma=\Sigma_p$ and $\Lambda=\emptyset$, we have $B_1=0$. Then, as $G \vdash \Sigma \Rightarrow \bigast \Sigma_p$ and $G \vdash (0 \Rightarrow \,)$, by $(L\!\to)$, we reach $G \vdash \Sigma, (\bigast \Sigma_p \to 0) \Rightarrow$ or equivalently $G \vdash \Sigma, (\bigast \Sigma_p \to B_1) \Rightarrow \Lambda$.  
Otherwise, $B_1=\bot$. As $G \vdash \Sigma_p \Rightarrow \bigast \Sigma_p$ and $G \vdash \Sigma- \Sigma_p, \bot \Rightarrow \Lambda$, by $(L\!\to)$, we get $G \vdash \Sigma, (\bigast \Sigma_p \to \bot) \Rightarrow \Lambda$ or equivalently $G \vdash \Sigma, (\bigast \Sigma_p \to B_1) \Rightarrow \Lambda$.

For the other disjuncts, if $B_i=\bot$ for any $i \in \{2, 3, 4\}$, then $G \vdash \Sigma, B_i \Rightarrow \Lambda$. Thus, it suffices to address the exceptional cases. For $B_2$, if $\Sigma=\emptyset$ and $\Lambda=\{\phi\}$ is $p$-free, then $B_2=\phi$. Clearly, $G \vdash \phi \Rightarrow \Lambda$. Hence, $G \vdash \Sigma, \phi \Rightarrow \Lambda$ or equivalently $G \vdash \Sigma, B_2 \Rightarrow \Lambda$. The case $\Sigma=\Lambda=\emptyset$ is similar. 

For $B_3$, if $\Sigma \Rightarrow \Lambda$ is an instance of an axiom in the form $(\mu \Rightarrow \mu)$,  $(\, \Rightarrow \bar{\nu})$, or $(\bar{\rho} \Rightarrow \,)$ in $\mathcal{A}$, then $B_3=1$. As $\mathcal{A} \subseteq G$, we have $G \vdash \Sigma \Rightarrow \Lambda$. Therefore, by $(1w)$, we get $G \vdash \Sigma, 1 \Rightarrow \Lambda$. Hence, $G \vdash \Sigma, B_3 \Rightarrow \Lambda$.  

For $B_4$, if $\Sigma \Rightarrow \Lambda$ is an instance of an axiom in the form $(\tGamma, \bar{\mu}\Rightarrow \tDelta)$ or $(\tGamma, \Rightarrow \bar{\mu}, \tDelta)$ in $\mathcal{A}$, then $B_4=\top$. If we add $\top$ to the antecedent of $\Sigma \Rightarrow \Lambda$, it is also an instance of the same axiom in $\mathcal{A}$. Thus, we get $G \vdash \Sigma, \top \Rightarrow \Lambda$ or equivalently $G \vdash \Sigma, B_4 \Rightarrow \Lambda$, since $\mathcal{A} \subseteq G$. This finishes the proof of $(iii)$.

For $(iv)$, we prove that for any $p$-free multiset $\bar{C}$ of formulas, if $\Sigma, \bar{C} \Rightarrow \Lambda$ is an instance of a meta-sequent in $\mathcal{A}$, then $G \vdash \bar{C} \Rightarrow \forall_{\!\mathcal{A}} p S$. 
Again, five cases arise, depending on the form of the meta-sequent used in $\mathcal{A}$:

\item[] $\bullet$ 
If $(\Sigma, \bar{C} \Rightarrow \Lambda) = (\phi \Rightarrow \phi)$ is an instance of the axiom $(\mu \Rightarrow \mu)$, then there are two cases to consider: either $\Sigma=\emptyset$ or $\Sigma=\{\phi\}$. If $\Sigma=\emptyset$ then $\bar{C}=\Lambda=\{\phi\}$. Thus, the formula $\phi$ is $p$-free and hence $B_2=\phi$. Since $G \vdash \bar{C} \Rightarrow \phi$, we have $G \vdash \bar{C} \Rightarrow B_2$ and hence $G \vdash \bar{C} \Rightarrow \forall_{\!\mathcal{A}} p S$, by $(R\vee)$. If $\Sigma = \Lambda=\{\phi\}$ then $\bar{C}=\emptyset$. Thus, $\Sigma \Rightarrow \Lambda$ is an instance of the axiom $(\mu \Rightarrow \mu)$ which implies $B_3=1$. Since $G \vdash (\, \Rightarrow 1)$ and $\bar{C}=\emptyset$, we have $G \vdash \bar{C} \Rightarrow \forall_{\!\mathcal{A}} p S$, by $(R\vee)$.

\item[] $\bullet$ 
If $(\Sigma, \bar{C} \Rightarrow \Lambda)=(\Rightarrow \bar{\psi})$ is an instance of the axiom $(\, \Rightarrow \bar{\nu}) \in \mathcal{A}$, then $\Sigma=\bar{C}=\emptyset$ and $\Lambda=\bar{\psi}$. Thus, $B_3=1$ and as in the prior case $G \vdash \bar{C} \Rightarrow \forall_{\!\mathcal{A}} p S$.

\item[] $\bullet$ 
If $(\Sigma, \bar{C} \Rightarrow \Lambda)=(\bar{\theta} \Rightarrow)$ is an instance of the axiom $(\bar{\rho} \Rightarrow) \in \mathcal{A}$, then there are three cases to consider: either $\Sigma=\emptyset$ or $\bar{C}=\emptyset$ or both $\Sigma$ and $\bar{C}$ are non-empty. If $\Sigma =\emptyset$, then $(\bar{C} \Rightarrow \,)=(\bar{\theta} \Rightarrow)$ is an instance of an axiom in $\mathcal{A}$. Since $\mathcal{A} \subseteq G$, we have $G \vdash (\bar{C} \Rightarrow \,)$, and hence $G \vdash \bar{C} \Rightarrow 0$ by $(0w)$. As $\Sigma=\emptyset$ and $\Lambda=\emptyset$, we have $B_2=0$ by definition. Thus, $G \vdash \bar{C} \Rightarrow B_2$, which implies $G \vdash \bar{C} \Rightarrow \forall_{\!\mathcal{A}} p S$ by $(R\vee)$.  

If $\bar{C}=\emptyset$, then $\Sigma \Rightarrow \Lambda$ is an instance of the axiom $(\bar{\rho} \Rightarrow \,) \in \mathcal{A}$. Thus, $B_3=1$. As $G \vdash (\, \Rightarrow 1)$, we get $G \vdash \bar{C} \Rightarrow B_3$ and by $(R\vee)$, $G \vdash \bar{C} \Rightarrow \forall_{\!\mathcal{A}} p S$.  

If $\Sigma$ and $\bar{C}$ are non-empty, then since any pair of elements in $\bar{\theta}= \Sigma \cup \bar{C}$ share the same variables and $\bar{C}$ is $p$-free, the multiset $\Sigma$ is also $p$-free. Hence, $\Sigma=\Sigma_p$. As $\Lambda=\emptyset$, we have $B_1=0$. Since $\Sigma_p \cup \bar{C}=\bar{\theta}$, we have $G \vdash \Sigma_p, \bar{C} \Rightarrow $, which implies $G \vdash \Sigma_p, \bar{C} \Rightarrow 0$ by $(0w)$. Hence, $G \vdash \bar{C} \Rightarrow \bigast \Sigma_p \to B_1$. Therefore, $G \vdash \bar{C} \Rightarrow \forall_{\!\mathcal{A}} p S$ by $(R\vee)$.

\item[] $\bullet$ 
Let $(\Sigma, \bar{C} \Rightarrow \Lambda)$ be the $\sigma$-instance of the axiom $(\tGamma, \bar{\mu} \Rightarrow \tDelta) \in \mathcal{A}$ and set $\sigma(\bar{\mu})=\bar{\phi}$. Then, there are two cases to consider: either $\bar{\phi} \subseteq \Sigma$ or $\bar{\phi} \nsubseteq \Sigma$. If $\bar{\phi} \subseteq \Sigma$, then $S=(\Sigma \Rightarrow \Lambda)$ is also an instance of the axiom $(\tGamma, \bar{\mu} \Rightarrow \tDelta) \in \mathcal{A}$ and hence $B_4=\top$. 
Now, as $G \vdash (\bar{C} \Rightarrow \top)$, we reach $G \vdash (\bar{C} \Rightarrow B_4)$. Therefore,  
$G \vdash \bar{C} \Rightarrow \forall_{\!\mathcal{A}} p S$, by the rule $(R\vee)$.  

If $\bar{\phi} \not\subseteq \Sigma$, then at least one of the elements of $\bar{\phi}$ is in $\bar{C}$ and hence $p$-free. As any pair of the elements in $\bar{\phi}$ have the same variables, $\bar{\phi}$ is $p$-free and hence $\bar{\phi} \subseteq \Sigma_p \cup \bar{C}$. Therefore, $(\Sigma_p, \bar{C} \Rightarrow B_1)$ is an instance of the axiom $(\tGamma, \bar{\mu} \Rightarrow \tDelta)$.  
Now, as $\mathcal{A} \subseteq G$, we reach $G \vdash \Sigma_p, \bar{C} \Rightarrow B_1$ which implies $G \vdash \bar{C} \Rightarrow (\bigast \Sigma_p \to B_1)$. Therefore, $G \vdash \bar{C} \Rightarrow \forall_{\!\mathcal{A}} p S$, by $(R\vee)$.  

\item[] $\bullet$ 
Let $(\Sigma, \bar{C} \Rightarrow \Lambda)$ be the $\sigma$-instance of the axiom $(\tGamma \Rightarrow \bar{\mu}) \in \mathcal{A}$ and set $\sigma(\bar{\mu})=\bar{\phi}$. Then, $\bar{\phi} = \Lambda$. Thus, $S=(\Sigma \Rightarrow \Lambda)$ is also an instance of the axiom $(\tGamma \Rightarrow \bar{\mu}) \in \mathcal{A}$ and hence $B_4=\top$.  
Now, similar to the first part of the previous case, we get $G \vdash \bar{C} \Rightarrow \forall_{\!\mathcal{A}} p S$. This concludes the proof for single-conclusion $G$. \\

Now, let $\mathcal{A}$ be a finite set of focused axioms and $G$ be a multi-conclusion calculus extending $\mathbf{CFL_e}+\mathcal{A}$. For any sequent $S=(\Sigma \Rightarrow \Lambda)$, define $\exists_{\!\mathcal{A}} p S$:
\begin{center}
    $\exists_{\!\mathcal{A}} p S:= [(\bigast \Sigma_p) * (\bigast \neg \Lambda_p) * A_1] \wedge A_2 \wedge A_3, $
\end{center}
where $\Sigma_p$ (resp. $\Lambda_p$) is the multiset of all $p$-free formulas in $\Sigma$ (resp. $\Lambda$) and for any $i \in \{1, 2, 3\}$, the formula $A_i$ is defined as $\top$ except in these cases:
\begin{equation*}
  \begin{cases}
   \text{$A_1=1$} & \text{if $\Sigma=\Sigma_p$ and $\Lambda=\Lambda_p$}\\
      \text{$A_2=0$} & \text{if $S$ is an instance of $(\mu \Rightarrow \mu)$ or some axiom in the form} \\
     &
     \text{$(\Rightarrow \bar{\nu})$ or $(\bar{\rho} \Rightarrow)$ in $\mathcal{A}$} \\
     \text{$A_3=\bot$} & \text{if $S$ is an instance of an axiom in the form} \\
     & \text{$(\tGamma, \bar{\mu}\Rightarrow \tDelta) \in \mathcal{A}$ or $(\tGamma, \Rightarrow \bar{\mu}, \tDelta) \in \mathcal{A}$}\\
  \end{cases}
\end{equation*}
\normalsize
Clearly, $\exists_{\!\mathcal{A}} p S$ is a $p$-free formula and $\mathcal{V}(\exists_{\!\mathcal{A}} p S) \subseteq \mathcal{V}(\Sigma \cup \Lambda)$. For $(i)$, i.e., $G \vdash \Sigma \Rightarrow \exists_{\!\mathcal{A}} p S, \Lambda$, it is enough to prove $G \vdash \Sigma \Rightarrow c, \Lambda$, for any conjunct $c$ of $\exists_{\!\mathcal{A}} p S$. For the first conjunct, if $\Sigma = \Sigma_p$ and $\Lambda=\Lambda_p$, then $A_1=1$. As $\mathbf{CFL_e} \subseteq G$, then $G \vdash \Sigma \Rightarrow \bigast \Sigma_p * 1$ and $G \vdash \, \Rightarrow \bigast (\neg \Lambda_p), \Lambda$. Thus, $G \vdash \Sigma \Rightarrow (\bigast \Sigma_p) * (\bigast \neg\Lambda_p) * A_1, \Lambda$. If $\Sigma \neq \Sigma_p$ or $\Lambda \neq \Lambda_p$, then $A_1=\top$. We have  $G \vdash \Sigma_p \Rightarrow (\bigast \Sigma_p) * (\bigast \neg \Lambda_p), \Lambda_p$ and $G \vdash \Sigma-\Sigma_p \Rightarrow \top, (\Lambda-\Lambda_p)$. Thus, $G \vdash \Sigma \Rightarrow (\bigast \Sigma_p) * (\bigast \neg\Lambda_p) * \top, \Lambda$. For the conjunct $A_i$, where $i \in \{2, 3\}$, if $A_i=\top$, clearly $G \vdash \Sigma \Rightarrow A_i, \Lambda$. For the exceptional cases, for $A_2$, if $S$ is an instance of the axiom $(\mu \Rightarrow \mu)$ or an axiom in the form $(\, \Rightarrow \bar{\nu})$ or $(\bar{\rho} \Rightarrow \,)$ in $\mathcal{A}$, then $A_2=0$ and as $\mathcal{A} \subseteq G$, we have $G \vdash \Sigma \Rightarrow \Lambda$. Therefore, $G \vdash \Sigma \Rightarrow 0, \Lambda$, by $(0w)$. For $A_3$, if $S$ is an instance of an axiom in the form $(\tGamma, \bar{\mu} \Rightarrow \tDelta)$ or $(\tGamma \Rightarrow \bar{\mu}, \tDelta)$ in $\mathcal{A}$, then $A_3=\bot$ and if we add $\bot$ to the succedent of $S$, it is still an instance of the same axiom. Thus, as $\mathcal{A} \subseteq G$, we have $G \vdash \Sigma \Rightarrow \bot, \Lambda$. Therefore, $G \vdash \Sigma \Rightarrow A_3, \Lambda$.

For $(ii)$, we prove that for any $p$-free multisets $\bar{C}$ and $\bar{D}$ of formulas, if $\Sigma, \bar{C} \Rightarrow \Lambda, \bar{D}$ is an instance of a meta-sequent in $\mathcal{A}$, then $G \vdash \exists_{\!\mathcal{A}} p S, \bar{C} \Rightarrow \bar{D}$. 
Five cases arise, depending on the form of the meta-sequent used in $\mathcal{A}$:

\item[] $\bullet$ 
If $(\Sigma, \bar{C} \Rightarrow \Lambda, \bar{D}) = (\phi \Rightarrow \phi)$ is an instance of the axiom $(\mu \Rightarrow \mu)$, then there are four cases to consider depending on whether $\Sigma$ or $\Lambda$ are empty or not. We only investigate the cases $\Sigma=\Lambda=\{\phi\}$ and ($\Sigma=\emptyset$ and $\Lambda=\{\phi\}$). The other two are similar.

In the first case, as $\Sigma=\Lambda=\{\phi\}$, the sequent $\Sigma \Rightarrow \Lambda$ is an instance of the axiom $(\mu \Rightarrow \mu)$ and hence $A_2=0$. As $\bar{C}=\bar{D}=\emptyset$ and $G \vdash (0 \Rightarrow \,)$, we get $G \vdash 0, \bar{C} \Rightarrow \bar{D}$ and hence $G \vdash \exists_{\!\mathcal{A}} p S, \bar{C} \Rightarrow \bar{D}$, by $(L\wedge)$. 

In the second case, $\bar{C}=\{\phi\}$ and $\bar{D}=\emptyset$. Thus, $\phi$ is $p$-free which implies $\Lambda=\Lambda_p$. Therefore, as $\Sigma=\Sigma_p=\emptyset$, we have $A_1=1$. 
As $G \vdash (\phi, 1* \neg \phi * 1 \Rightarrow \,)$ and $\Sigma_p=\emptyset$ and $\Lambda_p=\{\phi\}$, we have $G \vdash \bar{C}, (\bigast \Sigma_p) * (\bigast \neg \Lambda_p)*A_1 \Rightarrow \bar{D}$. Hence, $G \vdash \bar{C}, \exists_{\!\mathcal{A}}p S \Rightarrow \bar{D}$, by $(L\wedge)$.

\item[] $\bullet$ 
If $(\Sigma, \bar{C} \Rightarrow \Lambda, \bar{D})=(\Rightarrow \bar{\psi})$ is an instance of the axiom $(\, \Rightarrow \bar{\nu}) \in \mathcal{A}$, then $\Sigma=\bar{C}=\emptyset$. There are two cases to consider: either $\Lambda=\bar{\psi}$ or $\Lambda \neq \bar{\psi}$. If $\Lambda=\bar{\psi}$, then $S=(\Sigma \Rightarrow \Lambda)$ is an instance of the axiom $(\, \Rightarrow \bar{\nu}) \in \mathcal{A}$ and hence $A_2=0$. Since $\bar{D}=\emptyset$ and $G \vdash (0 \Rightarrow \,)$, we get $G \vdash \bar{C}, A_2 \Rightarrow \bar{D}$. and hence $G \vdash \bar{C}, \exists_{\!\mathcal{A}} p S \Rightarrow \bar{D}$, by $(L\wedge)$.

If $\Lambda \neq \bar{\psi}$, then $\bar{D} \subseteq \bar{\psi}$ is non-empty. Thus, there exists a $p$-free formula in $\bar{\psi}$. Since the variables of any pair in $\bar{\psi}$ are equal, $\bar{\psi}$ is $p$-free. Hence, $\Lambda \subseteq \bar{\psi}$ is $p$-free which implies $\Lambda=\Lambda_p$. As $\Sigma=\Sigma_p=\emptyset$, we have $A_1=1$. As $\mathcal{A} \subseteq G$ and $(\, \Rightarrow \bar{\psi})$ is an instance of an axiom in $\mathcal{A}$, we get $G \vdash (\, \Rightarrow \Lambda, \bar{D})$ or equivalently $G \vdash (\, \Rightarrow \Lambda_p, \bar{D})$. Hence, $G \vdash 1*\bigast (\neg \Lambda_p) * 1 \Rightarrow \bar{D}$. As $A_1=1$, $\bigast \Sigma_p=1$ and $\bar{C}=\emptyset$, we reach $G \vdash \bar{C}, (\bigast \Sigma_p)* (\bigast \neg \Lambda_p)* A_1 \Rightarrow \bar{D}$ and hence $G \vdash \bar{C}, \exists_{\!\mathcal{A}} p S \Rightarrow \bar{D}$, by $(L\wedge)$.

\item[] $\bullet$ 
If $\Sigma, \bar{C} \Rightarrow \Lambda, \bar{D}$ is an instance of the axiom $(\bar{\rho} \Rightarrow) \in \mathcal{A}$, then $\Lambda= \bar{D}=\emptyset$. This case is similar to the previous case, switching the roles of $\bar{C}$ and $\Sigma$ with $\bar{D}$ and $\Lambda$, respectively.

\item[] $\bullet$ 
Let $(\Sigma, \bar{C} \Rightarrow \Lambda, \bar{D})$ be the $\sigma$-instance of the axiom $(\tGamma, \bar{\mu} \Rightarrow \tDelta) \in \mathcal{A}$ and set $\sigma(\bar{\mu})=\bar{\phi}$. Then, there are two cases: either $\bar{\phi} \subseteq \Sigma$ or $\bar{\phi} \nsubseteq \Sigma$. If $\bar{\phi} \subseteq \Sigma$, then $\Sigma \Rightarrow \Lambda$ is an instance of the axiom $(\tGamma, \bar{\mu} \Rightarrow \tDelta) \in \mathcal{A}$ and hence $A_3=\bot$. Thus, as $G \vdash \bot, \bar{C} \Rightarrow \bar{D}$, we get $G \vdash \exists_{\!\mathcal{A}} p S, \bar{C} \Rightarrow \bar{D}$, by $(L\wedge)$.

If $\bar{\phi} \nsubseteq \Sigma$, then $\bar{\phi} \cap \bar{C}$ is non-empty. Hence, $\bar{\phi}$ has a $p$-free element and as the variables of any pair in $\bar{\phi}$ are equal, $\bar{\phi}$ is $p$-free. So, $\bar{\phi} \subseteq \Sigma_p \cup \bar{C}$. Thus, $ A_1, \Sigma_p, \bar{C} \Rightarrow \Lambda_p, \bar{D}$ is an instance of the axiom $(\tGamma, \bar{\mu} \Rightarrow \tDelta) \in \mathcal{A}$.  Hence, $ A_1, \Sigma_p, \bar{C} \Rightarrow \Lambda_p, \bar{D}$ is the $\sigma'$-instance of the axiom $(\tGamma, \bar{\mu} \Rightarrow \tDelta) \in \mathcal{A}$. Now, as $\mathcal{A} \subseteq G$, we reach $G \vdash A_1, \Sigma_p, \bar{C} \Rightarrow \Lambda_p, \bar{D}$. Hence, $G \vdash \bigast \Sigma_p * (\bigast \neg \Lambda_p) *A_1, \bar{C} \Rightarrow \bar{D}$. Thus, $G \vdash \bar{C}, \exists_{\!\mathcal{A}} p S \Rightarrow \bar{D}$, by $(L\wedge)$.

\item[] $\bullet$ 
The case where $\Sigma, \bar{C} \Rightarrow \Lambda, \bar{D}$ is an instance of $(\tGamma \Rightarrow  \bar{\mu}, \tDelta) \in \mathcal{A}$ is similar to the previous case, switching the roles of $\bar{C}$ and $\Sigma$ with $\bar{D}$ and $\Lambda$, respectively.

This completes the proof for the properties of $\exists_{\!\mathcal{A}} p S$. For $\forall_{\!\mathcal{A}} p S$, it is enough to define $\forall_{\!\mathcal{A}} p S:=\neg \exists_{\!\mathcal{A}} p S$. Clearly, $\forall_{\!\mathcal{A}} p S$ is $p$-free and $\mathcal{V}(\forall_{\!\mathcal{A}} p S) \subseteq \mathcal{V}(S)$. Using the full duality between the antecedents and the succedents in $\mathbf{CFL_e}$, the conditions $(iii)$ and $(iv)$ are direct consequences of $(i)$ and $(ii)$.
\end{proof}

\subsection{Single-conclusion Case} \label{Subsection1}
In this subsection, we prove the single-conclusion part of Theorem \ref{MainTheorem}. First, we need a definition. 

\begin{dfn}
A \emph{non-trivial partition} of a sequent $S$ is a tuple $(S_1, \ldots, S_n)$ of non-empty sequents such that $n \geq 2$ and $S=S_1 \cdot S_2 \cdot \ldots  \cdot S_n$. 
\end{dfn}

Note that if $(S_1, S_2, \ldots, S_n)$ is a non-trivial partition of $S$, then as $n \geq 2$, each $S_i$ is a proper subsequent of $S$ for every $1 \leq i \leq n$. 

\begin{thm} \label{UniformInterpolation}
Let $G$ be a terminating single-conclusion semi-analytic calculus extending $\mathbf{FL_e}$. Then, $G$ has single-conclusion uniform interpolation.
\end{thm}

\begin{proof}
First, note that any axiom of $G$ is single-conclusion focused. The reason is that any axiom of $G$ is either single-conclusion focused or a single-conclusion semi-analytic rule with no premise, which is itself a single-conclusion focused axiom. Call this set of axioms $\mathcal{A}$.  
Second, since $G$ extends $\mathbf{FL_e}$, all instances of any of the axioms of $\mathbf{FL_e}$ are provable in $G$, and all the rules of $\mathbf{FL_e}$ are admissible in $G$. Hence, without loss of generality, we can assume that these axioms and rules are all available in $G$.  
Third, as $G$ is a terminating calculus, there exists a well-founded order $\preceq$ such that $G$ is terminating with respect to $\preceq$. Therefore, proper subsequents of $S$, premises of any instance of any rule in $G$ with the conclusion $S$, and if $S = (\Box \Gamma \Rightarrow \Box \Delta)$ such that $\Gamma \cup \Delta \neq \emptyset$, the sequent $\Gamma \Rightarrow \Delta$ is strictly below $S$ with respect to $\preceq$, for any single-conclusion sequent $S$.

Now, for any single-conclusion sequents $U$ and $V$ where $V^s=\emptyset$ and any atom $p$, we define $p$-free formulas $\exists p V$ and $\forall p U$ and we prove that they meet the conditions of Definition \ref{DfnRelativeInterpolation}. 
We define $\forall p U$ and $\exists p V$ by recursion on $\preceq$ and prove their properties by induction on $\preceq$. More precisely, we define $\exists p V$ using the value of $\forall p U'$ and $\exists p V'$, for any $U' \prec V$ and $V' \prec V$. Moreover, we define $\forall p U$ using the value of $\forall p U'$ and $\exists p V'$, for any $U' \prec U$ and $V' \preceq U$. 
Notice the equality sign in the last inequality $V' \preceq U$. This means that in the definition of $\forall p U$, if $U^s=\emptyset$, we also allow the use of $\exists p U$. This is unproblematic as $\exists p U$ will be defined prior to $\forall p U$ and the definition of $\exists p U$ only depends on the  sequents lower than $U$. 
After providing the recursive definitions, using the corresponding induction on $\preceq$ as will be explained later, we prove that the defined formulas have the claimed properties.

If $V$ is the empty sequent, define $\exists p V$ as $1$. Otherwise, define:
\begin{center}
$\exists p V:= \Big(\bigwedge_{par(V)} \bigast_i \exists p S_i \Big) \wedge \Big(\ER V\Big)
\wedge \Big(\Em V\Big) \wedge \Big(\EA V\Big)$ 
\end{center}
where:

\item[] $\bullet$
In the first conjunct, $\emph{par}(V)$ denotes the set of all non-trivial partitions $(S_1, \ldots, S_n)$ of $V$. Since the partitions are non-trivial, each $S_i$ is a proper subsequent of $V$. Therefore, as $G$ is terminating with respect to $\preceq$, we have $S_i \prec V$. Additionally, observe that the number of non-trivial partitions of any sequent is finite. Putting together, we can see that the use of the first conjunct in the recursive definition is allowed.

\item[] $\bullet$
$\ER V$ is defined as $\bigwedge_{(R, \sigma) \in \mathcal{L}(V)} F(V, R, \sigma)$, where $\mathcal{L}(V)$ is the set of all pairs $(R, \sigma)$ such that $R$ is a \emph{left} single-conclusion semi-analytic rule in $G$ and $\sigma$ is a local substitution for $R$ such that the conclusion of the $\sigma$-instance of $R$ is $V$. To define $F(V, R, \sigma)$, let $R$ be in the form:
\small\begin{center}
 \AxiomC{$\{\tPi_j , \bar{\nu}_{js} \Rightarrow \bar{\rho}_{js} \mid 1 \leq j \leq m, 1 \leq s \leq l_j \}$} 
 \AxiomC{$\{\tGamma_i , \bar{\mu}_{ir} \Rightarrow \tDelta_i \mid 1 \leq i \leq n, 1 \leq r \leq k_i \}$}
 \BinaryInfC{$\tPi_1, \dots, \tPi_m, \tGamma_1, \dots, \tGamma_n, \mu \Rightarrow \tDelta_1, \dots, \tDelta_n $}
 \DisplayProof
\end{center}
\normalsize
and $T_{js}=\sigma(\tPi_j , \bar{\nu}_{js} \Rightarrow \bar{\rho}_{js})$, for $ 1 \leq j \leq m$ and $ 1 \leq s \leq l_j $, and $ S_{ir} = \sigma(\tGamma_i , \bar{\mu}_{ir} \Rightarrow \tDelta_i)$, for $ 1 \leq i \leq n$ and $ 1 \leq r \leq k_i$. Then, if $n \neq 0$, i.e., if $R$ has a premise on the right branch, we define $F(V, R, \sigma)$ as:
\[
\bigwedge_{e=1}^n\Big[\Big(\Big(\bigast_j \bigwedge_s \forall p T_{js}\Big) * \Big(\bigast_{i \neq e} \bigwedge_r \forall p S_{ir}\Big)\Big) \to \bigvee_r \exists p S_{er}\Big]
\]
If $n=0$, define $F(V, R, \sigma)$ as $(\bigast_j \bigwedge_s \forall p T_{js}) \to 0$.
Note that by Remark \ref{BackwardFiniteness}, as the number of rules in $G$ is finite, the set $\mathcal{L}(V)$ is finite. Therefore, the conjunction on $\mathcal{L}(V)$ is well-defined. Moreover, as $G$ is terminating with respect to $\preceq$, the premises of any instance of any rule in $G$ are lower than its conclusion, and hence 
$T_{js} \prec V$ and $S_{ir} \prec V$ for all indices. Therefore, the use of $\forall p T_{js}$, $\forall p S_{ir}$, and $\exists p S_{er}$ in the recursive definition of $\exists p V$ is allowed.

\item[] $\bullet$
If the modal rule $(K)$ is or both the rules $(K)$ and $(D)$ are present in $G$ and $V$ is of the form $(\Box \Gamma \Rightarrow)$, for some non-empty $\Gamma$, then define $\Em V$ as $\Box \exists p V'$, where $V'=(\Gamma \Rightarrow \,)$. Otherwise, define $\Em V$ as $\top$. As $G$ is terminating with respect to $\preceq$ and $\Gamma$ is non-empty, in the first case, we have $V' \prec V$, and hence the use of $\exists p V'$ in the recursive definition of $\exists p V$ is allowed. Moreover, note that if the rule $(K)$ is or both $(K)$ and $(D)$ are present in $G$, then the language must be $\mathcal{L}_{\Box}$ and hence the use of $\Box$ in $\Box \exists p V'$ is also allowed.

\item[] $\bullet$
$\EA V$ is the formula constructed in Theorem \ref{UniformInterpolationForAxiom}. Recall that $\mathcal{A}$ is the set of all axioms in $G$, and we observed that it consists of single-conclusion focused axioms. Note that the conditions of Theorem \ref{UniformInterpolationForAxiom} are trivially satisfied.

This completes the definition of $\exists p V$. For $\forall p U$, for any single-conclusion sequent $U$, if $U$ is empty, define $\forall p U$ as $0$. Otherwise, define $\forall p U$ as:
\begin{center}
$\forall p U:= \bigvee_{par_{\bullet}(U)}((\bigast_{i \neq e} \exists p S_i) \to \forall p S_e) \vee  \ALR U \vee \ARR U \vee \Am U \vee \Aa U \vee \Aemp U$
\end{center}
where:

\item[] $\bullet$
In the first disjunct, $par_{\bullet}(U)$ denotes the set of all tuples $(S_1, \ldots, S_n, e)$, where $1 \leq e \leq n$ and $(S_1, \ldots, S_n)$ is a non-trivial partition of $U$ and $S_i^s = \emptyset$ for any $i \neq e$. Similar to the definition of $\exists p V$, the use of this disjunct is allowed in the recursive definition.

\item[] $\bullet$
$\ALR U$ is defined as $\bigvee_{(R, \sigma) \in \mathcal{L}(U)} F_l(U, R, \sigma)$, where $\mathcal{L}(U)$ is the set of all pairs $(R, \sigma)$ consisting of a \emph{left} single-conclusion semi-analytic rule $R$ in $G$ and a local substitution $\sigma$ for $R$ such that the conclusion of the $\sigma$-instance of $R$ is $U$. To define $F_l(U, R, \sigma)$, let $R$ have the form:
\small\begin{center}
 \AxiomC{$\{\tPi_j , \bar{\nu}_{js} \Rightarrow \bar{\rho}_{js} \mid 1 \leq j \leq m, 1 \leq s \leq l_j \}$} 
 \AxiomC{$\{\tGamma_i , \bar{\mu}_{ir} \Rightarrow \tDelta_i \mid 1 \leq i \leq n, 1 \leq r \leq k_i \}$}
 \BinaryInfC{$\tPi_1, \dots, \tPi_m, \tGamma_1, \dots, \tGamma_n, \mu \Rightarrow \tDelta_1, \dots, \tDelta_n $}
 \DisplayProof
\end{center}
\normalsize
and $T_{js} = \sigma(\tPi_j , \bar{\nu}_{js} \Rightarrow \bar{\rho}_{js})$, for $1 \leq j \leq m$ and $1 \leq s \leq l_j$, and $S_{ir} = \sigma(\tGamma_i , \bar{\mu}_{ir} \Rightarrow \tDelta_i)$, for $1 \leq i \leq n$ and $1 \leq r \leq k_i$. Then, define $F_l(U, R, \sigma)$ as:
\[
\Big(\bigast_j \bigwedge_s \forall p T_{js}\Big) * \Big(\bigast_i \bigwedge_r \forall p S_{ir}\Big).
\]
Like the definition of $\exists p V$, this disjunct is allowed in the recursive definition.

\item[] $\bullet$
$\ARR U$ is defined as $\bigvee_{(R, \sigma) \in \mathcal{R}(U)} F_r(U, R, \sigma)$, where $\mathcal{R}(U)$ is the set of all pairs $(R, \sigma)$ of \emph{right} single-conclusion semi-analytic rule $R$ in $G$ and local substitution $\sigma$ for $R$ such that the conclusion of the $\sigma$-instance of $R$ is $U$. To define $F_r(U, R, \sigma)$, if $R$ has the form:
\begin{center}
 \AxiomC{$\{\tGamma_i , \bar{\mu}_{ir} \Rightarrow \bar{\nu}_{ir} \mid 1 \leq i \leq n, 1 \leq r \leq k_i\}$}
 \UnaryInfC{$\tGamma_1, \dots, \tGamma_n \Rightarrow \mu $}
 \DisplayProof
\end{center}
and $S_{ir}=\sigma(\tGamma_i , \bar{\mu}_{ir} \Rightarrow \bar{\nu}_{ir})$, for any $ 1 \leq i \leq n$ and $ 1 \leq r \leq k_i$, we define $F_r(U, R, \sigma) = (\bigast_i \bigwedge_r \forall p S_{ir})$. Similar to the previous case, as $G$ is terminating with respect to $\preceq$, the use of this disjunct in the recursive definition is allowed.

\item[] $\bullet$
If either of the rules $(K)$ or $\{(K),(D)\}$ is present in $G$ and $U$ is the conclusion of an instance of $(K)$ or $(D)$ with the premise $U'$, then define $\Am U$ as $\Box \forall p U'$. Otherwise, define $\Am U$ as $\bot$. Note that, in the first case, since $G$ is terminating with respect to $\preceq$, we have $U' \prec U$. Hence, the formula $\forall p U'$ is allowed in the recursive definition. Moreover, the language must be $\mathcal{L}_{\Box}$, and hence the use of $\Box$ in $\Box \forall p U'$ is also allowed.

\item[] $\bullet$
$\Aa U$ is the formula constructed in Theorem \ref{UniformInterpolationForAxiom}.

\item[] $\bullet$
If $U^s=\emptyset$, define the formula $\Aemp U$ as $\exists p U \to 0$. Otherwise, define $\Aemp U$ as $\bot$. As discussed before, as the definition of $\exists p U$ is prior to the definition of $\forall p U$ and the definition of $\exists p U$ only refers to the sequents lower than $U$, the use of $\exists p U$ in the recursive definition of $\forall p U$ is allowed.\\

\noindent This completes the definition of $\exists p V$ and $\forall p U$, where $V^s = \emptyset$. We now prove that $\forall p U$ and $\exists p V$ satisfy the properties in Definition \ref{DfnRelativeInterpolation}. To this end, as promised earlier, we use induction on $\preceq$. Specifically, we establish \emph{each} property for $\exists p V$ based on the validity of \emph{all} the properties for $\forall p U'$ and $\exists p V'$, for any $U' \prec V$ and $V' \prec V$, where $V'^s = \emptyset$. Similarly, we prove \emph{each} property for $\forall p U$, assuming the validity of \emph{all} the properties for $\forall p U'$ and $\exists p V'$, for any $U' \prec U$ and $V' \preceq U$ such that $V'^s = \emptyset$. Note that in the second part $V' = U$ is possible. As in the earlier recursive definition, this causes no issue, as the properties for $\exists p U$ will have already been established before those for $\forall p U$ using the properties for the sequents lower than $U$.

We begin with the properties related to the variables. First, by looking into the recursive construction of $\exists p V$ and $\forall p U$ and using the induction hypothesis we explained above, it is clear that $\exists p V$ and $\forall p U$ are both $p$-free.

Second, we have to prove that $\mathcal{V}(\exists p V) \subseteq \mathcal{V}(V)$ and $\mathcal{V}(\forall p U) \subseteq \mathcal{V}(U)$, for any single-conclusion sequents $U$ and $V$ such that $V^s = \emptyset$. We only prove the first claim. The second is similar. If $V$ is the empty sequent, then $\exists p V = 1$ and hence there is nothing to prove. Otherwise, by the recursive construction of $\exists p V$, it is enough to show that the atomic formulas occurring in any of the conjuncts in the definition of $\exists p V$ appear in $V$. For the first conjunct, by the induction hypothesis and the fact that $S_i \prec V$, we have $\mathcal{V}(\exists p S_i) \subseteq \mathcal{V}(S_i)$, for any $1 \leq i \leq n$. As $(S_1, S_2, \ldots, S_n)$ is a partition of $V$, it is clear that $\mathcal{V}(S_i) \subseteq \mathcal{V}(V)$, for any $1 \leq i \leq n$. Hence, 
$\mathcal{V}(\bigwedge_{par(V)} \bigast_i \exists p S_i) \subseteq \mathcal{V}(V)$.
For the conjunct $\ER V$, let $(R, \sigma) \in \mathcal{L}(V)$. Then, as any single-conclusion semi-analytic rule is occurrence-preserving and the $\sigma$-instance of $R$ with premises $\{T_{js}\}_{js}$ and $\{S_{ir}\}_{ir}$ has the conclusion $V$, we have $\mathcal{V}(T_{js}) \cup \mathcal{V}(S_{ir}) \subseteq \mathcal{V}(V)$, for any $1 \leq j \leq m$, $1 \leq s \leq l_j$, $1 \leq i \leq n$ and $1 \leq r \leq k_i$. By the induction hypothesis and the fact that $T_{js} \prec V$ and $S_{ir} \prec V$, we have $\mathcal{V}(\forall p T_{js}) \subseteq \mathcal{V}(T_{js})$ and $\mathcal{V}(\forall p S_{ir}) \cup \mathcal{V}(\exists p S_{ir}) \subseteq \mathcal{V}(S_{ir})$. Therefore, $\mathcal{V}(F(V, R, \sigma)) \subseteq \mathcal{V}(V)$ which shows that $\mathcal{V}(\ER V) \subseteq \mathcal{V}(V)$.
For $\Em V$, let $V$ be in the form $(\Box \Gamma \Rightarrow \,)$, for some non-empty $\Gamma$ and set $V' = (\Gamma \Rightarrow \,)$. It is clear that $\mathcal{V}(V') \subseteq \mathcal{V}(V)$. By the induction hypothesis and the fact that $V' \prec V$, we have $\mathcal{V}(\exists p V') \subseteq \mathcal{V}(V')$. Hence, as $\Em V=\Box \exists p V'$, we have $\mathcal{V}(\Em V) \subseteq \mathcal{V}(V)$. In the otherwise case, we have $\Em V = \top$ and hence there is nothing to prove. Finally, for $\EA V$, the claim $\mathcal{V}(\EA V) \subseteq \mathcal{V}(V)$ is already proved in Theorem \ref{UniformInterpolationForAxiom}.\\

\noindent Now, we proceed to prove that $G \vdash V \cdot (\Rightarrow \exists p V)$, for any sequent $V$ such that $V^s=\emptyset$. If $V$ is the empty sequent, then by definition $\exists p V = 1$, and hence $V \cdot (\Rightarrow \exists p V)$ is $(\, \Rightarrow 1)$ which is provable in $G$. If $V$ is non-empty, it is enough to prove $G \vdash V \cdot (\, \Rightarrow c)$, for any conjunct $c$ in the definition of $\exists p V$. Then, by the rule $(R \wedge)$, we have $G \vdash V \cdot (\, \Rightarrow \exists p V)$. In the following, we will address each conjunct separately:

\item[] $\circ$ 
For the conjunct $(\bigwedge_{par(V)} \bigast_i \exists p S_i)$, for any non-trivial partition $(S_1, \dots, S_n)$ of $V=(\Sigma \Rightarrow)$, we prove $G \vdash \Sigma \Rightarrow \bigast_i \exists p S_i$. Let $S_i=(\Sigma_i \Rightarrow \,)$. As the partition is nontrivial, each $S_i$ is a proper subsequent of $V$. Hence, $S_i \prec V$, for each $1 \leq i \leq n$. 
By the induction hypothesis, $G \vdash \Sigma_i \Rightarrow \exists p S_i$, for any $1 \leq i \leq n$. Therefore, by the rule $(R*)$, we get $\Sigma_1, \dots, \Sigma_n \Rightarrow \bigast_i \exists p S_i$ or equivalently $\Sigma \Rightarrow \bigast_i \exists p S_i$ in $G$.

\item[] $\circ$ For the conjunct $\ER V$, it is enough to show the provability of $V \cdot (\, \Rightarrow F(V, R, \sigma))$ in $G$, for any $(R, \sigma) \in \mathcal{L}(V)$. Let $R$ be in the form:
\small\begin{center}
 \AxiomC{$\{\tPi_j , \bar{\nu}_{js} \Rightarrow \bar{\rho}_{js} \mid 1 \leq j \leq m, 1 \leq s \leq l_j \}$} 
 \AxiomC{$\{\tGamma_i , \bar{\mu}_{ir} \Rightarrow \tDelta_i \mid 1 \leq i \leq n, 1 \leq r \leq k_i \}$}
 \BinaryInfC{$\tPi_1, \dots, \tPi_m, \tGamma_1, \dots, \tGamma_n, \mu \Rightarrow \tDelta_1, \dots, \tDelta_n $}
 \DisplayProof
\end{center}
\normalsize
For simplicity, use the Convention \ref{convention} and moreover set $\sigma(\tGamma_i)=\Gamma_i$ and $\sigma(\tPi_j)=\Pi_j$. Then, 
$T_{js}=(\Pi_j , \bar{\psi}_{js} \Rightarrow \bar{\theta}_{js})$ for $ 1 \leq j \leq m$ and $ 1 \leq s \leq l_j $, $ S_{ir} = (\Gamma_i , \bar{\phi}_{ir} \Rightarrow ) $ for $ 1 \leq i \leq n$ and $ 1 \leq r \leq k_i $ are the premises of the $\sigma$-instance of $R$ with the conclusion $V=(\Pi_1, \ldots, \Pi_m, \Gamma_1, \ldots, \Gamma_n, \phi \Rightarrow \,)$. Let $\Pi=\bigcup_j \Pi_j$ and $\Gamma=\bigcup_i \Gamma_i$.
There are two cases to address: either $n=0$ or $n \neq 0$. If $n \neq 0$, using the definition of $F(V, R, \sigma)$, we have to show that:
\[
G \vdash V \cdot ( \Rightarrow\bigwedge_{e=1}^n\Big[\Big(\Big(\bigast_j \bigwedge_s \forall p T_{js}\Big) * \Big(\bigast_{i \neq e} \bigwedge_r \forall p S_{ir}\Big)\Big) \to \bigvee_r \exists p S_{er}\Big].
\]
As each premise of any instance of any rule in $G$ is lower than its conclusion, we have $T_{js} \prec V$ and $S_{ir} \prec V$ for all indices. Therefore, by the induction hypothesis, the following are all provable in $G$:\\

{\begin{math}
 \begin{tabular}{lll}
$\Pi_j , \forall p T_{js} , \bar{\psi}_{js} \Rightarrow \bar{\theta}_{js}$ \;\;\; & for $1 \leq j \leq m$ and $1 \leq s \leq l_j$,  & (1)\\
$\Gamma_i , \forall p S_{ir}, \bar{\phi}_{ir} \Rightarrow$  & for $1 \leq i \leq n$ and $1 \leq r \leq k_i$ that $i \neq e$, \;\;\; & (2)\\
$\Gamma_e, \bar{\phi}_{er}   \Rightarrow \exists p S_{er}$  & for $1 \leq r \leq k_e$. \;\;\; & (3)
 \end{tabular} 
\end{math}
}
\vspace{10pt}

\noindent Using the rule $(L \wedge)$ on (1) and (2) and the rule $(R \vee)$ on (3) we get:\\

{
\begin{math}
 \begin{tabular}{lll}
$G \vdash \Pi_j , \bigwedge_s  \forall p T_{js} , \bar{\psi}_{js} \Rightarrow \bar{\theta}_{js}$ \;\;\; & for $1 \leq j \leq m$, \\
$G \vdash \Gamma_i , \bigwedge_r \forall p S_{ir}, \bar{\phi}_{ir} \Rightarrow$ \;\;\; & for $1 \leq i \leq n$ that $i \neq e$, \\
$G \vdash \Gamma_e, \bar{\phi}_{er}   \Rightarrow \bigvee_r \exists p S_{er}$ \;\;\; &
 \end{tabular} 
\end{math}
}
\vspace{10pt}

\noindent These sequents can be premises of an instance of $R$. To see how, modify $\sigma$ to $\sigma'$ by setting $\sigma'(q)=\sigma(q)$, for any atom $q$ occurring in the rule $R$, $\sigma'(\tPi_j)=\Pi_j \cup \{\bigwedge_s  \forall p T_{js}\}$, for any $1 \leq j \leq m$, $\sigma'(\tGamma_i)=\Gamma_i \cup \{\bigwedge_r \forall p S_{ir}\}$ and $\sigma'(\tDelta_i)=\emptyset$, for any $i \in \{1, \ldots, n\}-\{e\}$, $\sigma'(\tGamma_e)=\Gamma_e$ and  $\sigma'(\tDelta_e)=\{\bigvee_r \exists p S_{er}\}$. Then, the above sequents form the $\sigma'$-instance of $R$. Applying this instance, we get:
\begin{center}
$\Pi, \Gamma, \{\bigwedge_s \forall p T_{js}\}_{j=1}^m , \{\bigwedge_r \forall p S_{ir}\}_{i=1, i \neq e}^n, \phi \Rightarrow \bigvee_r \exists p S_{er}$ 
\end{center}
in $G$. Now, by the rule $(L *)$, we get:
\begin{center}
$\Pi, \Gamma, (\bigast_j \bigwedge_s \forall p T_{js}) * (\bigast_{i \neq e} \bigwedge_r \forall p S_{ir}), \phi \Rightarrow \bigvee_r \exists p S_{er},$    
\end{center}
and finally, using the rule $(R\!\to)$, we obtain:
\begin{center}
$\Pi, \Gamma, \phi \Rightarrow [(\bigast_j \bigwedge_s \forall p T_{js}) * (\bigast_{i \neq e} \bigwedge_r \forall p S_{ir}) \to \bigvee_r \exists p S_{er}]$.   
\end{center}
This completes the proof of the case $n \neq 0$. For $n=0$, using a similar argument to the one above, we have: 
\begin{center}
$G \vdash \Pi_j , \bigwedge_s  \forall p T_{js} , \bar{\psi}_{js} \Rightarrow \bar{\theta}_{js}$,
\end{center}
for any $1 \leq j \leq m$ and $1 \leq s \leq l_j$. These sequents form an instance of $R$ similar to the previous case. Applying this instance, we get 
$\Pi, \{\bigwedge_s \forall p T_{js}\}_{j}, \phi \Rightarrow  $
which by $(0w)$, $(R\!\to)$ and $(L*)$ implies $G \vdash \Pi, \phi \Rightarrow (\bigast_j \bigwedge_s \forall p T_{js}\}_{j} \to 0) $.

\item[] $\circ$
For the conjunct $\Em V$, if it is defined as $\bot$, there is nothing to prove as $G \vdash V \cdot (\bot \Rightarrow \,)$. Therefore, we assume that either the rule $(K)$ is or both $(K)$ and $(D)$ are present in $G$ and $V$ is in the form $(\Box \Gamma \Rightarrow \,)$, for some non-empty multiset $\Gamma$. By definition, $\Em V=\Box \exists p V'$, where $V'=(\Gamma \Rightarrow \,)$. As $\Gamma$ is non-empty, by Definition \ref{Dfn:TerminatingCalculi}, we have $V' \prec V$. Hence, by the induction hypothesis, $V' \cdot (\Rightarrow \exists p V')$ or equivalently $\Gamma \Rightarrow \exists p V'$ is provable in $G$. Applying the rule $(K)$, available in $G$ by assumption, we get $\Box \Gamma \Rightarrow \Box \exists p V'$ or equivalently $V \cdot (\Rightarrow \Box \exists p V')$ in $G$. Hence, $G \vdash V \cdot (\Rightarrow \Em V)$.

\item[] $\circ$
For the conjunct $\EA V$, by Theorem \ref{UniformInterpolationForAxiom}, we have $G \vdash V \cdot (\Rightarrow \EA V )$.\\

This completes the proof of $G \vdash V \cdot (\, \Rightarrow \exists p V)$, for any sequent $V$ such that $V^s=\emptyset$. Now, we prove that $G \vdash U \cdot (\forall p U \Rightarrow \,)$, for any single-conclusion sequent $U$. If $U$ is the empty sequent, by definition $\forall p U=0$. Therefore, $U \cdot (\forall p U \Rightarrow \,)=(0 \Rightarrow \,)$ which is provable in $G$. If $U$ is non-empty, it is enough to show $G \vdash U \cdot (d \Rightarrow \,)$, for any disjunct $d$ in the definition of $\forall p U$. Then, by the rule $(L \vee)$, we can get $G \vdash U \cdot (\forall p U \Rightarrow \,)$. In the following, we investigate each disjunct separately:

\item[] $\circ$
For the disjunct $\bigvee_{par_{\bullet}(U)} ((\bigast_{i \neq e} \exists p S_i) \to \forall p S_e)$, we prove that for any non-trivial partition $(S_1, \ldots, S_n)$ of $U$ and any $1 \leq e \leq n$ such that $S_i^s = \emptyset$ for any $i \neq e$, we have $G \vdash U \cdot ((\bigast_{i \neq e} \exists p S_i) \to \forall p S_e \Rightarrow \,)$. Let $U=(\Gamma \Rightarrow \Delta)$. As the partition is non-trivial, each $S_i$ is a proper subsequent of $U$ and hence $S_i \prec U$.  
As $S_i^s = \emptyset$, for any $i \neq e$, we have $S_i = (\Gamma_i \Rightarrow)$ for some multiset $\Gamma_i$ and $S_e = (\Gamma_e \Rightarrow \Delta)$, for some multiset $\Gamma_e$. Note that $\Gamma=\bigcup_i \Gamma_i$. By the induction hypothesis, $G \vdash (\Gamma_i \Rightarrow \exists p S_i)$ for any $i \neq e$, and by $(R *)$ we obtain $G \vdash (\bigcup_{i \neq e} \Gamma_i \Rightarrow \bigast_{i \neq e} \exists p S_i)$.  
For $S_e = (\Gamma_e \Rightarrow \Delta)$, by the induction hypothesis, $G \vdash \Gamma_e, \forall p S_e \Rightarrow \Delta$. By $(L\!\to)$, we get $G \vdash \Gamma_1, \dots, \Gamma_n, (\bigast_{i \neq e} \exists p S_i) \to \forall p S_e \Rightarrow \Delta$ or equivalently $G \vdash \Gamma, (\bigast_{i \neq e} \exists p S_i) \to \forall p S_e \Rightarrow \Delta$. 

\item[] $\circ$
For the disjunct $\ALR U$, it is enough to prove $U \cdot (F_l(U, R, \sigma) \Rightarrow \,)$ for any $(R, \sigma) \in \mathcal{L}(U)$. Let $R$ be of the form:
\small\begin{center}
 \AxiomC{$\{\tPi_j , \bar{\nu}_{js} \Rightarrow \bar{\rho}_{js} \mid 1 \leq j \leq m, 1 \leq s \leq l_j \}$} 
 \AxiomC{$\{\tGamma_i , \bar{\mu}_{ir} \Rightarrow \tDelta_i \mid 1 \leq i \leq n, 1 \leq r \leq k_i \}$}
 \BinaryInfC{$\tPi_1, \dots, \tPi_m, \tGamma_1, \dots, \tGamma_n, \mu \Rightarrow \tDelta_1, \dots, \tDelta_n $}
 \DisplayProof
\end{center}
\normalsize
For simplicity, use Convention \ref{convention} and moreover set $\sigma(\tPi_j) = \Pi_j$, $\sigma(\tGamma_i) = \Gamma_i$, and $\sigma(\tDelta_i) = \Delta_i$. Then,  
$T_{js} = (\Pi_j , \bar{\psi}_{js} \Rightarrow \bar{\theta}_{js})$ for $1 \leq j \leq m$ and $1 \leq s \leq l_j$, and  
$S_{ir} = (\Gamma_i , \bar{\phi}_{ir} \Rightarrow \Delta_i)$ for $1 \leq i \leq n$ and $1 \leq r \leq k_i$  
are the premises of the $\sigma$-instance of $R$, with the conclusion  
\[
U = (\Pi_1, \ldots, \Pi_m, \Gamma_1, \ldots, \Gamma_n, \phi \Rightarrow \Delta_1, \ldots, \Delta_n).
\]  
Let $\Pi = \bigcup_j \Pi_j$, $\Gamma = \bigcup_i \Gamma_i$, and $\Delta = \bigcup_i \Delta_i$. Using the definition of $F_l(U, R, \sigma)$, we have to prove:
\begin{center}
$G \vdash \Pi, \Gamma, (\bigast_j \bigwedge_s \forall p T_{js}) * (\bigast_i \bigwedge_r \forall p S_{ir}), \phi \Rightarrow \Delta$.
\end{center}
Since $U$ is the conclusion of an instance of a rule in $G$ with premises $T_{js}$'s and $S_{ir}$'s, we have $T_{js} \prec U$ and $S_{ir} \prec U$ for all indices. Therefore, by the induction hypothesis, we obtain:  \\

\begin{math}
 \begin{tabular}{ll}
$G \vdash \Pi_j , \forall p T_{js} , \bar{\psi}_{js} \Rightarrow \bar{\theta}_{js}$ \;\;\; & for $1 \leq j \leq m$ and $1 \leq s \leq l_j$\\
$G \vdash \Gamma_i , \forall p S_{ir}, \bar{\phi}_{ir} \Rightarrow \Delta_i$ \;\;\; & for $1 \leq i \leq n$ and $1 \leq r \leq k_i$
 \end{tabular} 
\end{math}
\vspace{10pt}

\noindent Hence, using the rule $(L \wedge)$, we get:\\

\begin{math}
 \begin{tabular}{ll}
$G \vdash \Pi_j , \bigwedge_s \forall p T_{js} , \bar{\psi}_{js} \Rightarrow \bar{\theta}_{js}$ \;\;\; & for $1 \leq j \leq m$\\
$G \vdash \Gamma_i , \bigwedge_r \forall p S_{ir}, \bar{\phi}_{ir} \Rightarrow \Delta_i$ \;\;\; & for $1 \leq i \leq n$
 \end{tabular} 
\end{math}
\vspace{10pt}

\noindent  These sequents can serve as premises of an instance of $R$. It is sufficient to modify $\sigma$ to $\sigma'$ by setting $\sigma'(q) = \sigma(q)$ for any atomic formula $q$ occurring in the rule $R$,  
$\sigma'(\tPi_j) = \Pi_j \cup \{\bigwedge_s \forall p T_{js}\}$ for any $1 \leq j \leq m$,  
$\sigma'(\tGamma_i) = \Gamma_i \cup \{\bigwedge_r \forall p S_{ir}\}$, and  
$\sigma'(\tDelta_i) = \Delta_i$ for any $1 \leq i \leq n$.  
Therefore, by applying the $\sigma'$-instance of the rule $R$ to the above sequents, we obtain:  

\begin{center}
$G \vdash \Pi, \Gamma, \{\bigwedge_s \forall p T_{js}\}_{j=1}^m, \{\bigwedge_r \forall p S_{ir}\}_{i=1}^n, \phi \Rightarrow \Delta$    
\end{center} 
and using the rule $(L *)$, we get:
\begin{center}
$G \vdash \Pi, \Gamma, (\bigast_j \bigwedge_s \forall p T_{js}) * (\bigast_i \bigwedge_r \forall p S_{ir}), \phi \Rightarrow \Delta$
\end{center}
which is what we wanted.
 
\item[$\circ$] 
For the disjunct $\ARR U$, it is enough to prove $U \cdot (F_r(U, R, \sigma) \Rightarrow \,)$ for any $(R, \sigma) \in \mathcal{R}(U)$. Let $R$ be of the form:  

\begin{center}
 \AxiomC{$\{\tGamma_i , \bar{\mu}_{ir} \Rightarrow \bar{\nu}_{ir} \mid 1 \leq i \leq n, 1 \leq r \leq k_i\}$}
 \UnaryInfC{$\tGamma_1, \dots, \tGamma_n \Rightarrow \mu $}
 \DisplayProof
\end{center}
Again, use Convention \ref{convention} and moreover set $\sigma(\tGamma_i) = \Gamma_i$. Then,  
$S_{ir} = (\Gamma_i , \bar{\phi}_{ir} \Rightarrow \bar{\psi}_{ir})$ for $1 \leq i \leq n$ and $1 \leq r \leq k_i$  
are the premises of the $\sigma$-instance of $R$, with the conclusion:  
\[
U = (\Gamma_1, \ldots, \Gamma_n \Rightarrow \phi).
\]  
Let $\Gamma = \bigcup_i \Gamma_i$. Using the definition of $F_r(U, R, \sigma)$, we need to prove:  
\begin{center}
$G \vdash \Gamma, (\bigast_i \bigwedge_r \forall p S_{ir}) \Rightarrow \phi$.
\end{center}
Since $U$ is the conclusion of an instance of a rule in $G$ with premises $S_{ir}$'s, we have $S_{ir} \prec U$ for all indices. Thus, by the induction hypothesis, we have:  
\begin{center}
$G \vdash \Gamma_i , \forall p S_{ir}, \bar{\phi}_{ir} \Rightarrow \bar{\psi}_{ir}$ \quad for $1 \leq i \leq n$ and $1 \leq r \leq k_i$
\end{center}
Using the rule $(L \wedge)$, we obtain:
\begin{center}
$G \vdash \Gamma_i , \bigwedge_r\forall p S_{ir}, \bar{\phi}_{ir} \Rightarrow \bar{\psi}_{ir}$ \quad for $1 \leq i \leq n$
\end{center}
Similar to the previous case, these sequents can serve as the premises of an instance of $R$: it is sufficient to modify $\sigma$ to $\sigma'$ by setting $\sigma'(q) = \sigma(q)$ for any atom $q$ occurring in the rule $R$, and  $\sigma'(\tGamma_i) = \Gamma_i \cup \{\bigwedge_r \forall p S_{ir}\}$ for any $1 \leq i \leq n$.  
Applying the $\sigma'$-instance of $R$ to the above sequents, we get:  
\begin{center}
$G \vdash \Gamma_1, \ldots, \Gamma_n,  \bigwedge_r \forall p S_{1r} , \dots , \bigwedge_r \forall p S_{nr}  \Rightarrow \phi$.
\end{center}
Using the rule $(L *)$, we finally get $G \vdash \Gamma_1, \ldots \Gamma_n,  \bigast_i \bigwedge_r \forall p S_{ir} \Rightarrow \phi$.

\item[] $\circ$ 
For the disjunct $\Am U$, if either $(K)$ is or both $(K)$ and $(D)$ are present in $G$ and $U$ is the conclusion of $(K)$ or $(D)$ with the premise $U'$, we defined $\Am U = \Box \forall p U'$. By the form of the rules $(K)$ or $(D)$, the sequent $U$ must be of the form $(\Box \Gamma \Rightarrow \Box \Delta)$, where $\Delta$ contains at most one formula. Moreover, $U'$ must be of the form $(\Gamma \Rightarrow \Delta)$. As $U$ is the conclusion of an instance of a rule in $G$ with premise $U'$, we have $U' \prec U$. Thus, by the induction hypothesis, we get $G \vdash (\Gamma, \forall p U' \Rightarrow \Delta)$. By the same modal rule that derived $U$ from $U'$, we conclude $G \vdash (\Box \Gamma, \Box \forall p U' \Rightarrow \Box \Delta)$ or equivalently $G \vdash U \cdot (\Box \forall p U' \Rightarrow \,)$.
In the other case, we defined $\Am U = \bot$. In this case, $U \cdot (\bot \Rightarrow \,)$ is clearly provable in $G$.

\item[] $\circ$ 
For the disjunct $\Aa U$, we have $G \vdash U \cdot (\Aa U \Rightarrow)$, by Theorem \ref{UniformInterpolationForAxiom}. 

\item[] $\circ$
For the disjunct $\Aemp U$, if $U^s = \emptyset$, we defined $\Aemp U = (\exists p U \to 0)$. Otherwise, $\Aemp U = \bot$. In the former case, by the induction hypothesis, we have $G \vdash U \cdot (\, \Rightarrow \exists p U)$. As $G \vdash (0 \Rightarrow \,)$, by $(L\!\to)$, we have $G \vdash U \cdot (\exists p U \to 0 \Rightarrow)$.  
In the latter case, as $\Aemp U = \bot$, it is clear that  $G \vdash U \cdot (\Aemp U \Rightarrow)$.\\

This completes the proof of $G \vdash U \cdot (\forall p U \Rightarrow \,)$ for any single-conclusion sequent $U$. Now, we proceed to prove that for any sequent $V$ such that $V^s = \emptyset$, and for any $G$-provable sequent of the form $V \cdot (\bar{C} \Rightarrow \bar{D})$, where $\bar{C}$ and $\bar{D}$ are $p$-free multisets and $|\bar{D}| \leq 1$, we have $G \vdash \exists p V, \bar{C} \Rightarrow \bar{D}$.

Again, we will prove this claim by induction on $\preceq$ in the style explained before. However, this time, in any inductive step, we will also use induction on the depth of the proof of $V \cdot (\bar{C} \Rightarrow \bar{D})$. More precisely, we will show:

\begin{itemize}

\item[$(*)$]
For any number $d \in \mathbb{N}$ and any proof $\pi$ in $G$ with depth at most $d$ of a sequent in the form $V \cdot (\bar{C} \Rightarrow \bar{D})$, where $\bar{C}$ and $\bar{D}$ are $p$-free multisets and $|\bar{D}| \leq 1$, we have $G \vdash \exists p V, \bar{C} \Rightarrow \bar{D}$.
\end{itemize}

\noindent To prove $(*)$, if $V$ is the empty sequent, then the provability of $V \cdot (\bar{C} \Rightarrow \bar{D})$ in $G$ implies $G \vdash \bar{C} \Rightarrow \bar{D}$. Thus, by the rule $(1 w)$, we have $G \vdash \bar{C}, 1 \Rightarrow \bar{D}$. Since $\exists p V = 1$ in this case, $(*)$ is proved.
If $V$ is non-empty, we use induction on $d$. For the base case, if $d = 0$, then $V \cdot (\bar{C} \Rightarrow \bar{D})$ is an instance of an axiom in $\mathcal{A}$. Therefore, by Theorem \ref{UniformInterpolationForAxiom}, we have $G \vdash \EA V, \bar{C} \Rightarrow \bar{D}$, and hence $G \vdash \exists p V, \bar{C} \Rightarrow \bar{D}$, as $\EA V$ is a conjunct in $\exists p V$.

For the induction step, let $R$ be the rule and $\sigma$ be the local substitution for $R$ such that the $\sigma$-instance of $R$ is the last rule applied in $\pi$. Now, there are several cases to consider: either $R$ is a left single-conclusion semi-analytic rule, a right single-conclusion semi-analytic rule, the rule $(K)$, or the rule $(D)$. Moreover, in the first case, either the main formula appears in $\bar{C}$ or it does not. In the following, we carefully analyze each of these cases separately. Moreover, recall that we follow Convention \ref{convention} to name substituted formulas:

\item[] $\circ$
Let $R$ be the following left single-conclusion semi-analytic rule:
\small\begin{center}
 \AxiomC{$\{\tPi_j , \bar{\nu}_{js} \Rightarrow \bar{\rho}_{js} \mid 1 \leq j \leq m, 1 \leq s \leq l_j \}$} 
 \AxiomC{$\{\tGamma_i , \bar{\mu}_{ir} \Rightarrow \tDelta_i \mid 1 \leq i \leq n, 1 \leq r \leq k_i \}$}
 \BinaryInfC{$\tPi_1, \dots, \tPi_m, \tGamma_1, \dots, \tGamma_n, \mu \Rightarrow \tDelta_1, \dots, \tDelta_n $}
 \DisplayProof
\end{center}
\normalsize
and $\sigma(\mu) = \phi \in \bar{C}$. Therefore, $\phi$ is $p$-free, and since $R$ is semi-analytic and hence occurrence-preserving, $\bar{\psi}_{js}$, $\bar{\theta}_{js}$, and $\bar{\phi}_{ir}$ are also $p$-free, for all indices.  
Now, since the conclusion of the $\sigma$-instance of $R$ is
\[
\sigma(\tPi_1, \dots, \tPi_m, \tGamma_1, \dots, \tGamma_n, \mu \Rightarrow \tDelta_1, \dots, \tDelta_n)=V \cdot (\bar{C} \Rightarrow \bar{D})
\]
and $\phi \in \bar{C}$, the rest of $\bar{C}$ must be split among $\sigma(\tPi_j)$'s and $\sigma(\tGamma_i)$'s. Let $\bar{X}_j$ and $\bar{Y}_i$ be the parts of $\bar{C}$ in $\sigma(\tPi_j)$ and $\sigma(\tGamma_i)$, respectively. Therefore, we must have $\sigma(\tPi_j) = \Pi_j \cup \bar{X}_j$ and $\sigma(\tGamma_i) = \Gamma_i \cup \bar{Y}_i$, for some multisets $\Pi_j$ and $\Gamma_i$. Moreover, let $\bar{D}_i$ be the part of $\bar{D}$ in $\sigma(\tDelta_i)$. This means that the $\sigma$-instance of $R$ has the form:
\begin{center}
\begin{tabular}{c}
\AxiomC{$\{\Pi_j, \bar{X}_j, \bar{\psi}_{js} \Rightarrow \bar{\theta}_{js}\}_{j,s}$}
\AxiomC{$\{\Gamma_i, \bar{Y}_i, \bar{\phi}_{ir} \Rightarrow \bar{D}_i\}_{i,r}$}
\BinaryInfC{$\Pi, \Gamma, \bar{X}, \bar{Y}, \phi \Rightarrow \bar{D}$}
\DisplayProof
\end{tabular}
\end{center}
where $\Pi = \bigcup_j \Pi_j$, $\Gamma = \bigcup_i \Gamma_i$, $\bar{X} = \bigcup_j \bar{X}_j$, and $\bar{Y} = \bigcup_i \bar{Y}_i$. Note that $V = (\Pi, \Gamma \Rightarrow \,)$ and $\bar{C} = \bar{X} \cup \bar{Y} \cup \{\phi\}$. We want to prove $G \vdash \exists p V, \bar{X}, \bar{Y}, \phi \Rightarrow \bar{D}$.
Denote $T_j = (\Pi_j \Rightarrow)$ and $S_i = (\Gamma_i \Rightarrow)$, for any $1 \leq i \leq n$ and $1 \leq j \leq m$. Let $I = \{i \in \{1, \dots, n\} \mid \Gamma_i \neq \emptyset\}$ and $J = \{j \in \{1, \dots, m\} \mid \Pi_j \neq \emptyset\}$.

First, note that it is impossible to have $I = J = \emptyset$, as this would imply that $V = (\Pi, \Gamma \Rightarrow \,)$ is the empty sequent, which contradicts our assumption. Thus, there are two cases: either the disjoint union of $I$ and $J$ is a singleton, or it has at least two elements. We investigate these two cases one by one.

In the first case, either $I = \emptyset$ and $J$ is a singleton, or $J = \emptyset$ and $I$ is a singleton. We will investigate only the first case; the other is similar. Let $J = \{e\}$. Therefore, $T_e = V = (\Pi_e \Rightarrow \,)$. Then, $\pi$ is of the following form:
\begin{center}
\AxiomC{$\pi_{es}$}
\noLine
\UnaryInfC{$\{\Pi_e, \bar{\psi}_{es}, \bar{X}_e \Rightarrow \bar{\theta}_{es}\}_{s}$}
\AxiomC{$\{\bar{\psi}_{js}, \bar{X}_j \Rightarrow \bar{\theta}_{js}\}_{j \neq e,s}$}
\AxiomC{$\{\bar{\phi}_{ir}, \bar{Y}_i \Rightarrow \bar{D}_i\}_{i,r}$}
\TrinaryInfC{$\Pi_e, \bar{X}, \bar{Y}, \phi \Rightarrow \bar{D}$}
\DisplayProof
\end{center}
Then, $V \cdot (\bar{\psi}_{es}, \bar{X}_e \Rightarrow \bar{\theta}_{es})$ is the conclusion of the proof $\pi_{es}$ for each $s$, and the depth of $\pi_{es}$ is smaller than $d$. 
Since $\bar{\psi}_{es}$ and $\bar{\theta}_{es}$ are $p$-free, by the induction hypothesis on $d$, we have $G \vdash \exists p V, \bar{\psi}_{es}, \bar{X}_e \Rightarrow \bar{\theta}_{es}$. 
Now, note that the following is an instance of $R$:
\begin{center}
\AxiomC{$\{\exists p V, \bar{\psi}_{es}, \bar{X}_e \Rightarrow \bar{\theta}_{es}\}_{s}$}
\AxiomC{$\{\bar{\psi}_{js}, \bar{X}_j \Rightarrow \bar{\theta}_{js}\}_{j \neq e,s}$}
\AxiomC{$\{\bar{\phi}_{ir}, \bar{Y}_i \Rightarrow \bar{D}_i\}_{i,r}$}
\TrinaryInfC{$ \exists p V, \bar{X}, \bar{Y}, \phi \Rightarrow \bar{D}$}  	\DisplayProof
\end{center}
To see why, it is enough to define $\sigma'$ similarly to $\sigma$ for the atomic formulas occurring in $R$ and set $\sigma'(\tPi_e) = \{\exists p V\} \cup \bar{X}_e$, $\sigma'(\tPi_j) = \bar{X}_j$ for $j \in \{1, \dots, m\} - \{e\}$, $\sigma'(\tGamma_i) = \bar{Y}_i$ and $\sigma'(\tDelta_i) = \bar{D}_i$ for any $1 \leq i \leq n$. 
Applying the $\sigma'$-instance of $R$, we obtain $G \vdash (\exists p V, \bar{X}, \bar{Y}, \phi \Rightarrow \bar{D})$ or, equivalently, $G \vdash (\exists p V, \bar{C} \Rightarrow \bar{D})$.

In the second case, we assume that the disjoint union of $I$ and $J$ has at least two elements. Then, the sequents $\{T_j\}_{j \in J}$ together with $\{S_i\}_{i \in I}$ form a non-trivial partition of $V$. The reason is that the family has at least two elements, and each of these sequents is non-empty. This implies that $S_i \prec V$ and $T_j \prec V$, for any $i \in I$ and $j \in J$. It also implies that the formula $(\bigast_{j \in J} \exists p T_{j}) \land (\bigast_{i \in I} \exists p S_i)$ appears as a conjunct in the first conjunct of $\exists p V$.
As $\bar{\phi}_{ir}$, $\bar{\psi}_{js}$, and $\bar{\theta}_{js}$ are all $p$-free, by the induction hypothesis on $\preceq$, we get:
\begin{center}
$G \vdash \exists p T_{j}, \bar{\psi}_{js}, \bar{X}_j \Rightarrow \bar{\theta}_{js}$
\qquad \qquad
$G \vdash \exists p S_{i}, \bar{\phi}_{ir}, \bar{Y}_i \Rightarrow \bar{D}_i$
\end{center}
for any $i \in I$ and $j \in J$. For any $i \notin I$ and $j \notin J$, since $\Pi_j$ and $\Gamma_i$ are empty, we have:
\begin{center}
$G \vdash \bar{\psi}_{js}, \bar{X}_j \Rightarrow \bar{\theta}_{js}$
\qquad \qquad
$G \vdash \bar{\phi}_{ir}, \bar{Y}_i \Rightarrow \bar{D}_i$
\end{center}
The above four families of sequents form an instance of the rule $R$. It is enough to use the local substitution $\sigma'$ for $R$, defined similarly to $\sigma$ on the atomic formulas, and set $\sigma'(\tPi_j)=\{\exists p T_{j}\} \cup \bar{X}_j$, for $j \in J$; $\sigma'(\tPi_j)=\bar{X}_j$, for $j \notin J$; $\sigma'(\tGamma_i)=\{\exists p S_{i}\} \cup \bar{Y}_i$, for any $i \in I$; $\sigma'(\tGamma_i)=\bar{Y}_i$, for any $i \notin I$; and $\sigma'(\tDelta_i)=\bar{D}_i$. Thus, by applying the $\sigma'$-instance of $R$ to the above sequents, we obtain:
\begin{center}
$G \vdash \{\exists p T_{j}\}_{j \in J},\{\exists p S_i\}_{i \in I}, \bar{X}, \bar{Y}, \phi \Rightarrow \bar{D}$.
\end{center}
Therefore, by applying the rule $(L *)$, we obtain:
\begin{center}
$G \vdash (\bigast_{j \in J} \exists p T_{j})*(\bigast_{i \in I} \exists p S_i), \bar{X}, \bar{Y}, \phi \Rightarrow \bar{D}$.
\end{center}
As $(\bigast_{j \in J} \exists p T_{j}) \land (\bigast_{i \in I} \exists p S_i)$ appears as a conjunct in $\exists p V$, by the rule $(L \wedge)$, we finally get $G \vdash (\exists p V, \bar{X}, \bar{Y}, \phi \Rightarrow \bar{D})$ or equivalently $G \vdash (\exists p V, \bar{C} \Rightarrow \bar{D})$.

\item[] $\circ$
Let $R$ be the following left single-conclusion semi-analytic rule:
\small\begin{center}
 \AxiomC{$\{\tPi_j , \bar{\nu}_{js} \Rightarrow \bar{\rho}_{js} \mid 1 \leq j \leq m, 1 \leq s \leq l_j \}$} 
 \AxiomC{$\{\tGamma_i , \bar{\mu}_{ir} \Rightarrow \tDelta_i \mid 1 \leq i \leq n, 1 \leq r \leq k_i \}$}
 \BinaryInfC{$\tPi_1, \dots, \tPi_m, \tGamma_1, \dots, \tGamma_n, \mu \Rightarrow \tDelta_1, \dots, \tDelta_n $}
 \DisplayProof
\end{center}
\normalsize
and $\phi = \sigma(\mu) \notin \bar{C}$. Since the conclusion of the $\sigma$-instance of $R$ is:
\[
\sigma(\tPi_1, \dots, \tPi_m, \tGamma_1, \dots, \tGamma_n, \mu \Rightarrow \tDelta_1, \dots, \tDelta_n)=V \cdot (\bar{C} \Rightarrow \bar{D})
\]
and $\phi = \sigma(\mu) \notin \bar{C}$, the multiset $\bar{C}$ must be split among $\sigma(\tPi_j)$'s and $\sigma(\Gamma_i)$'s. Let $\bar{X}_j$ and $\bar{Y}_i$ denote the parts of $\bar{C}$ in $\sigma(\tPi_j)$ and $\sigma(\tGamma_i)$, respectively. Therefore, we must have $\sigma(\tPi_j) = \Pi_j \cup \bar{X}_j$ and $\sigma(\tGamma_i) = \Gamma_i \cup \bar{Y}_i$, for some multisets $\Pi_j$ and $\Gamma_i$. 

Now, there are two cases to consider: either $n = 0$ or $n \neq 0$. We will investigate each case separately. If $n \neq 0$, as $\bar{D}$ has at most one element, there must exist $1 \leq e \leq n$ such that $\sigma(\tDelta_e) = \bar{D}$ and $\sigma(\tDelta_i) = \emptyset$ for $i \neq e$. If $\bar{D} = \emptyset$, we pick $e$ arbitrarily. Therefore, the $\sigma$-instance of $R$ has the form:
\begin{center}
\AxiomC{$\{\Pi_j, \bar{X}_j, \bar{\psi}_{js} \Rightarrow \bar{\theta}_{js}\}_{j,s}$}
\AxiomC{$\{\Gamma_i, \bar{Y}_i, \bar{\phi}_{ir} \Rightarrow\}_{i \neq e, r}$}
\AxiomC{$\{\Gamma_e, \bar{Y}_e, \bar{\phi}_{er} \Rightarrow \bar{D}\}_r$}
\RightLabel{$(\dagger)$}
\TrinaryInfC{$\Pi, \Gamma, \bar{X}, \bar{Y}, \phi \Rightarrow \bar{D}$}
\DisplayProof
\end{center}
where $\Pi=\bigcup_j \Pi_j$, $\Gamma=\bigcup_i \Gamma_i$, $\bar{X}=\bigcup_j \bar{X}_j$ and $\bar{Y}=\bigcup_i \bar{Y}_i$. Therefore, $V=(\Pi, \Gamma , \phi \Rightarrow )$ and $\bar{C}= \bar{X} \cup \bar{Y}$. We want to prove $G \vdash (\exists p V, \bar{X}, \bar{Y} \Rightarrow \bar{D})$. 

Define the local substitution $\sigma'$ for $R$ similarly to $\sigma$ on the atomic formulas occurring in $R$ and set $\sigma'(\tPi_j) = \Pi_j$, for any $1 \leq j \leq m$, $\sigma'(\tGamma_i) = \Gamma_i$, and $\sigma'(\tDelta_i) = \emptyset$, for any $1 \leq i \leq n$. Then, the $\sigma'$-instance of $R$ has the form:
\begin{center}
\AxiomC{$\{\Pi_j, \bar{\psi}_{js} \Rightarrow \bar{\theta}_{js}\}_{j,s}$}
\AxiomC{$\{\Gamma_i, \bar{\phi}_{ir} \Rightarrow\}_{i \neq e, r}$}
\AxiomC{$\{\Gamma_e, \bar{\phi}_{er} \Rightarrow \}_r$}
\TrinaryInfC{$\Pi, \Gamma, \phi \Rightarrow $}
\DisplayProof
\end{center}
Therefore, $(R, \sigma') \in \mathcal{L}(V)$.
Let $T_{js} = (\Pi_j, \bar{\psi}_{js} \Rightarrow \bar{\theta}_{js})$ and $S_{ir} = (\Gamma_i, \bar{\phi}_{ir} \Rightarrow )$. Note that $T_{js}$ and $S_{ir}$ are the premises of an instance of a rule in $G$ with the conclusion $V$. Therefore, $T_{js} \prec V$ and $S_{ir} \prec V$, for all indices. Moreover, by $(\dagger)$, we have $G \vdash T_{js} \cdot (\bar{X}_j \Rightarrow)$, $G \vdash S_{er} \cdot (\bar{Y}_e \Rightarrow \bar{D})$, and $G \vdash S_{ir} \cdot ( \bar{Y}_i \Rightarrow )$ for $i \neq e$. Therefore, by the fact that $\bar{X}_j$, $\bar{Y}_i$ and $\bar{D}$ are $p$-free, the induction hypothesis on $\preceq$ shows that for all possible indices:
\begin{center}
$G \vdash (\bar{X}_j \Rightarrow \forall p T_{js})$
\hspace{5pt}
	  	,
	  	\hspace{5pt}
$G \vdash (\bar{Y}_i \Rightarrow \forall p S_{ir})_{i \neq e}$
\hspace{5pt}
	  	,
	  	\hspace{5pt}
$G \vdash (\bar{Y}_e , \exists p S_{er} \Rightarrow \bar{D})$.
\end{center}
Now, using the rules $(R \wedge)$ and $(L \vee)$, we have: 
\begin{center}
$G \vdash (\bar{X}_j \Rightarrow \bigwedge_s \forall p T_{js})$
,
$G \vdash (\bar{Y}_i \Rightarrow \bigwedge_r \forall p S_{ir})_{i \neq e}$
	  	,
$G \vdash (\bar{Y}_e, \bigvee_r \exists p S_{er} \Rightarrow \bar{D})$.
\end{center}
Denote $L_j=(\bigwedge_s \forall p T_{js})$ for any $j$, $M_i=(\bigwedge_r \forall p S_{ir})$ for $i \neq e$, and $N=(\bigvee_r \exists p S_{er})$. Thus,
\begin{center}
\AxiomC{$\{\bar{X}_j \Rightarrow L_j\}_j$}
\RightLabel{{\scriptsize $(R *)$}}
\UnaryInfC{$\bar{X} \Rightarrow \bigast_j L_j$}
\AxiomC{$\{\bar{Y}_i \Rightarrow M_i \}_{i \neq e}$}
\RightLabel{{\scriptsize $(R *)$}}
\UnaryInfC{$\{\bar{Y}_i\}_{i=1, i \neq e}^n \Rightarrow \bigast_{i \neq e} M_i$}
\RightLabel{{\scriptsize $(R *)$}}
\BinaryInfC{$\bar{X}, \{\bar{Y}_i\}_{i=1, i \neq e}^n \Rightarrow (\bigast_j L_j) * (\bigast_{i \neq e} M_i)$}
\AxiomC{$\bar{Y}_e, N \Rightarrow \bar{D} $}
\RightLabel{{\scriptsize $(L\!\to)$}}
\BinaryInfC{$ \bar{X}, \bar{Y}, (\bigast_j L_j) * (\bigast_{i \neq e} M_i) \to N \Rightarrow \bar{D} $}
\DisplayProof
\end{center}
As $(R, \sigma') \in \mathcal{L}(V)$, the formula $(\bigast_j L_j) * (\bigast_{i \neq e} M_i) \to N$ appears as a conjunct in $F(V, R, \sigma')$ and hence in $\exists p V$. Therefore, by $(L \wedge)$, we finally get $G \vdash (\exists p V, \bar{X}, \bar{Y} \Rightarrow \bar{D})$ or equivalently $G \vdash (\exists p V, \bar{C} \Rightarrow \bar{D})$.

Now, we investigate the case $n=0$. In this case, the rule $R$ has no right branches. Hence, $\bar{D}=\emptyset$. Moreover, $V=(\Pi, \phi, \Rightarrow \,)$ and $\bar{C}=\bar{X}$. Thus, we must prove $G \vdash (\exists p V, \bar{X} \Rightarrow \,)$. Similar to the above argument, we define the local substitution $\sigma'$ for $R$ similarly to $\sigma$ on the atomic formulas but $\sigma'(\tPi_j)=\Pi_j$, for any $1 \leq j \leq m$. Then, the $\sigma'$-instance of $R$ has the form:
\begin{center}
\AxiomC{$\{\Pi_j, \bar{\psi}_{js} \Rightarrow \bar{\theta}_{js}\}_{j,s}$}
\UnaryInfC{$\Pi, \phi \Rightarrow $}
\DisplayProof
\end{center}
Therefore, $(R, \sigma') \in \mathcal{L}(V)$. Set $T_{js}=(\Pi_j, \bar{\psi}_{js} \Rightarrow \bar{\theta}_{js})$. In a similar fashion to what we did in the previous case, using the induction hypothesis and the rules $(L \wedge)$ and $(L*)$, we have $G \vdash (\bar{X} \Rightarrow \bigast_j \bigwedge_s \forall p T_{js})$. Using $G \vdash (0 \Rightarrow \,)$ and the rule $(L \to)$, we get
\[
G \vdash \bar{X}, (\bigast_j \bigwedge_s \forall p T_{js}) \to 0  \Rightarrow \,
\]
As $(R, \sigma') \in \mathcal{L}(V)$, the formula $F(V, R, \sigma')=(\bigast_j \bigwedge_s \forall p T_{js} \to 0)$ appears as a conjunct in $\exists p V$. Hence, by $(L \wedge)$, we finally get $G \vdash (\exists p V, \bar{X} \Rightarrow \,)$ or equivalently $G \vdash (\exists p V, \bar{C} \Rightarrow \bar{D})$.

\item[] $\circ$
Let $R$ be the following right single-conclusion semi-analytic rule:
\begin{center}
 \AxiomC{$\{\tGamma_i , \bar{\mu}_{ir} \Rightarrow \bar{\nu}_{ir} \mid 1 \leq i \leq n, 1 \leq r \leq k_i\}$}
 \UnaryInfC{$\tGamma_1, \dots, \tGamma_n \Rightarrow \mu $}
 \DisplayProof
\end{center}
Thus, $V \cdot (\bar{C} \Rightarrow \bar{D})= \sigma(\tGamma \Rightarrow \mu)$. Therefore, $\bar{C}$ must be split among $\sigma(\tGamma_i)$'s. Let $\bar{C}_i$ be the part of $\bar{C}$ in $\sigma(\tGamma_i)$. Hence, $\sigma(\tGamma_i)=\Gamma_i \cup \bar{C}_i$, for some multiset $\Gamma_i$. Thus, the $\sigma$-instance of $R$ has the form:
\begin{center}
\AxiomC{$\{\Gamma_i, \bar{C}_i, \bar{\phi}_{ir} \Rightarrow \bar{\psi}_{ir}\}_{i,r}$}
\UnaryInfC{$ \Gamma, \bar{C}\Rightarrow \phi$}
\DisplayProof
\end{center}
where $\Gamma=\bigcup_i \Gamma_i$. Note that $V=(\Gamma \Rightarrow \,)$ and we want to prove $G \vdash (\exists p V, \bar{C} \Rightarrow \phi)$. Moreover, as $\bar{D}=\{\phi\}$, the formula $\phi$ is $p$-free and as $R$ is occurrence-preserving, $\bar{\psi}_{ir}$ and $\bar{ \phi}_{ir}$ are also $p$-free. 
Denote $S_{i}=(\Gamma_i \Rightarrow \,)$ and let $I=\{i \in \{1, \ldots, n\} \mid \Gamma_i \neq \emptyset\}$. Note that $I$ is non-empty, because otherwise, $V=(\Gamma \Rightarrow \,)$ would be empty which contradicts with our assumption. Therefore, there are two cases to consider: either $I$ is a singleton or it has at least two elements. We will investigate each case separately.

In the first case, $I$ is a singleton. Let $I=\{e\}$. Then, $V=(\Gamma_e \Rightarrow \,)$ and $\pi$ has the form:
\begin{center}
\AxiomC{$\pi_{er}$}
\noLine
\UnaryInfC{$\{\Gamma_e, \bar{C}_e, \bar{\phi}_{er} \Rightarrow \bar{\psi}_{er}\}_r$}
\AxiomC{$\{ \bar{C}_i, \bar{\phi}_{ir} \Rightarrow \bar{\psi}_{ir} \}_{i \neq e,r}$}
\BinaryInfC{$ \Gamma_e, \bar{C}\Rightarrow \phi$}
\DisplayProof
\end{center}
Note that the depth of the proof $\pi_{er}$ for $\Gamma_e, \bar{C}_e, \bar{\phi}_{er} \Rightarrow \bar{\psi}_{er}$ is less than $d$. Thus, by the induction hypothesis on $d$, we get $G \vdash \exists p V, \bar{C}_e, \bar{\phi}_{er} \Rightarrow \bar{\psi}_{er}$. Define the local substitution $\sigma'$ for $R$ similar to $\sigma$ on the atomic formulas occurring in $R$ and set $\sigma'(\tGamma_i)=\bar{C}_i$, for $i \in \{1, \ldots, n\}-\{e\}$ and $\sigma'(\tGamma_e)=\{\exists p V\} \cup \bar{C}_e$. Therefore, the $\sigma'$-instance of $R$ has the form:
\begin{center}
	  	\AxiomC{$\{\exists p V, \bar{C}_e, \bar{\phi}_{er} \Rightarrow \bar{\psi}_{er}\}_r$}
	  	\AxiomC{$\{\bar{C}_i, \bar{\phi}_{ir} \Rightarrow \bar{\psi}_{ir}\}_{i \neq e, r}$}
  		\BinaryInfC{$ \exists p V, \bar{C}\Rightarrow \phi$}
  		\DisplayProof
\end{center}
Hence, we get $G \vdash  \exists p V, \bar{C} \Rightarrow \phi$, as desired.

In the second case, we assume that $I$ has at least two elements. Then, the sequents $\{S_i\}_{i \in I}$ form a non-trivial partition of $V$ as the family has at least two elements and all of them are non-empty. Moreover, it implies that $S_i \prec V$, for any $i \in I$, and $\bigast_{i \in I} \exists p S_i$  
appears as a conjunct in $\exists p V$. Now, by the induction hypothesis on $\preceq$, we get $G \vdash \exists p S_{i}, \bar{C}_i, \bar{\phi}_{ir} \Rightarrow \bar{\psi}_{ir}$, for any $i \in I$. For $i \notin I$, as $\Gamma_i =\emptyset$, we already have $\bar{C}_i, \bar{\phi}_{ir} \Rightarrow \bar{\psi}_{ir}$. All these sequents form an instance of $R$:
\begin{center}
	  	\AxiomC{$\{\exists p S_i, \bar{C}_i, \bar{\phi}_{ir} \Rightarrow \bar{\psi}_{ir}\}_{i \in I,r}$}
	  	\AxiomC{$\{\bar{C}_i, \bar{\phi}_{ir} \Rightarrow \bar{\psi}_{ir}\}_{i \notin I, r}$}
  		\BinaryInfC{$ \{\exists p S_i\}_{i \in I}, \bar{C}_1, \dots, \bar{C}_n  \Rightarrow \phi$}
  		\DisplayProof
\end{center}
To see why, it is enough to define $\sigma'$ similarly to $\sigma$ on the atomic formulas occurring in $R$ and set $\sigma'(\tGamma_i)=\{\exists p S_i\} \cup \bar{C}_i$ for $i \in I$ and $\sigma'(\tGamma_i)=\bar{C}_i$, for $i \notin I$. 
Now, by applying the $\sigma'$-instance of $R$, we have $G \vdash \{\exists p S_i\}_{i \in I}, \bar{C} \Rightarrow \phi$. By $(L *)$, we get $G \vdash \bigast_{i \in I} \exists p S_i, \bar{C} \Rightarrow \phi$. As $\bigast_{i \in I} \exists p S_i$ appears as a conjunct in $\exists p V$, by $(L \wedge)$, we finally get $G \vdash (\exists p V, \bar{C} \Rightarrow \phi)$.

\item[] $\circ$
Let $R$ be either $(K)$ or $(D)$. We only explain the former case. The latter is similar. As $V \cdot (\bar{C} \Rightarrow \bar{D})$ is the conclusion of an instance of $(K)$, it must have the form $\Box \Sigma \Rightarrow \Box \alpha$, for some multiset $\Sigma$ and some formula $\alpha$. Thus, $\bar{C}=\overline{\Box C'}$ and $V=(\Box\Gamma  \Rightarrow )$ for some multisets $\bar{C'}$ and $\Gamma$ and $\bar{D}=\{\Box \alpha\}$. Moreover, the last rule in $\pi$ is in the form:
\begin{center}
  	\begin{tabular}{c}
	  	\AxiomC{$\Gamma, \bar{C'} \Rightarrow \alpha$}
  		\UnaryInfC{$\Box\Gamma, \overline{\Box C'} \Rightarrow \Box \alpha$}
  		\DisplayProof
\end{tabular}
\end{center}
As $V$ is non-empty, $\Gamma$ is non-empty. Therefore, defining $V'=(\Gamma \Rightarrow \,)$, we have $V' \prec V$. As $\bar{C'}$ and $\alpha$ are $p$-free, by the induction hypothesis, we have $G \vdash \exists p V', \bar{C'} \Rightarrow \alpha$. Using $(K)$, we get $G \vdash \Box\exists p V', \overline{\Box C'} \Rightarrow \Box \alpha$ or equivalently $G \vdash \Box\exists p V', \bar{C} \Rightarrow \bar{D}$. Now, notice that by definition, $\Em V=\Box\exists p V'$. Hence, $\Box \exists p V'$ appears as a conjunct in the definition of $\exists p V$. Therefore, by $(L \wedge)$, we get $G \vdash \exists p V, \bar{C} \Rightarrow \bar{D}$.\\

This completes the proof of $(*)$. Now, we prove that for any provable sequent in the form $U \cdot (\bar{C} \Rightarrow)$ in $G$, where $\bar{C}$ is a $p$-free multiset, we have $G \vdash \bar{C} \Rightarrow \forall p U$.
Again, we will prove this claim by induction on $\preceq$ in the style explained before. In the inductive step, we also use an induction on the depth of the proof of $U \cdot (\bar{C} \Rightarrow \,)$.
More precisely, we show:

\begin{itemize}
\item[$(**)$]
For any number $d \in \mathbb{N}$ and any proof $\pi$ in $G$ with depth at most $d$ of a sequent in the form $U \cdot (\bar{C} \Rightarrow)$, where $\bar{C}$ is a $p$-free multiset, we have $G \vdash \bar{C} \Rightarrow \forall p U$.
\end{itemize}

\noindent To prove $(**)$, if $U$ is the empty sequent, then the provability of $U \cdot (\bar{C} \Rightarrow \,)$ in $G$ implies $G \vdash \bar{C} \Rightarrow \,$. Thus, by the rule $(0w)$, we have $G \vdash \bar{C} \Rightarrow 0$. Since $\forall p U = 0$ in this case, $(**)$ is proved.
If $U$ is non-empty, we use induction on $d$. For the base case, if $d=0$, then $U \cdot (\bar{C} \Rightarrow \,)$ is an instance of an axiom in $\mathcal{A}$. Therefore, by Theorem \ref{UniformInterpolationForAxiom}, $G \vdash \bar{C} \Rightarrow \Aa U$ and hence $G \vdash \bar{C} \Rightarrow \forall p U$.

For the induction step, let $R$ be the rule, and let $\sigma$ be a local substitution for $R$ such that the $\sigma$-instance of $R$ is the last rule applied in $\pi$. There are several cases to consider: either $R$ is a left single-conclusion semi-analytic rule, a right single-conclusion semi-analytic rule, the rule $(K)$, or the rule $(D)$. In the first case, the main formula either appears in $\bar{C}$ or it does not. In the following, we carefully analyze each of these cases separately. Additionally, recall that we use Convention \ref{convention} to name the substituted formulas.

\item[] $\circ$
Let $R$ be the following left single-conclusion semi-analytic rule:
\small\begin{center}
 \AxiomC{$\{\tPi_j , \bar{\nu}_{js} \Rightarrow \bar{\rho}_{js} \mid 1 \leq j \leq m, 1 \leq s \leq l_j \}$} 
 \AxiomC{$\{\tGamma_i , \bar{\mu}_{ir} \Rightarrow \tDelta_i \mid 1 \leq i \leq n, 1 \leq r \leq k_i \}$}
 \BinaryInfC{$\tPi_1, \dots, \tPi_m, \tGamma_1, \dots, \tGamma_n, \mu \Rightarrow \tDelta_1, \dots, \tDelta_n $}
 \DisplayProof
\end{center}
\normalsize
and $\phi = \sigma(\mu) \in \bar{C}$. Since $\phi$ is $p$-free and $R$ is semi-analytic and hence occurrence-preserving, $\bar{\psi}_{js}$, $\bar{\theta}_{js}$, and $\bar{\phi}_{ir}$ are also $p$-free.
Now, as the conclusion of the $\sigma$-instance of $R$ is 
\[
\sigma(\tPi_1, \dots, \tPi_m, \tGamma_1, \dots, \tGamma_n, \mu \Rightarrow \tDelta_1, \dots, \tDelta_n)=U \cdot (\bar{C} \Rightarrow \,)
\]
and $\phi = \sigma(\mu) \in \bar{C}$, the rest of $\bar{C}$ must be split among $\sigma(\tPi_j)$'s and $\sigma(\tGamma_i)$'s. Let $\bar{X}_j$ and $\bar{Y}_i$ be the parts of $\bar{C}$ in $\sigma(\tPi_j)$ and $\sigma(\tGamma_i)$, respectively. Therefore, we must have $\sigma(\tPi_j) = \Pi_j \cup \bar{X}_j$ and $\sigma(\tGamma_i) = \Gamma_i \cup \bar{Y}_i$, for some multisets $\Pi_j$ and $\Gamma_i$. This implies that the $\sigma$-instance of $R$ has the form:
\begin{center}
\AxiomC{$\{ \Pi_j, \bar{X}_j, \bar{\psi}_{js} \Rightarrow \bar{\theta}_{js}\}_{j,s}$}
\AxiomC{$\{\Gamma_i, \bar{Y}_i, \bar{\phi}_{ir} \Rightarrow \Delta_i \}_{i,r}$}
\BinaryInfC{$\Pi, \Gamma, \bar{X}, \bar{Y}, \phi \Rightarrow \Delta$}
\DisplayProof
\end{center}
where $\Pi = \bigcup_j \Pi_j$, $\Gamma = \bigcup_i \Gamma_i$, $\bar{X} = \bigcup_j \bar{X}_j$, $\bar{Y} = \bigcup_i \bar{Y}_i$, $\Delta = \bigcup_i \Delta_i$, and $\Delta_i = \sigma(\tDelta_i)$. Therefore, $U = (\Pi, \Gamma \Rightarrow \Delta)$, and $\bar{C} = \bar{X} \cup \bar{Y} \cup \{\phi\}$. We aim to prove that $G \vdash \bar{X}, \bar{Y}, \phi \Rightarrow \forall p U$.
There are three cases to consider: either there exists $1 \leq e \leq n$ such that $U = (\Gamma_e \Rightarrow \Delta_e)$, or there exists $1 \leq e' \leq m$ such that $U = (\Pi_{e'} \Rightarrow \,)$, or such $e$ and $e'$ do not exist.

In the first case, we have $\Pi_j = \Gamma_i = \Delta_i = \emptyset$ for any $i \in \{1, \ldots, n\} - \{e\}$ and $1 \leq j \leq m$. Hence, $\pi$ is in the form:
\begin{center}
\AxiomC{$\{ \bar{X}_j, \bar{\psi}_{js} \Rightarrow \bar{\theta}_{js}\}_{j,s}$}
\AxiomC{$\{ \bar{Y}_i, \bar{\phi}_{ir} \Rightarrow \}_{i \neq e,r}$}
\AxiomC{$\pi_{er}$}
\noLine
\UnaryInfC{$\Gamma_e, \bar{Y}_e, \bar{\phi}_{er} \Rightarrow \Delta_e$}
\TrinaryInfC{$\Gamma_e, \bar{X}, \bar{Y}, \phi \Rightarrow \Delta_e$}
\DisplayProof
\end{center}
Now, as the depth of the proof $\pi_{er}$ is smaller than $d$ and by the fact that the $\bar{\phi}_{er}$'s are all $p$-free, by the induction hypothesis on $d$, we have $G \vdash \bar{Y}_e, \bar{\phi}_{ir} \Rightarrow \forall p U$. 
Define the local substitution $\sigma'$ similarly to $\sigma$ on the atomic formulas occurring in $R$ and set $\sigma'(\tPi_j) = \bar{X}_j$, for any $1 \leq j \leq m$, $\sigma'(\tGamma_i) = \bar{Y}_i$, for any $1 \leq i \leq n$, $\sigma'(\tDelta_i) = \emptyset$, for any $i \in \{1, \ldots, n\}-\{e\}$, and $\sigma'(\tDelta_e) = \{\forall p U\}$. Thus, the $\sigma'$-instance of $R$ has the form:
\begin{center}
\AxiomC{$\{ \bar{X}_j, \bar{\psi}_{js} \Rightarrow \bar{\theta}_{js}\}_{j,s}$}
\AxiomC{$\{ \bar{Y}_i, \bar{\phi}_{ir} \Rightarrow  \}_{i \neq e,r}$}
\AxiomC{$\{\bar{Y}_e, \bar{\phi}_{er} \Rightarrow \forall p U \}_{r}$}
\TrinaryInfC{$\bar{X}, \bar{Y}, \phi \Rightarrow \forall p U$}
\DisplayProof
\end{center}
Therefore, we reach $G \vdash \bar{X}, \bar{Y}, \phi \Rightarrow \forall p U$.

In the second case, there exists $1 \leq e' \leq m$ such that $U=(\Pi_{e'} \Rightarrow \,)$. Therefore, $\Pi_j=\Gamma_i=\Delta_i=\emptyset$ for any $1 \leq i \leq n$ and $j \in \{1, \ldots, m\}-\{e'\}$. Hence, $\pi$ has the form:
\begin{center}
\AxiomC{$\pi_{e's}$}
\noLine
\UnaryInfC{$\{ \Pi_{e'}, \bar{X}_{e'}, \bar{\psi}_{e's} \Rightarrow \bar{\theta}_{e's}\}_{s}$}
\AxiomC{$\{ \bar{X}_j, \bar{\psi}_{js} \Rightarrow \bar{\theta}_{js}\}_{j \neq e',s}$}
\AxiomC{$\{ \bar{Y}_i, \bar{\phi}_{ir} \Rightarrow \}_{i,r}$}
\TrinaryInfC{$\Pi_{e'},\bar{X}, \bar{Y}, \phi \Rightarrow $}
\DisplayProof
\end{center}
Recall that when $U^s=\emptyset$, our induction hypothesis for proving the properties of $\forall p U$ can rely on the claims for $\exists p U$. Therefore, by the induction hypothesis, we have $G \vdash \exists p U, \bar{X}_{e'}, \bar{\psi}_{e's} \Rightarrow \bar{\theta}_{e's}$. Similar to the previous case, these sequents form an instance of $R$ in the following way:
\begin{center}
\AxiomC{$\{ \exists p U, \bar{X}_{e'}, \bar{\psi}_{e's} \Rightarrow \bar{\theta}_{e's}\}_{s}$}
\AxiomC{$\{ \bar{X}_j, \bar{\psi}_{js} \Rightarrow \bar{\theta}_{js}\}_{j \neq e',s}$}
\AxiomC{$\{ \bar{Y}_i, \bar{\phi}_{ir} \Rightarrow \}_{i,r}$}
\TrinaryInfC{$\exists p U,\bar{X}, \bar{Y}, \phi \Rightarrow $}
\DisplayProof
\end{center}
Hence, $G \vdash (\exists p U,\bar{X}, \bar{Y}, \phi \Rightarrow \,)$. Thus,
by $(0w)$ and $(R\!\to)$, we get $G \vdash \bar{X}, \bar{Y}, \phi \Rightarrow \exists p U \to 0$. As $U^s=\emptyset$, by definition $\Aemp U=(\exists p U \to 0)$. Therefore, $\exists p U \to 0$ appears as a disjunct in the definition of $\forall p U$. Hence, by $(R \vee)$, we get $G \vdash \bar{X}, \bar{Y}, \phi \Rightarrow \forall p U$.

For the third case, $(\Gamma_i \Rightarrow \Delta_i)$ and $(\Pi_j \Rightarrow \,)$ are proper subsequents of $U$, for any $1 \leq i \leq n$ and $1 \leq j \leq m$. Here, there are two cases to consider: either $n=0$ or $n \neq 0$. We first examine the case $n \neq 0$ before proceeding to the other.
If $n \neq 0$, since $\Delta$ contains at most one formula, there exists $1 \leq e \leq n$ such that $\Delta = \Delta_e$ and $\Delta_i = \emptyset$ for any $i \in \{1, \dots, n\} - \{e\}$. Note that if $\Delta = \emptyset$, we can choose $e$ arbitrarily. Define $T_j = (\Pi_j \Rightarrow \,)$, $S_e = (\Gamma_e \Rightarrow \Delta_e)$, and $S_i = (\Gamma_i \Rightarrow \,)$ for $i \in \{1, \dots, n\} - \{e\}$.
Let $I = \{ i \in \{1, \dots, n\} - \{e\} \mid \Gamma_i \neq \emptyset \}$ and $J = \{ j \in \{1, \dots, m\} \mid \Pi_j \neq \emptyset \}$. Clearly, either $I$ or $J$ must be non-empty; otherwise, $S_e = U$, contradicting our assumption. Therefore, $S_e$, together with $\{T_j\}_{j \in J}$ and $\{S_i\}_{i \in I}$, forms a non-trivial partition of $U$ since this family has at least two elements, all of which are non-empty. This also implies that $S_e$, $T_j$'s, and $S_i$'s are all strictly below $U$ with respect to $\preceq$, for any $i \in I$ and $j \in J$. Moreover, as the tuple $(\{S_i\}_{i \in I \cup \{e\}}, \{T_j\}_{j \in J}, e)$ is in $par_{\bullet}(U)$, the formula $ (\bigast_{j \in J} \exists p T_{j})*(\bigast_{i \in I} \exists p S_i) \to \forall p S_e$ is a disjunct in the definition of $\forall p U$.

Now, since $\bar{\psi}_{js}$, $\bar{\theta}_{js}$, and $\bar{\phi}_{ir}$ are $p$-free, applying the induction hypothesis for $\preceq$, we obtain:

\begin{center}
	  	$G \vdash \exists p T_{j}, \bar{\psi}_{js}, \bar{X}_j \Rightarrow \bar{\theta}_{js}$
	  	\hspace{5pt}
	  	,
	  	\hspace{5pt}
	  	$G \vdash \exists p S_{i}, \bar{\phi}_{ir}, \bar{Y}_i \Rightarrow $
	  \hspace{5pt}
	  	,
	  	\hspace{5pt}
	  	$G \vdash \bar{\phi}_{er}, \bar{Y}_e \Rightarrow \forall p S_{e}$. 
\end{center}
for any $i \in I$ and $j \in J$. Moreover, note that for $i \notin I \cup \{e\}$ and $j \notin J$, as $\Pi_j=\Gamma_i=\Delta_i=\emptyset$, we have:
\begin{center}
	  	$G \vdash \bar{\psi}_{js}, \bar{X}_j \Rightarrow \bar{\theta}_{js}$
	  	\hspace{5pt}
	  	,
	  	\hspace{5pt}
	  	$G \vdash  \bar{\phi}_{ir}, \bar{Y}_i \Rightarrow $
	  \hspace{5pt}
\end{center}
Now, define the local substitution $\sigma'$ similarly to $\sigma$ on the atomic formulas occurring in $R$ and set $\sigma'(\tPi_j)=\{\exists p T_{j}\} \cup \bar{X}_j$ for any $j \in J$, $\sigma'(\tPi_j)=\bar{X}_j$ for any $j \notin J$, $\sigma'(\tGamma_i)=\{\exists p S_{i}\} \cup \bar{Y}_i$ for any $i \in I$, $\sigma'(\tGamma_i)=\bar{Y}_i$ for any $i \notin I$, $\sigma'(\tDelta_i)=\emptyset$ for any $i \in \{1, \ldots, n\}-\{e\}$, and $\sigma'(\tDelta_e)=\{\forall p S_e\}$. Therefore, using the $\sigma'$-instance of $R$, we have:
\[
\{\exists p T_{j}\}_{j \in J},  \{\exists p S_i\}_{i \in I}, \bar{X}, \bar{Y}, \phi \Rightarrow \forall p S_e.
\]
By $(L*)$, we get $G \vdash (\bigast_{j \in J} \exists p T_{j})*(\bigast_{i \in I} \exists p S_i), \bar{X}, \bar{Y}, \phi \Rightarrow \forall p S_e$ and by $(R\!\to)$, $G \vdash \bar{X}, \bar{Y}, \phi \Rightarrow (\bigast_{j \in J} \exists p T_{j})*(\bigast_{i \in I} \exists p S_i) \to \forall p S_e$. As $ (\bigast_{j \in J} \exists p T_{j})*(\bigast_{i \in I} \exists p S_i) \to \forall p S_e$ is a disjunct in  the definition of $\forall p U$, we get $G \vdash \bar{X}, \bar{Y}, \phi \Rightarrow \forall p U$ by $(R \vee)$.

For $n=0$, recall that we are in the third case, so each $T_{j}=(\Pi_j \Rightarrow \,)$ is a proper subsequent of $U$, for all $1 \leq j \leq m$. Let $J=\{j \in \{1, \ldots, m\} \mid \Pi_j \neq \emptyset \}$. Since $U$ is non-empty and each $T_j$ is a proper subsequent of $U$, there must be at least two elements in $J$. 
Hence, $\{T_j\}_{j \in J}$ form a non-trivial partition of $U$. This also implies that $T_j \prec U$ for any $j \in J$. Now, since $\bar{\psi}_{js}$ and $\bar{\theta}_{js}$ are $p$-free, by the induction hypothesis for $\preceq$, we have $G \vdash \exists p T_{j}, \bar{\psi}_{js}, \bar{X}_j \Rightarrow \bar{\theta}_{js}$,
for any $j \in J$. Moreover, for $j \notin J$, as $\Pi_j=\emptyset$, we have $G \vdash \bar{\psi}_{js}, \bar{X}_j \Rightarrow \bar{\theta}_{js}$.
These sequents form an instance of $R$:
\begin{center}
\AxiomC{$\{ \exists p T_j, \bar{X}_{j}, \bar{\psi}_{js} \Rightarrow \bar{\theta}_{js}\}_{j \in J, s}$}
\AxiomC{$\{\bar{X}_j, \bar{\psi}_{js} \Rightarrow \bar{\theta}_{js}\}_{j \notin J,s}$}
\BinaryInfC{$\{\exists p T_j\}_{j \in J} ,\bar{X},  \phi \Rightarrow \,$}
\DisplayProof
\end{center}
To see why, it is enough to define the local substitution $\sigma'$ for $R$ to coincide with $\sigma$ on the atomic formulas occurring in $R$ and set $\sigma'(\tPi_j)=\{\exists p T_{j}\} \cup \bar{X}_j$ for $j \in J$, and $\sigma'(\tPi_j)=\bar{X}_j$ for $j \notin J$. Therefore, using the $\sigma'$-instance of $R$, we obtain
$G \vdash (\{\exists p T_{j}\}_{j \in J}, \bar{X}, \phi \Rightarrow \,)$.
Pick an arbitrary $e' \in J$ and set $J'=J-\{e'\}$. By $(L *)$, $(0w)$, and $(R\!\to)$, we conclude:
\begin{center}
$G \vdash \bigast_{j \in J'} \exists p T_{j}, \bar{X}, \phi \Rightarrow (\exists p T_{e'} \to 0)$.
\end{center}
As the succedent of $T_{e'}$ is empty, by definition, $\Aemp T_{e'}=(\exists p T_{e'} \to 0)$. Hence, as $T_{e'}$ is non-empty, $\Aemp T_{e'}$ occurs as a disjunct in $\forall p T_{e'}$ which implies: 
\begin{center}
$G \vdash \bigast_{j \in J'} \exists p T_{j}, \bar{X}, \phi \Rightarrow \forall p T_{e'}$.
\end{center}
Then, by $(R\!\to)$, we get: 
\begin{center}
$G \vdash \bar{X}, \phi \Rightarrow (\bigast_{j \in J'} \exists p T_{j}) \to \forall p T_{e'}. \quad \quad (\dagger)$
\end{center}
As the tuple $(\{T_j\}_{j \in J}, e')$ is in $par_{\bullet}(U)$, the succedent of $(\dagger)$ is a disjunct in the definition of $\forall p U$. Hence, by $(R \vee)$, we obtain $G \vdash \bar{X}, \phi \Rightarrow \forall p U$.

\item[] $\circ$
Let $R$ be the left single-conclusion semi-analytic rule as described in the previous case, but assume that $\phi = \sigma(\mu) \notin \bar{C}$. Since the conclusion of the $\sigma$-instance of $R$ is:
\[
\sigma(\tPi_1, \dots, \tPi_m, \tGamma_1, \dots, \tGamma_n, \mu \Rightarrow \tDelta_1, \dots, \tDelta_n)=U \cdot (\bar{C} \Rightarrow \,)
\]
and $\phi = \sigma(\mu) \notin \bar{C}$, the multiset $\bar{C}$ must be split among $\sigma(\tPi_j)$'s and $\sigma(\Gamma_i)$'s. Let $\bar{X}_j$ and $\bar{Y}_i$ denote the parts of $\bar{C}$ in $\sigma(\tPi_j)$ and $\sigma(\tGamma_i)$, respectively. Therefore, we must have $\sigma(\tPi_j) = \Pi_j \cup \bar{X}_j$ and $\sigma(\tGamma_i) = \Gamma_i \cup \bar{Y}_i$, for some multisets $\Pi_j$ and $\Gamma_i$.  This means that the $\sigma$-instance of $R$ has the form:
\begin{center}
\AxiomC{$\{ \Pi_j, \bar{X}_j, \bar{\psi}_{js} \Rightarrow \bar{\theta}_{js} \}_{j,s}$}
\AxiomC{$\{ \Gamma_i, \bar{Y}_i, \bar{\phi}_{ir} \Rightarrow \Delta_i \}_{i,r}$}
\RightLabel{$\;\; (\ddagger)$}
\BinaryInfC{$\Pi, \Gamma, \bar{X}, \bar{Y}, \phi \Rightarrow \Delta$}
\DisplayProof
\end{center}
where $\Pi=\bigcup_j \Pi_j$, $\Gamma=\bigcup_i \Gamma_i$, $\bar{X}=\bigcup_j \bar{X}_j$, $\bar{Y}=\bigcup_i \bar{Y}_i$, $\Delta=\bigcup_i \Delta_i$ and $\Delta_i=\sigma(\tDelta_i)$.  Note that $U=(\Pi, \Gamma, \phi \Rightarrow \Delta)$ and $\bar{C}= \bar{X} \cup \bar{Y}$. We want to prove $G \vdash \bar{X}, \bar{Y} \Rightarrow \forall p U$. Define the local substitution $\sigma'$ for $R$ to coincide with $\sigma$ on the atoms occurring in $R$ and set $\sigma'(\tPi_j)=\Pi_j$, $\sigma'(\tGamma_i)=\Gamma_i$, $\sigma'(\tDelta_i)=\Delta_i$, for $1 \leq i \leq n$ and $1 \leq j \leq m$. Then, the $\sigma'$-instance of $R$ has the form:
\begin{center}
\AxiomC{$\{\Pi_j,  \bar{\psi}_{js} \Rightarrow \bar{\theta}_{js} \}_{j,s}$}
\AxiomC{$\{\Gamma_i,  \bar{\phi}_{ir} \Rightarrow \Delta_i \}_{i,r}$}
\BinaryInfC{$\Pi, \Gamma,  \phi \Rightarrow \Delta$}
\DisplayProof
\end{center}
Therefore, $(R, \sigma') \in \mathcal{L}(U)$.
Let $T_{js}= (\Pi_j, \bar{\psi}_{js} \Rightarrow \bar{\theta}_{js})$ and $S_{ir} = (\Gamma_i, \bar{\phi}_{ir} \Rightarrow \Delta_i)$. Note that $T_{js}$ and $S_{ir}$ are premises of an instance of a rule in $G$ with the  conclusion $U$. Thus, we have $T_{js} \prec U$ and $S_{ir} \prec U$, for all indices. Moreover, by $(\ddagger)$, we have $G \vdash T_{js}  \cdot (\bar{X}_j \Rightarrow)$ and $G \vdash S_{ir} \cdot ( \bar{Y}_i \Rightarrow )$. Thus, by the fact that $\bar{X}_j$ and $\bar{Y}_i$ are $p$-free, the induction hypothesis on $\preceq$ shows that for all possible indices:
\begin{center}
	  	$G \vdash \bar{X}_j \Rightarrow \forall p T_{js}$
	  	\hspace{5pt}
	  	,
	  	\hspace{5pt}
	  	$G \vdash \bar{Y}_i \Rightarrow \forall p S_{ir}$
\end{center}
By $(R \wedge)$,  $G \vdash \bar{X}_j \Rightarrow \bigwedge_s \forall p T_{js}$
and $G \vdash \bar{Y}_i \Rightarrow \bigwedge_r \forall p S_{ir}$ for any $i$ and $j$. By $(R *)$, $G \vdash \bar{X}, \bar{Y} \Rightarrow (\bigast_j \bigwedge_s \forall p T_{js}) * (\bigast_i \bigwedge_r \forall p S_{ir})$.
As $(R, \sigma') \in \mathcal{L}(U)$, we have $F_l(U, R, \sigma')=(\bigast_j \bigwedge_s \forall p T_{js}) * (\bigast_i \bigwedge_r \forall p S_{ir})$. Hence, it is a disjunct in the definition of $\forall p U$. Hence, by $(R \vee)$, we get $G \vdash \bar{X}, \bar{Y} \Rightarrow \forall p U$.

\item[] $\circ$
Let $R$ be the following right single-conclusion semi-analytic rule:
\begin{center}
 \AxiomC{$\{\tGamma_i , \bar{\mu}_{ir} \Rightarrow \bar{\nu}_{ir} \mid 1 \leq i \leq n, 1 \leq r \leq k_i\}$}
 \UnaryInfC{$\tGamma_1, \dots, \tGamma_n \Rightarrow \mu $}
 \DisplayProof
\end{center}
As the conclusion of the $\sigma$-instance of $R$ is $U \cdot (\bar{C} \Rightarrow \,)= \sigma(\tGamma_1, \ldots, \tGamma_n \Rightarrow \mu)$, the multiset $\bar{C}$ must be split among $\sigma(\tGamma_i)$'s. Let $\bar{C}_i$ be its part in $\sigma(\tGamma_i)$, so $\sigma(\tGamma_i)=\Gamma_i \cup \bar{C}_i$, for some multiset $\Gamma_i$. So, the $\sigma$-instance of $R$ has the form:
\begin{center}
\AxiomC{$\{\Gamma_i, \bar{C}_i, \bar{\phi}_{ir} \Rightarrow \bar{\psi}_{ir}\}_{i,r}$}
\UnaryInfC{$ \Gamma, \bar{C}\Rightarrow \phi$}
\DisplayProof
\end{center}
where $\Gamma=\bigcup_i \Gamma_i$. Note that $U=(\Gamma \Rightarrow \phi)$ and we want to prove $G \vdash \bar{C} \Rightarrow \forall p U$.
Define the local substitution $\sigma'$ for $R$ to coincide with $\sigma$ on the atoms occurring in $R$ and set $\sigma'(\tGamma_i)=\Gamma_i$. Thus, the $\sigma'$-instance of $R$ has the form:
\begin{center}
\AxiomC{$\{ \Gamma_i, \bar{\phi}_{ir} \Rightarrow \bar{\psi}_{ir} \}_{i,r}$}
\UnaryInfC{$ \Gamma  \Rightarrow \phi$}
\DisplayProof
\end{center}
Therefore, $(R, \sigma') \in \mathcal{R}(U)$.
Let $S_{ir}= (\Gamma_i , \bar{\phi}_{ir} \Rightarrow \bar{\psi}_{ir})$. 
Note that $S_{ir}$'s are the premises of an instance of a rule in $G$ with the conclusion $U$. Hence, $S_{ir} \prec U$, for all indices. Thus, by the induction hypothesis for $\preceq$, we get $G \vdash \bar{C}_i \Rightarrow \forall p S_{ir}$, for all indices. Therefore, by $(R \wedge)$, we get $G \vdash \bar{C}_i \Rightarrow \bigwedge_r \forall p S_{ir}$ and by $(R *)$, we have $G \vdash \bar{C} \Rightarrow \bigast_i \bigwedge_r \forall p S_{ir}$. As $F_r(U, R, \sigma')=\bigast_i \bigwedge_r \forall p S_{ir}$, this formula appears as one of the disjuncts in $\ARR U$ and hence in the definition of $\forall p U$. Therefore, by $(R \vee)$, we get $G \vdash \bar{C} \Rightarrow \forall p U$.

\item[] $\circ$
Let $R$ be either $(K)$ or $(D)$. We only explain the former case. The latter is similar. As $U \cdot (\bar{C} \Rightarrow \,)$ is the conclusion of an instance of $(K)$, it has the form $\Box \Sigma \Rightarrow \Box \alpha$, for some multiset $\Sigma$ and some formula $\alpha$. Thus, there are multisets $\bar{C'}$ and $\Gamma$ such that $\bar{C}=\Box \bar{C'}$ and $\Sigma=\Gamma \cup \bar{C'}$. Therefore, $U=(\Gamma \Rightarrow \alpha)$ and the last step in $\pi$ has the following form:
\begin{center}
\AxiomC{$\Gamma, \bar{C'} \Rightarrow \alpha$}
\UnaryInfC{$\Box\Gamma, \overline{\Box C'} \Rightarrow \Box \alpha$}
\DisplayProof
\end{center}
Let $U'=(\Gamma \Rightarrow \alpha)$. Clealy, $U' \prec U$. As $\bar{C'}$ is $p$-free, the induction hypothesis for $\preceq$ gives $G \vdash \bar{C'} \Rightarrow \forall p U'$. 
Using $(K)$, we get $G \vdash \overline{\Box C'} \Rightarrow \Box \forall p U'$. As $(K)$ is present in $G$ and $U'$ is the premise of an instance of $(K)$ with the conclusion $U$, by definition $\Am U = \Box \forall p U'$. Hence, $\Box \forall p U'$ appears as a disjunct in the definition of $\forall p U$, yielding $G \vdash \bar{C} \Rightarrow \forall p U$, by $(R \vee)$. This completes the inductive proof of the properties claimed for $\exists p V$ and $\forall p U$.
\end{proof}

\begin{cor}\label{MainCorforSemiAnalytic}
If $L \supseteq \mathsf{FL_e}$ is a logic with a terminating single-conclusion semi-analytic calculus, then $L$ has the uniform interpolation property.
\end{cor}
\begin{proof}
Let $G$ be a terminating single-conclusion semi-analytic calculus for $L$. As $L \supseteq \mathsf{FL_e}$, by Lemma \ref{FLAdmissibility}, the calculus $G$ extends $\mathbf{FL_e}$. Now, it is enough to use Theorem \ref{UniformInterpolation} and Theorem \ref{SequentImpliesLogic}.
\end{proof}

\subsection{Multi-conclusion Case}\label{subsection2}
In this subsection, we prove the multi-conclusion part of Theorem \ref{MainTheorem}.

\begin{thm}\label{StrongUniformInterpolation}
Let $G$ be a terminating multi-conclusion semi-analytic calculus extending $\mathbf{CFL_e}$. Then, $G$ has multi-conclusion uniform interpolation.
\end{thm}
\begin{proof}
Similar to the proof of Theorem \ref{UniformInterpolation}, we begin with some preliminary remarks.  First, every axiom of $G$ is focused, since each is either a focused axiom or a multi-conclusion semi-analytic rule without premises, which counts as focused; let this set be $\mathcal{A}$.
Second, since $G$ extends $\mathbf{CFL_e}$, we may assume, without loss of generality, that the axioms and rules of $\mathbf{CFL_e}$ are all available in $G$. Third, as $G$ is a terminating calculus, there exists a well-founded order $\preceq$ such that $G$ is terminating with respect to $\preceq$. Therefore, proper subsequents of $S$, premises of any instance of any rule in $G$ with the conclusion $S$, and if $S = (\Box \Gamma \Rightarrow \Box \Delta)$ such that $\Gamma \cup \Delta \neq \emptyset$, the sequent $\Gamma \Rightarrow \Delta$, all are strictly below $S$ with respect to $\preceq$, for any sequent $S$.

Now, for any sequent $S$ and any atom $p$, we define a $p$-free formula $\forall p S$ and we prove that it meets the conditions of Definition \ref{DfnRelativeInterpolation}.  
As $G$ is a terminating calculus, there is a well-founded order $\preceq$ on sequents such that $G$ is terminating with respect to $\preceq$.  
We define $\forall p S$ by recursion on $\preceq$ and prove its properties by induction on $\preceq$. More precisely, we define $\forall p S$ using the value of $\forall p S'$, for any $S' \prec S$. Then, using the corresponding induction on $\preceq$, we prove that the formula $\forall p S$ has the claimed properties.  
Later, at the end of the proof, we will define $\exists p S = \neg \forall p S$ and prove the required properties for $\exists p S$ from the ones for $\forall p S$.

If $S$ is the empty sequent, 
set $\forall p S$ as $0$. Otherwise, define $\forall p S$ as:
\begin{center}
$\forall p S:= \bigvee_{\!\!par(S)} (\bigplus_{i} \forall p S_i)
\vee \ALR S \vee \ARR S
\vee \Am S \vee \Ab S \vee \Aa S $   
\end{center}
where:
\item[] $\bullet$
in the first disjunct, $par(S)$ denotes the set of all non-trivial partitions $(S_1, \ldots, S_n)$ of $S$. Similar to the argument in the proof of Theorem \ref{UniformInterpolation}, this disjunct is allowed in the recursive definition.

\item[] $\bullet$
$\ALR S$ is defined as $\bigvee_{(R, \sigma) \in \mathcal{L}(S)} F_l(S, R, \sigma)$, where $\mathcal{L}(S)$ is the set of all pairs $(R, \sigma)$ of left multi-conclusion semi-analytic rule $R$ in $G$ and a local substitution $\sigma$ for $R$ such that the conclusion of the $\sigma$-instance of $R$ is $S$. To define $F_l(S, R, \sigma)$, let $R$ be of the form:
\small\begin{center}
 \AxiomC{$\{\tGamma_i , \bar{\mu}_{ir} \Rightarrow \bar{\nu}_{ir}, \tDelta_i \mid 1 \leq i \leq n, 1 \leq r \leq k_i \}$}
 \UnaryInfC{$\tGamma_1, \dots, \tGamma_n, \mu \Rightarrow \tDelta_1, \dots, \tDelta_n $}
 \DisplayProof
\end{center}
\normalsize
and $ S_{ir} = \sigma(\tGamma_i , \bar{\mu}_{ir} \Rightarrow \bar{\nu}_{ir}, \tDelta_i)$, for $ 1 \leq i \leq n$ and $ 1 \leq r \leq k_i$. Then, define $F_l(S, R, \sigma)$ as $(\bigast_i \bigwedge_r \forall p S_{ir})$.
Similar to the argument in the proof of Theorem \ref{UniformInterpolation}, this disjunct is allowed in the recursive definition.
\item[] $\bullet$
$\ARR S$ is defined as $\bigvee_{(R, \sigma) \in \mathcal{R}(S)} F_r(S, R, \sigma)$, where $\mathcal{R}(S)$ is the set of all pairs $(R, \sigma)$ of the right multi-conclusion semi-analytic rule $R$ in $G$ and local substitution $\sigma$ for $R$ such that the conclusion of the $\sigma$-instance of $R$ is $S$. To define $F_r(S, R, \sigma)$, let $R$ be of the form:
\small\begin{center}
 \AxiomC{$\{\tGamma_i , \bar{\mu}_{ir} \Rightarrow \bar{\nu}_{ir}, \tDelta_i \mid 1 \leq i \leq n, 1 \leq r \leq k_i \}$}
 \UnaryInfC{$\tGamma_1, \dots, \tGamma_n \Rightarrow \mu, \tDelta_1, \dots, \tDelta_n $}
 \DisplayProof
\end{center}
\normalsize
and $ S_{ir} = \sigma(\tGamma_i , \bar{\mu}_{ir} \Rightarrow \bar{\nu}_{ir}, \tDelta_i)$, for $ 1 \leq i \leq n$ and $ 1 \leq r \leq k_i$. Then, define $F_r(S, R, \sigma)$  as $(\bigast_i \bigwedge_r \forall p S_{ir})$.  Similar to the previous case, this disjunct in the recursive definition is allowed.

\item[] $\bullet$
If either $(K)$ is or both $(K)$ and $(D)$ are present in $G$ and $S$ is the conclusion of an instance of one of the rules $(K)$ or $(D)$ with the premise $S'$, we define $\Am S$ as $\Box \forall p S'$. Otherwise, $\Am S$ is defined as $\bot$. The well-definedness is again in place.

\item[] $\bullet$
If either $(K)$ is or both $(K)$ and $(D)$ are present in $G$ and $S$ is of the form $(\Box \Gamma \Rightarrow \,)$, for some non-empty multiset $\Gamma$, then $\Ab S$ is defined as $\neg \Box \neg \forall p S''$, where  $S''=(\Gamma \Rightarrow)$; otherwise, $\Ab S$ is defined as $\bot$.

\item[] $\bullet$
$\Aa S$ is the formula constructed in Theorem \ref{UniformInterpolationForAxiom}.  Recall that $\mathcal{A}$ is the set of all axioms in $G$, and we observed that it consists of focused axioms. Moreover, note that the conditions of Theorem \ref{UniformInterpolationForAxiom} are trivially satisfied.\\

\noindent This completes the definition of $\forall p S$. We now proceed to show that $\forall p S$ satisfies the properties listed in Definition \ref{DfnRelativeInterpolation}. To this end, we use induction on $\preceq$. More precisely, we prove \emph{any} of the properties of $\forall p S$, assuming the validity of \emph{all} the properties for $\forall p S'$, for all $S' \prec S$. This format is not crucial as it was for Theorem \ref{UniformInterpolation}. However, we use it to be similar to the proof of that Theorem.

The proof that $\forall p S$ is $p$-free is easy and similar to the corresponding claim in the proof of Theorem \ref{UniformInterpolation}. Similarly, it is easy to prove $\mathcal{V}(\forall p S) \subseteq \mathcal{V}(S)$. Therefore, we start by showing that $G \vdash S \cdot (\forall p S \Rightarrow \,)$, for any sequent $S$.
If $S$ is empty, then $\forall p S=0$ and as $G \vdash S \cdot (\forall p S \Rightarrow \,)=(0 \Rightarrow \,)$, there is nothing to prove. Otherwise, by the recursive definition of $\forall p S$, it is enough to prove $G \vdash S \cdot (d \Rightarrow \,)$, for any disjunct $d$ in the definition of $\forall p S$. Then, by $(L\vee)$, we can reach $G \vdash S \cdot (\forall p S \Rightarrow \,)$. In the following, we will address each disjunct, separately:

\item[] $\circ$
For the first disjunct, let $(S_1, \dots, S_n)$ be a non-trivial partition of $S$. We show that $G \vdash S \cdot (\bigplus_i \forall p S_i \Rightarrow)$. Let $S_i=(\Gamma_i \Rightarrow \Delta_i)$. Hence, $S=(\Gamma_1, \ldots, \Gamma_n \Rightarrow \Delta_1, \ldots, \Delta_n)$. Since $S_i \prec S$, for each $1 \leq i \leq n$, by the induction hypothesis on $\preceq$, we have $G \vdash (\Gamma_i, \forall p S_i \Rightarrow \Delta_i)$. Now, by $(L +)$ admissible in $G 
$, we get $G \vdash \Gamma_1, \dots, \Gamma_n, (\bigplus_i \forall p S_i) \Rightarrow \Delta_1, \dots, \Delta_n$.

\item[] $\circ$
For the disjunct $\ALR S$, we prove $G \vdash S \cdot (F_l(S, R, \sigma) \Rightarrow \,)$, for any $(R, \sigma) \in \mathcal{L}(S)$. Let $R$ be the following left multi-conclusion semi-analytic rule:
\begin{center}
 \AxiomC{$\{\tGamma_i , \bar{\mu}_{ir} \Rightarrow \bar{\nu}_{ir}, \tDelta_i \mid 1 \leq i \leq n, 1 \leq r \leq k_i\}$}
 \UnaryInfC{$\tGamma_1, \dots, \tGamma_n, \mu \Rightarrow \tDelta_1, \ldots, \tDelta_n $}
 \DisplayProof
\end{center}
For simplicity, use Convention \ref{convention} and set $\sigma(\tGamma_i)=\Gamma_i$ and $\sigma(\tDelta_i)=\Delta_i$. Then, $ S_{ir} = (\Gamma_i, \bar{\phi}_{ir} \Rightarrow \bar{\psi}_{ir}, \Delta_i) $ for $ 1 \leq i \leq n$ and $ 1 \leq r \leq k_i $ are the premises of the $\sigma$-instance of $R$ with the conclusion $S=(\Gamma_1, \ldots, \Gamma_n, \phi \Rightarrow \Delta_1, \ldots, \Delta_n)$. Let  $\Gamma=\bigcup_i \Gamma_i$ and $\Delta=\bigcup_i \Delta_i$. By the definition of $F_l(S, R, \sigma)$, we should prove:
\begin{center}
$G \vdash \Gamma, \phi, \bigast_i \bigwedge_r \forall p S_{ir} \Rightarrow \Delta$.
\end{center}
As the premise of any instance of any rule in $G$ is lower than its conclusion, we have $S_{ir} \prec S$, for all indices. Thus, by the induction hypothesis, we get:
\begin{center}
$G \vdash \Gamma_i , \forall p S_{ir}, \bar{\phi}_{ir} \Rightarrow \bar{\psi}_{ir}, \Delta_i$ \quad for $1 \leq i \leq n$ and $1 \leq r \leq k_i$
\end{center}
Using the rule $(L \wedge)$, we get:
\begin{center}
$G \vdash \Gamma_i, \bigwedge_r\forall p S_{ir}, \bar{\phi}_{ir} \Rightarrow \bar{\psi}_{ir}, \Delta_i$ \quad for $1 \leq i \leq n$.
\end{center}
These sequents can be the premises of an instance of $R$. To see why, it is enough to define the local substitution $\sigma'$ to coincide with $\sigma$ on the atomic formulas occurring in $R$ and set $\sigma'(\tGamma_i)=\Gamma_i \cup \{\bigwedge_r \forall p S_{ir}\}$ and $\sigma'(\tDelta_i)=\Delta_i$, for any $1 \leq i \leq n$. Applying the $\sigma'$-instance of $R$, we get:
\begin{center}
$G \vdash \Gamma, \bigwedge_r \forall p S_{1r} , \dots , \bigwedge_r \forall p S_{nr}, \phi  \Rightarrow \Delta$.
\end{center}
Using the rule $(L *)$, we get $G \vdash \Gamma,  \bigast_i \bigwedge_r \forall p S_{ir}, \phi \Rightarrow \Delta$, as desired.

\item[] $\circ$
The case for the disjunct $\ARR S$ is similar to the previous case.

\item[] $\circ$
The case for $\Am S$ is similar to the corresponding case in the proof of Theorem \ref{UniformInterpolation}.

\item[] $\circ$ 
For the disjunct $\Ab S$, if $(K)$ is or both $(K)$ and $(D)$ are present in $G$ and $S$ is of the form $S=(\Box \Gamma \Rightarrow \,)$, for some non-empty $\Gamma$, then $\Ab S$ is defined as $\neg \Box \neg \forall p S''$, where $S''=(\Gamma \Rightarrow \,)$. As $S$ is non-empty, we have $S'' \prec S$. Hence, by the induction hypothesis, we get $G \vdash (\Gamma, \forall p S'' \Rightarrow)$. By the rule $(0 w)$ and then $(R\!\to)$, we get $G \vdash (\Gamma \Rightarrow \neg \forall p S'')$. Thus, using the rule $(K)$, we have $G \vdash (\Box \Gamma \Rightarrow \Box \neg \forall p S'')$. By $G \vdash (0 \Rightarrow)$ and  $(L\!\to)$, we get $G \vdash (\Box \Gamma, \neg \Box \neg \forall p S'' \Rightarrow)$, as desired. In the otherwise case, we have $\Ab S=\bot$. Then, as $G \vdash S \cdot (\bot \Rightarrow \,)$, the claim is clear.

\item[] $\circ$
The case for the disjunct $\Aa S$ is clear from Theorem \ref{UniformInterpolationForAxiom}.\\

This completes the proof of $G \vdash S \cdot (\forall p S \Rightarrow \,)$, for any sequent $S$. Now, we want to prove that for any provable sequent in $G$ in the form $S \cdot (\bar{C} \Rightarrow \bar{D})$, where $\bar{C}$ and $\bar{D}$ are $p$-free multisets, we have $G \vdash \bar{C} \Rightarrow \bar{D}, \forall p S$.
Again, we prove this claim by induction on $\preceq$ and in any inductive step, we also use an induction on the depth of the proof of $S \cdot (\bar{C} \Rightarrow \bar{D})$.
More precisely, we show:

\begin{itemize}
\item[$(*)$]
For any number $d \in \mathbb{N}$ and any proof $\pi$ in $G$ with depth at most $d$ of a sequent in the form $S \cdot (\bar{C} \Rightarrow \bar{D})$, where $\bar{C}$ and $\bar{D}$ are $p$-free multisets, we have $G \vdash \bar{C} \Rightarrow \forall p S, \bar{D}$.
\end{itemize}

\noindent If $S$ is the empty sequent, then $\forall p S=0$. As $S \cdot (\bar{C} \Rightarrow \bar{D})$ is provable in $G$, we have $G \vdash (\bar{C} \Rightarrow \bar{D}, 0)$ by $(0w)$. If $S$ is non-empty, we proceed by induction on $d$. The base case $d=0$ is similar to the one in the proof of Theorem \ref{UniformInterpolation}. 
For the induction step, let $R$ be the rule and let $\sigma$ be a local substitution for $R$ such that the $\sigma$-instance of $R$ is the last rule applied in $\pi$. There are several cases to consider: either $R$ is  left multi-conclusion semi-analytic,  right multi-conclusion semi-analytic, the rule $(K)$, or the rule $(D)$. In the first (resp. second) case, either the main formula appears in $\bar{C}$ (resp. $\bar{D}$), or it does not. In the following, we carefully analyze each case. Additionally, recall that we use Convention \ref{convention} to name the substituted formulas:

\item[] $\circ$
Let $R$ be the following left multi-conclusion semi-analytic rule:
\begin{center}
 \AxiomC{$\{\tGamma_i , \bar{\mu}_{ir} \Rightarrow \bar{\nu}_{ir}, \tDelta_i \mid 1 \leq i \leq n, 1 \leq r \leq k_i\}$}
 \UnaryInfC{$\tGamma_1, \dots, \tGamma_n, \mu \Rightarrow \tDelta_1, \ldots, \tDelta_n$}
 \DisplayProof
\end{center}
and $\phi=\sigma(\mu) \in \bar{C}$. Since $\phi \in \bar{C}$ is $p$-free and $R$ is occurrence-preserving, $\bar{\phi}_{ir}$ and $\bar{\psi}_{ir}$ are also $p$-free. Now, as the conclusion of the $\sigma$-instance of $R$ is  
\[
\sigma(\tGamma_1, \ldots, \tGamma_n, \mu \Rightarrow  \tDelta_1, \ldots, \tDelta_n)=S \cdot (\bar{C} \Rightarrow \bar{D})
\]
and $\phi=\sigma(\mu) \in \bar{C}$, the multiset $\bar{C}-\{\phi\}$ (resp. $\bar{D}$) splits among $\sigma(\tGamma_i)$'s (resp. $\sigma(\tDelta_i)$'s). Let $\bar{X}_i$ (resp. $\bar{D}_i$) be the parts of $\bar{C}-\{\phi\}$ (resp. $\bar{D}$) in $\sigma(\tGamma_i)$ (resp. $\sigma(\tDelta_i)$). Hence, $\sigma(\tGamma_i)=\Gamma_i \cup \bar{X}_i$ and $\sigma(\tDelta_i)=\Delta_i \cup \bar{D}_i$, for some multisets $\Gamma_i$ and $\Delta_i$. Thus, the $\sigma$-instance of $R$ has the form:
\begin{center}
\AxiomC{$\{\Gamma_i, \bar{X}_i, \bar{\phi}_{ir} \Rightarrow \bar{\psi}_{ir}, \bar{D}_i, \Delta_i\}_{i,r}$}
\UnaryInfC{$ \Gamma, \bar{X}, \phi \Rightarrow \bar{D}, \Delta$}
\DisplayProof
\end{center}
where $\Gamma=\bigcup_i \Gamma_i$, $\bar{X}=\bigcup_i \bar{X}_i$, and $\Delta=\bigcup_i \Delta_i$. Note that $S=(\Gamma \Rightarrow \Delta)$ and we want to prove $G \vdash (\bar{X}, \phi \Rightarrow \bar{D}, \forall p S)$.
Let $I=\{i \in \{1, \ldots, n\} \mid \Gamma_i \cup \Delta_i \neq \emptyset\}$. Note that $I$ is non-empty, as otherwise $S$ would be empty, which is a contradiction. Thus, there are two cases to consider: either $I$ is a singleton or it has at least two elements. Denote $S_{i}=(\Gamma_i \Rightarrow \Delta_i)$, for $1 \leq i \leq n$.

If $I$ is a singleton, let $I=\{e\}$. Then, $S=(\Gamma_e \Rightarrow \Delta_e)$, $\Gamma_i=\Delta_i=\emptyset$ for any $i \in \{1, \ldots, n\}-\{e\}$, and $\pi$ is of the form:
\begin{center}
\AxiomC{$\pi_{er}$}
\noLine
\UnaryInfC{$\{\Gamma_e, \bar{X}_e, \bar{\phi}_{er} \Rightarrow \bar{\psi}_{er}, \bar{D}_e, \Delta_e
\}_r$}
\AxiomC{$\{ \bar{X}_i, \bar{\phi}_{ir} \Rightarrow \bar{\psi}_{ir}, \bar{D}_i\}_{i \neq e,r}$}
\BinaryInfC{$ \Gamma_e, \bar{X}, \phi \Rightarrow \bar{D}, \Delta_e$}
\DisplayProof
\end{center}
Now, as the depth of $\pi_{er}$ is smaller than $d$, by the induction hypothesis on $d$, we get $G \vdash \bar{X}_e, \bar{\phi}_{er} \Rightarrow \bar{\psi}_{er}, \bar{D}_e, \forall p S$.  
Define the local substitution $\sigma'$ for $R$ similarly to $\sigma$ on the atomic formulas occurring in $R$ and set $\sigma'(\tGamma_i) = \bar{X}_i$ and $\sigma'(\tDelta_i) = \bar{D}_i$, for $i \in \{1, \ldots, n\} - \{e\}$, and $\sigma'(\tGamma_e) = \bar{X}_e$ and $\sigma'(\tDelta_e) = \bar{D}_e \cup \{\forall p S\}$.  
Then, the $\sigma'$-instance of $R$ has the form:
\begin{center}
	  	\AxiomC{$\{\bar{X}_e, \bar{\phi}_{er} \Rightarrow \bar{\psi}_{er}, \bar{D}_e, \forall p S\}_r$}
	  	\AxiomC{$\{\bar{X}_i, \bar{\phi}_{ir} \Rightarrow \bar{\psi}_{ir}, \bar{D}_i\}_{i \neq e, r}$}
  		\BinaryInfC{$ \bar{X}, \phi \Rightarrow \bar{D}, \forall p S$}
  		\DisplayProof
\end{center}
Hence, $G \vdash  \bar{X}, \phi \Rightarrow \bar{D}, \forall p S$, as desired.

In the second case, if $I$ has at least two elements, the sequents $\{S_i\}_{i \in I}$ form a non-trivial partition of $S$ as it has at least two elements, and all of them are non-empty. This implies that $S_i \prec S$, for any $i \in I$. It also implies that the formula $\bigplus_{i \in I} \forall p S_i$ is a disjunct in the definition of $\forall p S$. Now, by the induction hypothesis on $\preceq$, we get $G \vdash  \bar{X}_i, \bar{\phi}_{ir} \Rightarrow \bar{\psi}_{ir}, \bar{D}_i, \forall p S_{i}$, for any $i \in I$. For $i \notin I$, as $\Gamma_i = \Delta_i = \emptyset$, we already have $G \vdash \bar{X}_i, \bar{\phi}_{ir} \Rightarrow \bar{\psi}_{ir}, \bar{D}_i$. All these sequents form an instance of $R$. It is enough to define the local substitution $\sigma'$ for $R$ to coincide with $\sigma$ on the atomic formulas occurring in $R$ and set $\sigma'(\tGamma_i) = \bar{X}_i$ for $1 \leq i \leq n$, $\sigma'(\tDelta_i) = \bar{D}_i \cup \{\forall p S_i\}$ for $i \in I$, and $\sigma'(\tDelta_i) = \bar{D}_i$ for $i \notin I$. Therefore, the $\sigma'$-instance of $R$ has the form:
\begin{center}
	  	\AxiomC{$\{\bar{X}_i, \bar{\phi}_{ir} \Rightarrow \bar{\psi}_{ir}, \bar{D}_i, \forall p S_i\}_{i \in I, r}$}
	  	\AxiomC{$\{\bar{X}_i, \bar{\phi}_{ir} \Rightarrow \bar{\psi}_{ir}, \bar{D}_i\}_{i \notin I, r}$}
  		\BinaryInfC{$ \bar{X}, \phi \Rightarrow \bar{D}, \{\forall p S_i\}_{i \in I}$}
  		\DisplayProof
\end{center}
Applying the $\sigma'$-instance of $R$ we get $G \vdash \bar{X}, \phi \Rightarrow \bar{D}, \{\forall p S_i\}_{i \in I}$. Then, by $(R+)$, admissible in $G$, we get $G \vdash \bar{X}, \phi \Rightarrow \bar{D}, \bigplus_{i \in I} \forall p S_i$. As $\bigplus_{i \in I} \forall p S_i$ appears as a disjunct in $\forall p S$, we get $G \vdash \bar{X}, \phi \Rightarrow \bar{D}, \forall p S$, by $(R\vee)$.

\item[] $\circ$
Let $R$ be the left multi-conclusion semi-analytic rule, as in the previous case, but assume that $\phi \notin \bar{C}$. Since the conclusion of the $\sigma$-instance of $R$ is:
\[
\sigma(\tGamma_1, \ldots, \tGamma_n, \mu \Rightarrow  \tDelta_1, \ldots, \tDelta_n)=S \cdot (\bar{C} \Rightarrow \bar{D})
\]
and $\phi = \sigma(\mu) \notin \bar{C}$, the multiset $\bar{C}$ (resp. $\bar{D}$) splits among the $\sigma(\tGamma_i)$'s (resp. $\sigma(\tDelta_i)$'s). Let $\bar{C}_i$ (resp. $\bar{D}_i$) be the parts of $\bar{C}$ (resp. $\bar{D}$) in $\sigma(\tGamma_i)$ (resp. $\sigma(\tDelta_i)$). Hence, $\sigma(\tGamma_i) = \Gamma_i \cup \bar{C}_i$ and $\sigma(\tDelta_i) = \Delta_i \cup \bar{D}_i$, for some multisets $\Gamma_i$ and $\Delta_i$. Therefore, the $\sigma$-instance of $R$ has the form:
\begin{center}
\AxiomC{$\{ \Gamma_i, \bar{C}_i, \bar{\phi}_{ir} \Rightarrow \bar{\psi}_{ir}, \bar{D}_i, \Delta_i \}_{i,r}$}
\UnaryInfC{$ \Gamma, \bar{C}, \phi \Rightarrow \bar{D}, \Delta$}
\DisplayProof
\end{center}
where $\Gamma=\bigcup_i \Gamma_i$ and $\Delta=\bigcup_i \Delta_i$. 
Note that $S=(\Gamma, \phi \Rightarrow \Delta)$ and we want to prove $G \vdash \bar{C} \Rightarrow \forall p S, \bar{D}$. Define the local substitution $\sigma'$ for $R$ to coincide with $\sigma$ on the atomic formulas occurring in $R$, and set $\sigma'(\tGamma_i) = \Gamma_i$ and $\sigma'(\tDelta_i) = \Delta_i$, for $1 \leq i \leq n$. Hence, the $\sigma'$-instance of $R$ has the form:
\begin{center}
\AxiomC{$\{\Gamma_i, \bar{\phi}_{ir} \Rightarrow \bar{\psi}_{ir}, \Delta_i \}_{i,r}$}
\UnaryInfC{$ \Gamma,  \phi \Rightarrow  \Delta$}
\DisplayProof
\end{center}
Therefore, $(R, \sigma') \in \mathcal{L}(S)$. Let $S_{ir} = (\Gamma_i, \bar{\phi}_{ir} \Rightarrow \bar{\psi}_{ir}, \Delta_i)$. Since the $S_{ir}$'s are the premises of an instance of a rule in $G$ with the conclusion $S$, we have $S_{ir} \prec S$, for all indices. Thus, by the induction hypothesis for $\preceq$, we get $G \vdash \bar{C}_i \Rightarrow \forall p S_{ir}, \bar{D}_i$. Then, by $(R \wedge)$, we obtain $G \vdash \bar{C}_i \Rightarrow \bigwedge_r \forall p S_{ir}, \bar{D}_i$, for any $1 \leq i \leq n$ and by $(R *)$, we reach $G \vdash \bar{C} \Rightarrow \bigast_i \bigwedge_r \forall p S_{ir}, \bar{D}$.
Since $(R, \sigma') \in \mathcal{L}(S)$ and $F_l(S, R, \sigma') = \bigast_i \bigwedge_r \forall p S_{ir}$, this formula appears as a disjunct in $\forall p S$. Therefore, by $(R \vee)$, we get $G \vdash \bar{C} \Rightarrow \forall p S, \bar{D}$.

\item[] $\circ$
The case where $R$ is a right multi-conclusion semi-analytic rule is similar to the previous cases.

\item[] $\circ$
Let $R$ be either $(K)$ or $(D)$. We will only explain the former case, as the latter is similar. Since $S \cdot (\bar{C} \Rightarrow \bar{D})$ is the conclusion of an instance of $(K)$, it has the form $(\Box \Sigma \Rightarrow \Box \alpha)$, for some multiset $\Sigma$ and some formula $\alpha$. Therefore, there exist multisets $\bar{C'}$ and $\Gamma$ such that 
$\bar{C} = \Box \bar{C'}$ and $\Sigma = \Gamma \cup \bar{C'}$.
Thus, the last step of $\pi$ has the following form:
\begin{center}
\AxiomC{$\Gamma, \bar{C'} \Rightarrow \alpha$}
\UnaryInfC{$\Box \Gamma, \overline{\Box C'} \Rightarrow \Box \alpha$}
\DisplayProof
\end{center}
Now, there are two cases to consider: either $\Box \alpha \in \bar{D}$ or $\Box \alpha \notin \bar{D}$.

In the first case, we must have $S=(\Box \Gamma \Rightarrow \,)$ and $\bar{D}=\{\Box \alpha\}$. Let $S'' = (\Gamma \Rightarrow \,)$. As $S$ is non-empty, $\Gamma$ is also non-empty. Hence, we have $S'' \prec S$, and as $\bar{C'}$ and $\alpha$ are both $p$-free, by the induction hypothesis for $\preceq$, we get
$G \vdash \bar{C'} \Rightarrow \forall p S'', \alpha$.
Thus, $G \vdash \bar{C'}, \neg \forall p S'' \Rightarrow \alpha$, by $G \vdash (0 \Rightarrow)$ and $(L\!\to)$. 
By $(K)$, we get $G \vdash \overline{\Box C'}, \Box \neg \forall p S'' \Rightarrow \Box \alpha$, and by $(0 w)$ and $(R\!\to)$, we reach $G \vdash \overline{\Box C'} \Rightarrow \neg \Box \neg \forall p S'', \Box \alpha$. Since $S$ is non-empty and has the form $(\Box \Gamma \Rightarrow \,)$, by definition we have $\Ab S=\neg \Box \neg \forall p S''$. Thus, this formula appears as a disjunct in $\forall p S$. Hence, we get $G \vdash \bar{C} \Rightarrow \forall p S, \bar{D}$, by $(R \vee)$.

In the second case, since $\Box \alpha \notin \bar{D}$, we must have $S=(\Box \Gamma \Rightarrow \Box \alpha)$ and $\bar{D}=\emptyset$.
Clearly, the following is an instance of the rule $(K)$:
\begin{center}
\AxiomC{$\Gamma \Rightarrow \alpha$}
\UnaryInfC{$\Box \Gamma \Rightarrow \Box \alpha$}
\DisplayProof
\end{center}
Let $S' = (\Gamma \Rightarrow \alpha)$. Since $S$ is non-empty, by definition, we have $\Am S = \Box \forall p S'$. As $S' \prec S$, by the induction hypothesis for $\preceq$, we get $G \vdash \bar{C'} \Rightarrow \forall p S'$. By applying $(K)$, we obtain $G \vdash \overline{\Box C'} \Rightarrow \Box \forall p S'$. Finally, since $\Am S = \Box \forall p S'$ is a disjunct in $\forall p S$, we conclude that $G \vdash \bar{C} \Rightarrow \forall p S$.\\

\noindent This completes the proof for $\forall p S$. For $\exists p S$, it is sufficient to define $\exists p S := \neg \forall p S$. We show that it satisfies all the conditions in Definition \ref{DfnRelativeInterpolation}. The conditions related to the variables are straightforward. For the other two conditions, let $S = (\Gamma \Rightarrow \Delta)$. Since $G \vdash \Gamma, \forall p S \Rightarrow \Delta$, by $(0w)$, we get $G \vdash \Gamma, \forall p S \Rightarrow \Delta, 0$, and by applying $(R\!\to)$, we obtain $G \vdash \Gamma \Rightarrow \Delta, \neg \forall p S$, which is equivalent to $G \vdash \Gamma \Rightarrow \Delta, \exists p S$. 
Second, for any $p$-free multisets $\bar{C}$ and $\bar{D}$, if $G \vdash \Gamma, \bar{C} \Rightarrow \bar{D}, \Delta$, then $G \vdash \bar{C} \Rightarrow \bar{D}, \forall p S$. Since $G \vdash (0 \Rightarrow \,)$, by $(L\!\to)$, we get $G \vdash \bar{C}, \neg \forall p S \Rightarrow \bar{D}$, or equivalently, $G \vdash \bar{C}, \exists p S \Rightarrow \bar{D}$.
\end{proof}

\begin{cor}\label{MainCorforMultiSemiAnalytic}
If $L \supseteq \mathsf{CFL_e}$ is a logic that has a terminating multi-conclusion semi-analytic calculus, then $L$ has uniform interpolation property.
\end{cor}
\begin{proof}
Let $G$ be a terminating multi-conclusion semi-analytic calculus for $L$. As $L \supseteq \mathsf{CFL_e}$, by Lemma \ref{FLAdmissibility}, the calculus $G$ extends $\mathbf{CFL_e}$. Now, it is enough to use Theorem \ref{StrongUniformInterpolation} and Theorem \ref{SequentImpliesLogic}.
\end{proof}

\section{Conclusion and Future Work}
In \cite{Craig}, we introduced a formalization of the informal notion of \emph{nice} proof systems by what we called (single-conclusion) multi-conclusion \emph{semi-analytic} sequent calculi. We demonstrated that any (modal) substructural logic extending $\mathsf{CFL_e}$ (resp. $\mathsf{FL_e}$) that admits a (single-conclusion) multi-conclusion semi-analytic calculus possesses the Craig interpolation property. In this paper, we extended this result to show that if the calculus is \emph{terminating}, the logic also enjoys the UIP.

This result generalizes the findings of \cite{Iemhoff1, Iemhoff}, which identified a connection between the existence of a special form of terminating sequent calculus, called focused calculi, and the UIP of the superintuitionistic logic it characterizes. Our approach provides a uniform and modular method for proving UIP for a broad class of (modal) substructural logics by examining only the syntactical structure of axioms and rules in their sequent calculi, alongside verifying termination.  
Using this method, we prove that the $K$, $D$, and $T$-type modal extensions of $\mathsf{FL_e}$, $\mathsf{FL_{ew}}$, $\mathsf{CFL_e}$, and $\mathsf{CFL_{ew}}$, as well as $\mathsf{CPC}$, $\mathsf{K}$, and $\mathsf{KD}$ satisfy UIP. On the negative side, since UIP is a rare property, our result implies that many (modal) substructural logics, including $\mathsf{K4}$ and $\mathsf{S4}$ cannot have a terminating single- or multi-conclusion semi-analytic calculus.  

For future work, it would be valuable to extend the methodology of the present paper to sequent calculi that permit analytic applications of the cut rule. Moreover, compared to propositional rules, our use of modal rules is quite restricted. Identifying a generic form of modal rules rendering the proof-theoretic argument for UIP remains an open and promising direction for further exploration.  \\

\noindent \textbf{Acknowledgment} We are grateful to Rosalie Iemhoff for drawing our attention to this intriguing line of research, generously sharing her insights on the topic, and engaging in valuable discussions with us. We also wish to thank Pavel Pudl\'{a}k, George Metcalfe, Revantha Ramanayake, and Masoud Memarzadeh for their thoughtful feedback on the initial draft and Hiroakira Ono and Silvio Ghilardi for the helpful discussions we had.

\bibliographystyle{plain}
\bibliography{Universal}

\begin{thebibliography}{10}

\bibitem{DP}
Amirhossein {Akbar Tabatabai} and Raheleh Jalali.
\newblock Universal proof theory: Feasible admissibility in intuitionistic modal logics.
\newblock {\em Annals of Pure and Applied Logic}, 176(2):103526, 2025.

\bibitem{Craig}
Amirhossein Akbar~Tabatabai and Raheleh Jalali.
\newblock Universal proof theory: Semi-analytic rules and {C}raig interpolation.
\newblock {\em Annals of Pure and Applied Logic}, 176(1):103509, 2025.

\bibitem{Aliz}
Majid Alizadeh, Farzaneh Derakhshan, and Hiroakira Ono.
\newblock Uniform interpolation in substructural logics.
\newblock {\em The Review of Symbolic Logic}, 7(3):455--483, 2014.

\bibitem{Bil}
Marta B{\'\i}lkov{\'a}.
\newblock Uniform interpolation and propositional quantifiers in modal logics.
\newblock {\em Studia Logica}, 85(1):1--31, 2007.

\bibitem{Chagrov}
Alexander Chagrov and Michael Zakharyaschev.
\newblock {\em Modal Logic}.
\newblock Oxford University Press, New York, 1997.

\bibitem{Cia}
Agata Ciabattoni, Nikolaos Galatos, and Kazushige Terui.
\newblock Algebraic proof theory for substructural logics: cut-elimination and completions.
\newblock {\em Annals of Pure and Applied Logic}, 163(3):266--290, 2012.

\bibitem{dosen}
Kosta Do{\v{s}}en.
\newblock Modal translations in substructural logics.
\newblock {\em Journal of Philosophical Logic}, 21:283--336, 1992.

\bibitem{Dyck}
Roy Dyckhoff.
\newblock Contraction-free sequent calculi for intuitionistic logic.
\newblock {\em The Journal of Symbolic Logic}, 57(3):795--807, 1992.

\bibitem{Ghil1}
Silvio Ghilardi and Marek Zawadowski.
\newblock Undefinability of propositional quantifiers in the modal system {S4}.
\newblock {\em Studia Logica}, 55(2):259--271, 1995.

\bibitem{girard1987linear}
Jean-Yves Girard.
\newblock Linear logic.
\newblock {\em Theoretical computer science}, 50(1):1--101, 1987.

\bibitem{LLL}
Jean-Yves Girard.
\newblock Light linear logic.
\newblock In {\em International Workshop on Logic and Computational Complexity}, pages 145--176. Springer, 1994.

\bibitem{affine}
Jean-Yves Girard.
\newblock Linear logic: its syntax and semantics.
\newblock {\em London Mathematical Society Lecture Note Series}, pages 1--42, 1995.

\bibitem{Iemhoff1}
Rosalie Iemhoff.
\newblock Uniform interpolation and sequent calculi in modal logic.
\newblock {\em Archive for Mathematical Logic}, 58(1):155--181, 2019.

\bibitem{Iemhoff}
Rosalie Iemhoff.
\newblock Uniform interpolation and the existence of sequent calculi.
\newblock {\em Annals of Pure and Applied Logic}, 170(11):102711, 2019.

\bibitem{kanovich}
Max Kanovich, Stepan Kuznetsov, Vivek Nigam, and Andre Scedrov.
\newblock Subexponentials in non-commutative linear logic.
\newblock {\em Mathematical Structures in Computer Science}, 29(8):1217--1249, 2019.

\bibitem{kiriyama}
Eiji Kiriyama and Hiroakira Ono.
\newblock The contraction rule and decision problems for logics without structural rules.
\newblock {\em Studia Logica}, 50:299--319, 1991.

\bibitem{elaine}
Bj{\"o}rn Lellmann, Carlos Olarte, and Elaine Pimentel.
\newblock A uniform framework for substructural logics with modalities.
\newblock In {\em LPAR}, pages 435--455, 2017.

\bibitem{Ono}
Tomasz~Kowalski Nikolaos~Galatos, Peter~Jipsen and Hiroakira Ono.
\newblock {\em Residuated Lattices: An Algebraic Glimpse at Substructural Logics}.
\newblock Elsevier, 2007.

\bibitem{Ono90}
Hiroakira Ono.
\newblock Structural rules and a logical hierarchy.
\newblock {\em Mathematical logic}, pages 95--104, 1990.

\bibitem{onoketab}
Hiroakira Ono.
\newblock Proof-theoretic methods in nonclassical logic--an introduction.
\newblock {\em Theories of types and proofs}, 2:207--254, 1998.

\bibitem{Pitts}
Andrew~M Pitts.
\newblock On an interpretation of second order quantification in first order intuitionistic propositional logic.
\newblock {\em The Journal of Symbolic Logic}, 57(1):33--52, 1992.

\bibitem{shavrukov1993subalgebras}
Vladimir~Yu Shavrukov.
\newblock {\em Subalgebras of diagonalizable algebras of theories containing arithmetic}.
\newblock Polska Akademia Nauk, Instytut Matematyczny Warsaw, 1993.

\bibitem{tro92}
Anne~Sjerp Troelstra.
\newblock Lectures on linear logic.
\newblock 1992.

\bibitem{troelstra}
Anne~Sjerp Troelstra and Helmut Schwichtenberg.
\newblock {\em Basic proof theory}.
\newblock Number~43. Cambridge University Press, 2000.

\bibitem{visser1996bisimulations}
Albert Visser.
\newblock {\em Bisimulations, Model Descriptions and Propositinal Quantifiers}, volume 161.
\newblock Citeseer, 1996.

\bibitem{visser1996uniform}
Albert Visser et~al.
\newblock Uniform interpolation and layered bisimulation.
\newblock {\em G{\"o}del}, 96(6):139--164, 1996.

\end{thebibliography}

\end{document}